\documentclass[aps, superscriptaddress,reprint,amsmath,amssymb,
pra,showkeys]{revtex4-2}

\usepackage{graphics}
\usepackage{dcolumn}
\usepackage{bm}
\usepackage{overpic} 
\usepackage{array}
\usepackage{booktabs}
\usepackage{multirow}
\usepackage{esint}
\usepackage{color}
\usepackage{graphicx}
\usepackage{epstopdf}
\usepackage[version=4]{mhchem}
\usepackage{verbatim}
\usepackage{subfigure}
\usepackage{fancyvrb}
\usepackage[export]{adjustbox}
\usepackage{svg} 
\usepackage{comment} 
\usepackage{ esint }

\usepackage{tikz}
\usepackage{tikz-3dplot} 
\usepackage{pgfplots}
\usepackage{pgfplotstable}
\pgfplotsset{compat = newest} 
\usepgfplotslibrary{ternary, units} 
\usetikzlibrary{shapes.geometric, arrows, decorations.pathmorphing, pgfplots.ternary, pgfplots.units}
\usepackage{etoolbox}
\definecolor{c0}{RGB}{31, 119, 180}
\definecolor{c1}{RGB}{255, 127, 14}
\definecolor{c3}{RGB}{214, 39, 40}
\definecolor{m0}{HTML}{0072BD}
\definecolor{m1}{HTML}{D95319}
\definecolor{9tavg}{HTML}{1d8fff}
\definecolor{9tmax}{HTML}{acd7ef}
\definecolor{9mavg}{HTML}{007f00}
\definecolor{9mmax}{HTML}{8fed8f}
\definecolor{9bavg}{HTML}{926fda}
\definecolor{9bmax}{HTML}{e5e5f9}
\definecolor{91e2}{HTML}{959595}
\definecolor{91e3}{HTML}{ffa400}
\definecolor{91e4}{HTML}{ffd600}
\definecolor{67fb}{HTML}{b0d0f6}
\definecolor{67db}{HTML}{2875bb}
\definecolor{67fo}{HTML}{feb06c}
\definecolor{67do}{HTML}{dc4c00}

\newrobustcmd*{\mycircle}[1]{\tikz{\filldraw[draw=#1,fill=#1] (0,0) circle [radius=0.1cm];}}
\newrobustcmd*{\mysquare}[1]{\tikz{\filldraw[draw=#1,fill=#1] (0,0) rectangle (0.4,0.2);}}
\DeclareRobustCommand\dashedgrey{\tikz[baseline=-0.6ex]\draw[91e2,thick,dashed] (0,0)--(0.54,0);}
\DeclareRobustCommand\dashedorange{\tikz[baseline=-0.6ex]\draw[91e3,thick,dashed] (0,0)--(0.54,0);}
\DeclareRobustCommand\dashedyellow{\tikz[baseline=-0.6ex]\draw[91e4,thick,dashed] (0,0)--(0.54,0);}

\usepackage{etoolbox}
\definecolor{c0}{RGB}{31, 119, 180}
\definecolor{c1}{RGB}{255, 127, 14}
\definecolor{c3}{RGB}{214, 39, 40}
\definecolor{e_error}{HTML}{ff0000} 
\definecolor{dens_error}{HTML}{0000ff} 

\DeclareRobustCommand\dashedred{\tikz[baseline=-0.6ex]\draw[e_error,thick,dashed] (0,0)--(0.54,0);}
\DeclareRobustCommand\dashedblue{\tikz[baseline=-0.6ex]\draw[dens_error,thick,dashed] (0,0)--(0.54,0);}
\DeclareRobustCommand\dottedblue{\tikz[baseline=-0.6ex]\draw[dens_error,thick,dotted] (0,0)--(0.54,0);}
\DeclareRobustCommand\dottedred{\tikz[baseline=-0.6ex]\draw[e_error,thick,dotted] (0,0)--(0.54,0);}
\DeclareRobustCommand\fullred{\tikz[baseline=-0.6ex]\draw[e_error,thick] (0,0)--(0.5,0);}
\DeclareRobustCommand\fullblue{\tikz[baseline=-0.6ex]\draw[dens_error,thick] (0,0)--(0.5,0);}
\DeclareRobustCommand\chainred{\tikz[baseline=-0.6ex]\draw[e_error, thick,dash dot ] (0,0)--(0.5,0);}
\DeclareRobustCommand\chainblue{\tikz[baseline=-0.6ex]\draw[dens_error, thick,dash dot ] (0,0)--(0.5,0);}

\usepackage{colortbl}

\usepackage{xr-hyper}

\usepackage{soul,xcolor}
\setstcolor{red}

\newcommand{\RNum}[1]{\uppercase\expandafter{\romannumeral #1\relax}}
\usepackage{textcomp, gensymb}
\DeclareUnicodeCharacter{0308}{HERE!HERE!}
\DeclareUnicodeCharacter{2212}{-}

\usepackage{scalerel}
\usepackage{tikz}
\usetikzlibrary{svg.path}
\definecolor{orcidlogocol}{HTML}{A6CE39}
\tikzset{
  orcidlogo/.pic={
    \fill[orcidlogocol] svg{M256,128c0,70.7-57.3,128-128,128C57.3,256,0,198.7,0,128C0,57.3,57.3,0,128,0C198.7,0,256,57.3,256,128z};
    \fill[white] svg{M86.3,186.2H70.9V79.1h15.4v48.4V186.2z}
                 svg{M108.9,79.1h41.6c39.6,0,57,28.3,57,53.6c0,27.5-21.5,53.6-56.8,53.6h-41.8V79.1z M124.3,172.4h24.5c34.9,0,42.9-26.5,42.9-39.7c0-21.5-13.7-39.7-43.7-39.7h-23.7V172.4z}
                 svg{M88.7,56.8c0,5.5-4.5,10.1-10.1,10.1c-5.6,0-10.1-4.6-10.1-10.1c0-5.6,4.5-10.1,10.1-10.1C84.2,46.7,88.7,51.3,88.7,56.8z};
  }
}
\newcommand\orcidicon[1]{\href{https://orcid.org/#1}{\mbox{\scalerel*{
\begin{tikzpicture}[yscale=-1,transform shape]
\pic{orcidlogo};
\end{tikzpicture}
}{|}}}} 

\begin{document}
%

\newcommand{\rz}{\mathbb{R}}
\newcommand{\gz}{\mathbb{Z}}
\newcommand{\cz}{\mathbb{C}}
\newcommand{\qz}{\mathbb{Q}}
\newcommand{\nz}{\mathbb{N}}

\newcommand{\bfa}{{\bf a}}
\newcommand{\bfb}{{\bf b}}
\newcommand{\bfc}{{\bf c}}
\newcommand{\bfd}{{\bf d}}
\newcommand{\bfe}{{\bf e}}
\newcommand{\bff}{{\bf f}}
\newcommand{\bfg}{{\bf g}}
\newcommand{\bfh}{{\bf h}}
\newcommand{\bfi}{{\bf i}}
\newcommand{\bfj}{{\bf j}}
\newcommand{\bfk}{{\bf k}}
\newcommand{\bfl}{{\bf l}}
\newcommand{\bfm}{{\bf m}}
\newcommand{\bfn}{{\bf n}}
\newcommand{\bfo}{{\bf o}}
\newcommand{\bfp}{{\bf p}}
\newcommand{\bfq}{{\bf q}}
\newcommand{\bfr}{{\bf r}}
\newcommand{\bfs}{{\bf s}}
\newcommand{\bft}{{\bf t}}
\newcommand{\bfu}{{\bf u}}
\newcommand{\bfv}{{\bf v}}
\newcommand{\bfw}{{\bf w}}
\newcommand{\bfx}{{\bf x}}
\newcommand{\bfy}{{\bf y}}
\newcommand{\tbfy}{{\tilde{\bfy}}}
\newcommand{\bfz}{{\bf z}}
\newcommand{\bfA}{{\bf A}}
\newcommand{\bfB}{{\bf B}}
\newcommand{\bfC}{{\bf C}}
\newcommand{\bfD}{{\bf D}}
\newcommand{\bfE}{{\bf E}}
\newcommand{\bfF}{{\bf F}}
\newcommand{\bfG}{{\bf G}}
\newcommand{\bfH}{{\bf H}}
\newcommand{\bfI}{{\bf I}}
\newcommand{\bfJ}{{\bf J}}
\newcommand{\bfK}{{\bf K}}
\newcommand{\bfL}{{\bf L}}
\newcommand{\bfM}{{\bf M}}
\newcommand{\bfN}{{\bf N}}
\newcommand{\bfO}{{\bf O}}
\newcommand{\bfP}{{\bf P}}
\newcommand{\bfQ}{{\bf Q}}
\newcommand{\bfR}{{\bf R}}
\newcommand{\bfS}{{\bf S}}
\newcommand{\bfT}{{\bf T}}
\newcommand{\bfU}{{\bf U}}
\newcommand{\bfV}{{\bf V}}
\newcommand{\bfW}{{\bf W}}
\newcommand{\bfX}{{\bf X}}
\newcommand{\bfY}{{\bf Y}}
\newcommand{\bfZ}{{\bf Z}}
\newcommand{\vphi}{{\varphi}}
\newcommand{\eps}{{\varepsilon}}
\newcommand{\Nhat}{\hat{\mbox{\tiny {\bf N}}}}
\newcommand{\ehat}{\hat{\bf e}}
\newcommand{\nhat}{\hat{\bf n}}
\newcommand{\uhat}{\hat{\bf u}}
\newcommand{\phihat}{\hat{\varphi}}
\newcommand{\xihat}{\hat{\xi}}
\newcommand{\fhat}{\hat{f}}
\newcommand{\Vhat}{\hat{V}}
\newcommand{\adj}{{\mbox{adj }}}
\newcommand{\beq}{\begin{equation}}
\newcommand{\eeq}{\end{equation}}
\newcommand{\beqs}{\begin{eqnarray}}
\newcommand{\eeqs}{\end{eqnarray}}
\newcommand{\beql}{\begin{equation} \label}

\newcommand{\normp}[1]{\| #1 \|}
\newcommand{\brho}{\boldsymbol{\rho}}
\newcommand{\expchar}[2]{\textrm{e}^{\frac{2\pi \textrm{i} #1}{#2}}}
\newcommand{\expcharconj}[2]{\textrm{e}^{-\frac{2\pi \textrm{i} #1}{#2}}}

\newcommand{\half}{\frac{1}{2}}
\newcommand{\calA}{{\cal A}}
\newcommand{\calB}{{\cal B}}
\newcommand{\calC}{{\cal C}}
\newcommand{\calD}{{\cal D}}
\newcommand{\calE}{{\cal E}}
\newcommand{\calF}{{\cal F}}
\newcommand{\calG}{{\cal G}}
\newcommand{\calH}{{\cal H}}
\newcommand{\calI}{{\cal I}}
\newcommand{\calJ}{{\cal J}}
\newcommand{\calK}{{\cal K}}
\newcommand{\calL}{{\cal L}}
\newcommand{\calM}{{\cal M}}
\newcommand{\calN}{{\cal N}}
\newcommand{\calO}{{\cal O}}
\newcommand{\calP}{{\cal P}}
\newcommand{\calQ}{{\cal Q}}
\newcommand{\calR}{{\cal R}}
\newcommand{\calS}{{\cal S}}
\newcommand{\calT}{{\cal T}}
\newcommand{\calU}{{\cal U}}
\newcommand{\calV}{{\cal V}}
\newcommand{\calW}{{\cal W}}
\newcommand{\calX}{{\cal X}}
\newcommand{\calY}{{\cal Y}}
\newcommand{\calZ}{{\cal Z}}

\newcommand{\hilb}{\mathsf{H}}
\newcommand{\thilb}{\tilde{\mathsf{H}}}
\newcommand{\hamil}{\mathfrak{H}}
\newcommand{\thrbyto}{\frac{3}{2}}
\newcommand{\Fhat}{\hat{F}}
\newcommand{\tlambda}{\tilde{\lambda}}
\newcommand{\fivbyto}{\frac{5}{2}}

%
%
%
%


\newenvironment{myproof}[1][Proof:]{\begin{trivlist}
\item[\hskip \labelsep {\bfseries #1}]}{\qed\end{trivlist}}

\newenvironment{myexample}[1][Example:]{\begin{trivlist}
\item[\hskip \labelsep {\bfseries #1}]}{\end{trivlist}}

\newenvironment{myremarks}[1][Remarks:]{\begin{trivlist}
\item[\hskip \labelsep {\bfseries #1}]}{\end{trivlist}}

\newenvironment{myremark}[1][Remark:]{\begin{trivlist}
\item[\hskip \labelsep {\bfseries #1}]}{\end{trivlist}}


\newcommand{\abs}[1]{\lvert#1\rvert}
\newcommand{\norm}[2]{\lVert#1\rVert_{#2}}
\newcommand{\innprod}[3]{\langle#1,#2\rangle_{#3}}
\newcommand{\biginnprod}[3]{\mathbf{\Big\langle}#1,#2\mathbf{\Big\rangle}_{#3}}
\newcommand{\esssup}[2]{\displaystyle\text{ess sup}_{#1}#2}
\newcommand{\Lpspc}[3]{\textsf{L}^{#1}_{#2}(#3)}
\newcommand{\infm}[2]{\displaystyle\inf_{#1}{#2}}
\newcommand{\supm}[2]{\displaystyle\sup_{#1}{#2}}
\newcommand{\argmin}[2]{\displaystyle\textrm{argmin}_{#1}{#2}}
\newcommand{\argmax}[2]{\displaystyle\textrm{argmax}_{#1}{#2}}
\newcommand{\mspan}[1]{\text{span}\{#1\}}
\newcommand{\pd}[2]{\frac{\partial#1}{\partial#2}}
\newcommand{\hpd}[3]{\frac{\partial^{#3}#1}{\partial#2^{#3}}}
\newcommand{\od}[2]{\frac{d#1}{d#2}}
\newcommand{\hod}[3]{\frac{d^{#3}#1}{d#2^{#3}}}
\newcommand{\isom}[2]{(#1|#2)}
\newcommand{\orb}[2]{\text{Orb}(#1,#2)}
\newcommand{\stab}[2]{\text{Stab}(#1,#2)}
\newcommand{\ball}[2]{\calB_{#1}(#2)}
\newcommand{\ccf}[1]{\mathsf{C}_{\mathsf{c}}(#1)}
\newcommand{\spt}[1]{\text{spt}.(#1)}
\newcommand{\mattr}[1]{\text{Tr.}(#1)}

\let\oldFootnote\footnote
\newcommand\nextToken\relax

\renewcommand\footnote[1]{%
    \oldFootnote{#1}\futurelet\nextToken\isFootnote}

\newcommand\isFootnote{%
    \ifx\footnote\nextToken\textsuperscript{,}\fi}
    
\newcommand{\dbltilde}[1]{\accentset{\approx}{#1}}

\title{Electronic structure prediction of medium and high entropy alloys across composition space}

\author{Shashank Pathrudkar \orcidicon{0000-0001-8546-8056}}
\affiliation{Department of Mechanical and Aerospace Engineering, Michigan Technological University}
\author{Stephanie Taylor \orcidicon{0000-0000-0000-0000}}
\affiliation{Department of Materials Science and Engineering, University of California, Los Angeles, CA 90095, USA}
\author{Abhishek Keripale \orcidicon{0009-0009-2127-6536}}
\affiliation{Department of Mechanical and Aerospace Engineering, Michigan Technological University}
\author{Abhijeet S. Gangan \orcidicon{0000-0000-0000-0000}}
\affiliation{Department of Materials Science and Engineering, University of California, Los Angeles, CA 90095, USA}
\author{Ponkrshnan Thiagarajan \orcidicon{0000-0003-3946-3902}}
\affiliation{Department of Mechanical and Aerospace Engineering, Michigan Technological University}
\author{Shivang Agarwal \orcidicon{0000-0001-9231-5461}}
\affiliation{Department of Electrical and Computer Engineering, University of California, Los Angeles, CA 90095, USA}
\author{Jaime Marian \orcidicon{0000-0001-9000-3405}}
\affiliation{Department of Materials Science and Engineering, University of California, Los Angeles, CA 90095, USA}
\author{Susanta Ghosh \orcidicon{0000-0002-6262-4121}}
\email{susantag@mtu.edu}
\affiliation{Department of Mechanical and Aerospace Engineering, Michigan Technological University}
\affiliation{Center for Artificial Intelligence, Michigan Technological University}
\author{Amartya S. Banerjee \orcidicon{0000-0001-5916-9167}}
\email{asbanerjee@ucla.edu}
\affiliation{Department of Materials Science and Engineering, University of California, Los Angeles, CA 90095, USA}
\date{October 7, 2024}

\begin{abstract}
\section*{Abstract}
We propose machine learning (ML) models to predict the electron density --- the fundamental unknown of a material's ground state --- across the composition space of concentrated alloys. From this, other physical properties can be inferred, enabling accelerated exploration. A significant challenge is that the number of descriptors and sampled compositions required for accurate prediction grows rapidly with species. To address this, we employ Bayesian Active Learning (AL), which minimizes  training data requirements by leveraging uncertainty quantification capabilities of Bayesian Neural Networks. Compared to strategic tessellation of the composition space, Bayesian-AL reduces the number of training data points by a factor of $2.5$ for ternary (\ce{SiGeSn}) and $1.7$ for quaternary (\ce{CrFeCoNi}) systems. We also introduce easy-to-optimize, body-attached-frame descriptors, which respect physical symmetries while keeping descriptor-vector size nearly constant as alloy complexity increases. Our ML models demonstrate high accuracy and generalizability in predicting both electron density and energy across composition space. 
\end{abstract}
\maketitle

\section{Introduction}
Electronic structure calculations, based on Kohn-Sham Density Functional Theory (KS-DFT) \cite{hohenberg1964inhomogeneous, kohn1965self, Martin_ES} serve as the workhorse of computational materials science simulations. The fundamental unknown in KS-DFT calculations is the ground state electron density, from which a wealth of material information --- including structural parameters, elastic constants, and material stability (e.g. phonon spectrum) --- may be inferred. Compared to more elaborate wave-function based quantum chemistry methods or simpler electronic structure techniques based on tight-binding, KS-DFT often offers a good balance between physical accuracy, transferability and computational efficiency, leading to its widespread use \cite{hafner2006toward}.

In spite of its many successes, KS-DFT is often practically limited by its cubic scaling computational cost with respect to the number of simulated atoms. While calculations involving just a few atoms within the computational unit cell can be executed with ease --- making high-throughput screening \cite{saal2013materials, emery2017high, choudhary2020high} and large scale materials data repositories possible (e.g. the Materials Project \cite{jain2013commentary, jain2020materials}) --- larger calculations often need to employ extensive high-performance computing resources or specialized solution techniques \cite{gavini2023roadmap, goedecker1999linear, banerjee2016chebyshev, banerjee2018two, motamarri2014subquadratic, lin2014siesta, dogan2023real}. Thus, routine calculations of a wide variety of important materials problems, e.g. the behavior of defects at realistic concentration \cite{gavini2007vacancy} and simulations of moir\`{e} superlattices \cite{carr2020electronic}, continue to be far from routine, or altogether computationally infeasible, with state-of-the-art KS-DFT implementations. Along these lines, simulations of disordered solids \cite{jaros1985electronic, wei1990electronic}, specifically, multi-element concentrated alloys featuring chemical disorder, represent a significant challenge. Indeed, the computational unit cell required to simulate medium and high entropy alloys at generic compositions, can get arbitrarily large, with the number of simulated atoms growing proportionally high. Thus, in spite of the technological relevance of such materials \cite{george2019high}, direct first-principles evaluation of their material properties over the entire composition space often remains computationally out of reach, unless approximations in KS-DFT calculations or special structural sampling techniques are used \cite{wang2021comparison, tian2016calculating, karabin2022ab, gao2016applications}. 

{Recently, electronic structure predictions using machine learning (ML) have gained a lot of attention and shown promise for various systems. The vast majority of such studies have focused on prediction of the electron density field \cite{lewis2021learning, jorgensen2022equivariant, zepeda2021deep, chandrasekaran2019solving, fiedler2023predicting, brockherde2017bypassing, del2023deep}, although a number of studies have also carried out predictions of the single and two particle density matrices \cite{tang2024improving, shao2023machine, hazra2024predicting, sager2022reducing}. In essence, ML techniques for field prediction serve as surrogate models for KS-DFT, enabling inexpensive evaluation of the electron density and related fields \cite{teh2021machine} from atomic configurations, once trained.} The predicted density can be used to compute various other downstream quantities, including the system's energy \cite{del2023deep}, electronic band diagrams \cite{pathrudkar2022machine} or properties of defects \cite{arora2022charge}. Some of these ML models use global system descriptors, e.g. strains commensurate with the system geometry \cite{teh2021machine, pathrudkar2022machine}, and are trained on KS-DFT data generated using specialized symmetry-adapted simulation techniques \cite{banerjee2021ab, yu2022density, ghosh2019symmetry}. The vast majority however, employ descriptors of the local atomic environment and are trained on KS-DFT data from standard codes, e.g. ones based on plane-waves. The output of the ML model, i.e., the electron density itself, can be represented in different ways. One strategy involves expanding the density as a sum of atom-centered basis functions \cite{grisafi2018transferable, fabrizio2019electron, jorgensen2022equivariant, fu2024recipe, qiao2022informing, del2023deep}, while another predicts the electron density at each grid point within a simulation cell \cite{pathrudkar2024electronic, jorgensen2022equivariant, gong2019predicting, chandrasekaran2019solving, pope2024towards, zhang2023design, zhang2023rational}. The first strategy is efficient but can be less accurate, as complex electron densities may not always be representable with a small number of basis functions. The second strategy is accurate but computationally expensive, as it requires ML model evaluation over a fine mesh of the simulation cell. However, it is amenable to easy parallelization based on domain decomposition and the evaluation process scales linearly with the system size \cite{pathrudkar2024electronic, fiedler2023predicting}. Yet another recent approach \cite{li2024image} predicts the entirety of the electron density field, using superposition of the atomic densities (SAD) as the input. This approach is efficient, since it can use a convolutional model to predict the electron density over a volume, avoiding  tedious grid point-wise inference. This approach is also accurate as it incorporates materials physics through the SAD. However, this method does not inherently accommodate the system's rotational symmetries, and integrating uncertainty quantification (UQ) features presents a challenge --- both aspects that are more readily addressed by the other approaches. {Finally, equivariant graph neural networks offer an elegant, end-to-end alternative that learns symmetry-preserving representations directly on atomistic graphs, and have been used for a variety of computational tasks, including electron density \cite{koker2023higher, jorgensen2022equivariant} and phonon-spectrum prediction \cite{okabe2024virtual}. In graph-based models the descriptors are not specified a priori but are learned during training. This flexibility often entails higher inference cost per structure --- particularly in high-throughput settings \cite{fung2021benchmarking, jiang2021could}.}

While previous studies have carried out ML-based electron density predictions for various molecular  systems, pure bulk metals and some specific alloys \cite{zepeda2021deep, lewis2021learning, chandrasekaran2019solving, fiedler2023predicting, pathrudkar2024electronic, eisenbach2020solidsolution, li2024image, medasani2016predicting}, the issue of electron density prediction for arbitrary compositions of concentrated multi-element alloys has not yet been addressed. Indeed, ML techniques have been applied to a variety of other properties of such systems \cite{greeley2024active, freitas2024equivariant, yang2024eutecticml, vazquez2023regressor, giles2022rhea, chen2022compdesign}, but the ability to predict their  electron density, i.e., the fundamental unknown of the material's ground state, remains an attractive unattained goal. Such predictive capabilities, if realized, may help overcome the aforementioned limitations of KS-DFT in simulating medium and high entropy alloys, and in turn, help accelerated exploration of new materials, e.g. alloys for next-generation microelectronics and novel magnetic storage systems \cite{gao2018high, kumari2022comprehensive}.  The key challenge to predicting fields such as the electron density for concentrated multi-element alloy systems, is that due to combinatorial reasons, the number of compositions which need to be sampled for development of accurate ML models can be very high. Hence the cost of data generation for developing ML models that work equally well across the composition space, also tends to be very high. Therefore, an open question is whether it is possible to produce accurate predictions for the entire composition space of multi-element alloys while limiting the data required to train the ML model. Indeed, compared to low-dimensional material parameters, such as elastic moduli or thermal expansion coefficients \cite{dai2020theoretical}, these data-related challenges can be far more severe for predicting fields.  

In recent years, significant progress has been made in using machine learning for high entropy alloys (HEAs), particularly, with the aid of machine learning interatomic potentials (MLIPs) \cite{deringer2019machine, kormann2021b2}. Many of these studies rely on highly exhaustive sets of training data \cite{byggmastar2021modeling, pandey2022machine, you2024principal}. Although these works present accurate MLIPs, the extensive training data required to achieve such accuracy is a limitation. For instance, the Mo–Nb–Ta–V–W training data set from Ref. \cite{byggmastar2021modeling} includes single isolated atoms, dimers, pure elements, binary to quinary \textit{bcc} alloys, equiatomic HEAs, and ordered/disordered structures. Additionally, the dataset covers liquid alloys, vacancies, and interstitial atoms. In \cite{pandey2022machine}, data for quaternary MoNbTaW is generated via \textit{ab initio} molecular dynamics (AIMD) for random alloy compositions at 500 K, 1000 K, and 1500 K, with 2\% variation in lattice parameters, and single point calculations involve random alloys with 2\% variation in volume and lattice angles. Along the same lines, in \cite{you2024principal}, in order to develop an interatomic potential for Lithium lanthanum zirconium oxide (LLZO) systems, the training set consisted of three components: (1) elemental materials and scaled structures for \ce{Li}, \ce{La}, \ce{Zr}, and \ce{O}; (2) structures from first-principles molecular dynamics simulations of LLZO crystals and amorphous phases at various temperatures; and (3) a two-body potential to constrain interatomic distances during molecular dynamics simulations. These different examples serve to highlight the fact that although it is possible to develop accurate interatomic potentials for medium to high entropy alloys, the training set often requires a large amount of static KS-DFT and AIMD simulations. Our work aims at accurately predicting the electron density of HEAs across the composition space while limiting the number of KS-DFT/AIMD simulations required to generate the training data.  

One major criticism of machine learning models is their lack of generalization, i.e., their inability to predict beyond the training data accurately. Indeed, the use of a large number of different configurations for generating  training data of MLIPs as described above, is also related to improving generalizability. In a recent work  \cite{pathrudkar2024electronic}, the authors demonstrated that utilization of data generated at high temperatures and the ensemble averaging nature of Bayesian Neural Networks can enhance the generalization ability of ML-based electron density prediction. This approach yielded highly accurate predictions for bulk aluminum (Al) and silicon germanium (\ce{SiGe}) systems. More importantly, it exhibited generalization capability by accurately predicting a variety of test systems with structural features not included in the training data, such as edge and screw dislocations, grain boundaries, and mono-vacancy and di-vacancy defects. This model was also shown to be capable of generalizing to  systems significantly larger than those used for training and can reliably predict the electron density for multi-million-atom systems using only modest computational resources. The potential of this ML electron density model to generalize to arbitrary alloy compositions is explored in this work. As a starting point, we found that for the \ce{SiGe} system, learning the binary alloy electron density at a fixed composition allows for reasonably accurate extrapolation to nearby compositions. This raises the question of whether such extrapolation applies to more complex systems, and if so, the minimum data needed to learn across composition space. We explore these questions here, in the context of ternary \ce{SiGeSn} and quaternary \ce{CrFeCoNi} systems.

Medium entropy alloy (MEA) and high entropy alloy (HEA) systems provide an opportunity to expose our models to a compositionally complex materials space. 
Thus, after investigating \ce{SiGe}, it was a natural choice to extend to the ternary system SiGeSn. Group IV alloys in the Si-Ge-Sn system are of great interest to the optoelectronics  industry, due to their utility for bandgap engineering. Notably, the addition of Sn is purported to lower the bandgap and produce a indirect-to-direct bandgap transition whose location is tunable within the SiGeSn composition space \cite{li2022sigesn, cao2020gesn}. The primary challenge related to the implementation and usage of ternary SiGeSn is that it is difficult to synthesize many of the compositions experimentally \cite{wirths2016sigesn}. The \ce{SiGe} phase diagram shows that Si and Ge are fully soluble in each other \cite{zhuang2019sigesn, olesinski1984sige}. In contrast, Sn is barely soluble in Si or Ge; it can be difficult to obtain compositions above a few percentage. Despite this, recent research developments have continued to push the limit of Sn incorporation \cite{buca2023sigesn}. In light of the experimental progress towards synthesizing such systems, there is interest in predicting the composition windows to aim for with respect to obtaining desired property targets, and this continues to be an active area of  research \cite{wang2019semiconducting, wang2020electrical} --- thus motivating our choice. In addition to \ce{SiGeSn}, we also wished to test how our methodology performs against a more challenging bulk metallic alloy system. Given the Cantor alloy's status as the most well-studied HEA to date, we selected a quaternary Cantor alloy variant \ce{CrFeCoNi}, and explored it across composition space. {We also investigated a more traditional quinary HEA, \ce{AlCrFeCoNi}, near equiatomic composition, for the sake of completeness.} The quaternary alloy system is much easier to experimentally synthesize, as it forms solid solution phases more readily. Its mechanical properties --- notably the high ductility and fracture toughness --- have led to a large volume of research studies focusing on this system \cite{cantor2021multicomponent}. Furthermore, \ce{CrFeCoNi} has also received interest in the field of nuclear materials for its high damage tolerance under irradiation; for instance, defect growth in \ce{CrFeCoNi} is over 40 times slower compared to pure Ni \cite{he2016irrad}.  Interestingly, despite the vast quantity of HEA research, the overwhelming majority of studies have tended to solely focus on equiatomic compositions (such as  $\text{Cr}_{0.25}\text{Fe}_{0.25}\text{Co}_{0.25}\text{Ni}_{0.25}$).  This is a bit surprising, considering that the idea of exploiting the high degree of freedom in compositional space for improved property design has been around since the beginning of the field. Yet, as case studies have emerged demonstrating that improved mechanical properties can be obtained with non-equiatomic HEA systems, interest along this direction has grown.  Currently, there exists a great deal of research momentum towards moving beyond equiatomic compositions and exploring material property maps across composition space, ultimately motivating our choice of this alloy system.

To address these complex alloy systems, we employed the following three key strategies to achieve highly accurate and reliable predictions across composition space while minimizing the required training data. The schematics of our proposed ML model is shown in Figure \ref{fig:model_schematic}.  

\textit{First}, we developed an uncertainty quantification (UQ)-based Active Learning (AL) approach for the electron density to select the most informative compositions and add them to the training data in each iteration, aiming to minimize the overall training data. The UQ capability of the Bayesian Neural Network is utilized to efficiently quantify uncertainty; hence, this AL approach is referred to as Bayesian Active Learning (Bayesian-AL). The compositions corresponding to the highest uncertainty are considered the most informative for the next iteration of AL.

\textit{Second}, we introduced novel descriptors for which the descriptor-vector size does not increase significantly with the number of alloy elements. The sizes of many existing descriptors rapidly increases with the number of distinct chemical elements in the system, which is a key challenge for multi-element alloy systems \cite{lei2022universal, Timmerman2024JCTC}. Our descriptors are position vectors in a body-attached frame and incorporate species information through the atomic number. Thus they do not depend on the number of distinct chemical elements that may be present, for a fixed number of atoms in the neighborhood. Furthermore, our descriptors also facilitate the selection of the optimal set of descriptors.

\textit{Third}, we trained our model on the difference between total densities and atomic densities, rather than solely on total densities. Observing that a model trained just on superposition of atomic densities (SAD) can obtain nearly $85$\% accuracy in density prediction \cite{li2024image}, we presumed that using the difference between total densities and atomic densities will allow for a higher resolution description of the chemical bonding in our model. In other words, if the complexity of the quantum-mechanical chemical bonding environment contributes about only about $15$\% accuracy overall, then training the model on the difference between total and atomic densities should help to improve its sensitivity to the fundamental chemistry present in a given system. In light of this,  we have trained a separate ML model to predict the difference between the electron density and the SAD, which we refer to as the 
$\delta\rho$ ML model. This model is found to be more accurate in energy predictions for {\ce{CrFeCoNi}} systems (which involve elements with hard pseudopotentials and semi-core states), in line with the above reasoning. 

These three methodological innovations ultimately resulted in highly accurate ML models, generalizable across the full composition space of the respective alloy systems, as demonstrated in the following Results section. {Additional results involving a high entropy quinary system (\ce{AlCrFeCoNi}) are presented in the Supplemental Materials.} We also note that our contribution is quite exhaustive, in that a whole plethora of ML models --- involving different materials systems (i.e., binary, ternary and quaternary alloys), different levels of Bayesian Active Learning, different levels of tessellation based training, and different predicted quantities (i.e., $\rho$ and $\delta \rho$ based models)  --- were carefully developed and extensively tested. The high quality predictions obtained by our ML models give us confidence that the techniques described above can be easily extended to other bulk high-entropy materials, or emergent low-dimensional functional materials featuring chemical complexity and disorder, e.g. high entropy MXenes \cite{nemani2021high, nemani2023functional} and high entropy 2D transition metal dichalcogenides \cite{deshpande2022entropy}. 

\textit{Finally}, when required, we accelerated the data generation process by judiciously integrating ML interatomic potentials with KS-DFT calculations, in lieu of full \textit{ab initio} molecular dynamics simulations. This further accelerates the development of our ML models. 

\section{Results}

This section evaluates the accuracy of the proposed machine learning (ML) model in comparison to the ground-truth, i.e., KS-DFT. Since the focus of this work is on electron density prediction for alloys, three systems have been considered as prototypical examples: a binary alloy --- \ce{Si_xGe_{1-x}}, a medium entropy ternary alloy --- \ce{Si_{x}Ge_{y}Sn_{1-x-y}}, and a high entropy quaternary alloy --- \ce{Cr_{x}Fe_{y}Co_{z}Ni_{1-x-y-z}}.  Though the developed ML framework should be applicable to any alloy with any number of elemental species, we present results for the aforementioned technologically important alloys \cite{buca2023sigesn, li2022sigesn, zhuang2019sigesn, wirths2016sigesn}, \cite{cantor2021crfeconi, wang2017crfeconi, sen2024crfeconi, chen2022crfeconi, cai2022crfeconi, zhong2021crfeconi, ramamurty2021crfeconi, tuomisto2020crfeconi, dugdale2020crfeconi, kim2019crfeconi, kim2018crfeconi, irving2016crfeconi}. 
The error in electron density prediction is measured using two metrics: Normalized Root Mean Squared Error (NRMSE) and relative $L_\text{1}$ error (\% $L_1$) \cite{zepeda2021deep} (see  Supplemental Material for further details).

%
At the onset, we made an attempt to develop an ML model that is accurate for all compositions of a binary alloy. It is found that a model trained with equiatomic \ce{SiGe} (Si$_x$Ge$_{1-x}$ with $x = 0.5$) achieves high accuracy in the vicinity of the training composition ($x = 0.5$), as illustrated in Figure \ref{fig:sige_rel_l1}(a). 
However, the error grows as the distance between the training and testing compositions increases in the composition space. If only two compositions that have the highest error are added to the training data the accuracy increases across the entire composition space, as shown in Figure \ref{fig:sige_rel_l1}(b). 
This experiment demonstrates that retraining the ML model with the addition of a few compositions with the highest error enables accurate prediction across the entire composition space.
However, as the number of alloying elements increases, the number of possible compositions in the composition space grows rapidly, making it challenging to simulate all compositions through KS-DFT. Therefore, the errors for all compositions will not be available to identify the most erroneous compositions to include in the next round of training.
To address the aforementioned challenge, we propose two systematic iterative training approaches for selecting optimal compositions for training the model: (i) an Uncertainty Quantification (UQ)-based Active Learning technique (referred to as Bayesian Active Learning) and (ii) a Tessellation-based iterative training technique.

\subsection{Minimizing the Training Data: Bayesian Active Learning and Tessellation}  
In this section, we compare the performance of the Bayesian Active Learning approach and the Tessellation-based iterative training approach. 
The Tessellation approach involves a systematic, progressively refined discretization of the composition space to obtain training compositions. In contrast, the Bayesian Active Learning, uses uncertainty measures to identify the most informative training compositions, thereby bypassing the need for knowledge of errors at all compositions. 

The training compositions obtained through progressively refined tessellation-based discretization of the composition space are shown in Figure~\ref{fig:T1toT4}. 
For the tessellation-based ML models, T1, T2, and T4 contain 3, 6, and 15 training compositions for ternary (e.g. SiGeSn) systems, and 4, 11, and 34 training compositions for quaternary (e.g. CrFeCoNi) systems, respectively. 

In the case of Bayesian Active Learning, we iteratively add alloy compositions to the training set.
For the ternary system, Bayesian AL starts with three training compositions as shown by white circles in Figure \ref{fig:AL4figs}\,a. This model is referred to as the AL1 and the errors in  the $\rho$ and energy for model AL1 are shown in Figures~\ref{fig:AL4figs}\,a,\,b. Based on the Uncertainty measure, shown in Figure \ref{fig:AL4figs}\,c, three additional training compositions corresponding to the highest uncertainty are chosen and are added to the training set. The model trained with these six training compositions, is referred to as AL2 and the errors  in  the $\rho$ and energy for model AL2 are shown in Figures~\ref{fig:AL4figs}\,d,\,e. Further details on these error are given in  Figure S6
\,d,\,e of the Supplemental Material. Similarly, for quaternary system, the training compositions used in Bayesian Active Learning models AL1, AL2 and AL3 are shown in Figure \ref{fig:AL_queries_quat}. 
Note that the errors in the electron density are computed for all compositions to illustrate the variation across the composition space. Although all compositions are simulated for error calculation, only a fraction of them are used for training, as shown in  
 Figure~\ref{fig:T1toT4} and Figure~\ref{fig:AL_queries_quat}.
Detailed explanations of both the Bayesian Active Learning and Tessellation approaches can be found in the Methods section.  

For the ternary SiGeSn alloy, errors in the electron density across the composition space for each iteration of both approaches are presented in Figure~\ref{fig:max_nrmse_al_vs_ts}. 
The initial iteration for both the Bayesian Active Learning (AL1) and Tessellation (T1) approaches is identical, as they each begin with 3 training compositions containing the pure elements silicon, germanium, and tin. Bayesian Active Learning requires only 6 training compositions (in AL2) to achieve slightly greater accuracy compared to the 15 needed by the Tessellation approach (in T4). The Tessellation approach performs well, requiring only 15 compositions to accurately predict across the composition space. However, the AL approach demonstrates superior efficiency compared to the systematic Tessellation method.
The error in energy for each iteration of both approaches are shown in Figure~\ref{fig:ModelPlots}. The Bayesian Active Learning based model trained on 6 compositions (AL2) is enough to obtain chemically accurate energy predictions. Thus, for the ternary system, Bayesian Active Learning achieves a reduction by factor of 2.5 in the cost of data generation compared to Tessellation. 

Similarly, the results for the quaternary alloy, CrFeCoNi, are shown in Figure~\ref{fig:max_nrmse_al_vs_ts}. 
The initial iteration for both the Bayesian Active Learning (AL1) and Tessellation (T1) approaches is identical, as they each begin with 4 training compositions containing the pure elements chromium, iron, cobalt, and nickel. Bayesian Active Learning requires only 20 training compositions (in AL3) to achieve much better accuracy compared to the 34 needed by the Tessellation approach (in T4).  The error in energy for each iteration of both approaches are shown in Figure~\ref{fig:ModelPlots}. For Bayesian Active Learning, 20 compositions (AL3) are sufficient to achieve energy predictions as accurate as those obtained with the Tessellation approach using 34 compositions (T4). These results for the quaternary system further demonstrate that while Tessellation is a reasonable approach, Bayesian Active Learning offers a significant advantage, reducing the cost of data generation by a factor of $1.7$ compared to Tessellation. 
Even though only 34 out of 69 points are on the boundary, the training points in AL2 and AL3 for the quaternary system are mostly positioned on the boundary of the composition space, with the exception of one point, see Figure~\ref{fig:AL_queries_quat}. This suggests that the points on the boundaries contain more valuable information for the ML model to learn from. 

\subsection{Generalization}

To showcase generalization capabilities of the model we tested the model on various test cases that are not used in the training and often significantly different form the training data,  including (i) systems with compositions not used in training, (ii) systems with vacancy defects,  (iii) `checkerboard' systems with clusters of atoms from the same species. 
For all these test systems, we assess the error in density prediction, as well as the error in energy obtained by postprocessing the predicted densities. 
Relative $L_1$ errors in prediction of $\rho$ for these testing cases are shown in Figure \ref{fig:KeyResultsAccuracy} for both ternary and quaternary alloys. For the ternary alloy, the model was trained on 64-atom systems, whereas for the quaternary alloy, the model was trained on 32-atom systems. 

\underline{Generalization across composition space}:
The prime objective of the ML model is to accurately predict electron density across the composition space while using only a small fraction of compositions for training. If successful, this approach would allow for the estimation of any property of interest for a given alloy at any composition. By leveraging fast ML inference, the vast composition space of multi-principal element alloys can be explored much more quickly than with conventional KS-DFT methods. 

To demonstrate the generalizability of the model beyond the training composition, the electron  density for a 64-atom \ce{SiGeSn} system is predicted across 45 distinct compositions spanning the entire composition space. The AL2 model uses only 6 out of these 45 SiGeSn compositions for training. The prediction errors for the 64-atom system are shown in Figure~\ref{fig:AL4figs} (and Figure S6 
of the Supplemental Material). For better readability, the values of the density and energy errors are shown for each composition in Figure S10 
of the Supplemental Material. The average energy error is $4.3 \times 10^{-4}$ Ha/atom, which  is well within chemical accuracy. To evaluate compositions that are not feasible to simulate with the 64-atom system, additional test compositions were generated using a 216-atom \ce{SiGeSn} system, as shown in Figure S1 
of the Supplemental Material. The errors in the electron density and energy for the 216-atom \ce{SiGeSn} system are presented in Figure S4 
of the Supplemental Material. The energy errors for these systems too are well within chemical accuracy, on average. Additionally, the errors in the electron density and energy for these 216-atom systems are of the same magnitude as those for the 64-atom systems, indicating generalizability to systems of larger size. 

The generalizability of the ML model beyond training compositions is also tested for the quaternary system, \ce{CrFeCoNi}, by evaluating the error in electron density predictions across the composition space, as shown in Figure \ref{fig:Quat_SAD_EnergyErrors}(a). 
Note that the AL2 model uses only 10 out of these 69 \ce{CrFeCoNi} compositions for training.
The error in the energy obtained from the predicted electron density for \ce{CrFeCoNi} system across the composition space are shown in Figure \ref{fig:Quat_SAD_EnergyErrors}(b). The AL3 model displays further improvement; for better readability, the values for energy errors are shown in Figure S11 
of the Supplemental Material. The average energy errors ($2.3 \times 10^{-3}$ Ha/atom) are very close to chemical accuracy, and ``worst case'' predictions ($3.5 \times 10^{-3}$ Ha/atom) are only slightly worse. A visualization of the difference between the KS-DFT-calculated and ML-predicted electron densities for the \ce{SiGeSn} and \ce{CrFeCoNi} systems are shown in Figure \ref{fig:cubeplots}.

The aggregated electron density and energy errors for the \ce{SiGeSn} and \ce{CrFeCoNi} systems  are shown in Figure \ref{fig:KeyResultsAccuracy}. On average, the errors in energy per atom for the quaternary systems are somewhat higher compared to the predictions of the ternary alloy cases. However, the atoms involved in the ternary system also have significantly more electrons per atom. Upon normalizing the energy errors in terms of the number of electrons in the simulation, the energy errors for the quaternary system ($\rho - \text{SAD}$ or $\delta \rho$ model) is found to be comparable to the errors for ternary systems (of the order of $10^{-4}$ Ha/electron, on average), as shown in Figure \ref{fig:KeyResultsAccuracy}. Overall, the low errors in prediction of electron density and energy for binary, ternary and quaternary alloy across the entire composition space demonstrate the generalization capacity of the proposed ML model.

\underline{Generalization to systems with defects}: 
We assess the performance of the ML model on systems containing localized defects, such as mono-vacancies and di-vacancies, even though the training was conducted exclusively on defect-free systems. The electron density fields predicted by the ML model match remarkably  well with the KS-DFT calculations, with error magnitudes for defective systems comparable to those for pristine systems, as shown in  Figure~\ref{fig:KeyResultsAccuracy}. Further details on the match between the ML-predicted and KS-DFT-obtained $\rho$ fields are provided in Figure S7 
of the Supplemental Material.  In addition to accurately predicting electron density, the energy errors remain within chemical accuracy. Note that for these systems, the atomic configurations away from the defects are quite close to the equilibrium configuration (see Figure S7 
of the Supplemental Material), resulting in very low errors in the ML predictions away from the defects. Consequently, the overall error remains low.


\underline{Generalization to handcrafted systems featuring} \underline{species segregation}: 
In multi-element alloys, species segregation naturally occurs, leading to the formation of element-enriched regions within the alloy \cite{wirths2016sigesn, middleburgh2014segregatio}. Therefore, it is important to evaluate model for these systems. Towards this, handcrafted systems featuring species segregation are created. Cubic simulation cells of 64 and 216 atoms occupying diamond lattice sites are divided up into smaller cubic sub-regions, i.e. either 8 bins ($2\,\times\,2\,\times\,2$) for the 64-atom and 216-atom cells, or 27 bins ($3\,\times\,3\,\times\,3$) for the 216-atom cell. Elemental labels are then assigned to each bin, such that no two neighboring bins contained the atoms of the same element, with periodic boundaries taken into consideration as well. In the 8-bin case, three compositions were considered: $\text{Si}_{0.25}\text{Ge}_{0.375}\text{Sn}_{0.375}$, $\text{Si}_{0.375}\text{Ge}_{0.25}\text{Sn}_{0.375}$, and $\text{Si}_{0.375}\text{Ge}_{0.375}\text{Sn}_{0.25}$. In the 27-bin case, just the equiatomic \ce{SiGeSn} case was considered (e.g. $\text{Si}_{0.33}\text{Ge}_{0.33}\text{Sn}_{0.33}$). The errors in electron density predicted by the ML model as well as in the corresponding energy for these handcrafted systems featuring species segregation are  shown in Figure~\ref{fig:KeyResultsAccuracy}, i.e. `checkerboard SiGeSn'. The errors for these unseen systems featuring species segregation are quite low asserting the generalizability of the ML model.

\subsection{Comparison of ML Models Trained on $\rho$ and $\delta \rho$}
The performance of the ML model on the \ce{CrFeCoNi} system lagged behind that of the \ce{SiGeSn} system in terms of energy predictions as shown in Figure~\ref{fig:KeyResultsAccuracy}\,(middle and bottom). In order to address that, we trained a separate ML model, only for the quaternary system   \ce{CrFeCoNi}. This model predicts the $\delta\rho$, which is the difference between the electron density $\rho$ and the superposition of atomic densities (SAD), denoted $\rho_{\text{SAD}}$, i.e., $\delta\rho = \rho - \rho_{\text{SAD}}$. We refer to this ML model as the `$\delta\rho$ ML model' to distinguish from the ML model described previously. To obtain the $\rho$ while using the $\delta\rho$ model, the $\rho_{\text{SAD}}$ needs to be added to its prediction. The energy computation through post-processing of $\rho$ remains the same. 
The $\delta\rho$ ML model performs better than the ML model for both the density and energy predictions as shown in Figure~\ref{fig:KeyResultsAccuracy}. 
The error in the energy predicted by the $\delta\rho$ ML model is presented in Figure \ref{fig:Quat_SAD_EnergyErrors} for various compositions of the \ce{CrFeCoNi} system. 
The $\delta\rho$ ML model reduced the maximum error in energy by a factor of two, compared to the $\rho$ ML model. 

In the following, we explain the superior performance of the $\delta\rho$ ML model for the \ce{CrFeCoNi} system. In contrast to the quadrivalent, softer \ce{Si}, \ce{Ge}, and \ce{Sn} pseudopotentials that were used in producing the electron density data of the \ce{SiGeSn} systems, the pseudopotentials for \ce{Cr}, \ce{Fe}, \ce{Co} and \ce{Ni} all included semi-core  states and were significantly harder. Each pseudoatom of the elements involved in the \ce{CrFeCoNi} system involved $14$ or more electrons, and \ce{CrFeCoNi} calculations generally  involved a mesh that was twice as fine as the \ce{SiGeSn} systems. Unlike the valence electrons, the semi-core states are not as active in bonding, yet the individual densities of these atoms have large contributions from their semi-core states. Thus, even in the presence of chemical bonding, as it happens in the alloy, the electron density field tends to concentrate around the nuclei,  due to which, it can be well approximated in terms of the superposition of the atomic densities i.e., $\rho_{\text{SAD}}$. Hence, by training the ML model on the \emph{difference}, i.e., $\delta\rho = \rho - \rho_{\text{SAD}}$, better accuracy can be achieved. These issues pertaining to semi-core states can become particularly important while computing energies from the electron density. The grounds-state KS-DFT energy has a large contribution from the electrostatic interactions \cite{ghosh2017sparc_2}, and the $\delta\rho$ ML model captures the contribution to this energy from the atomic sites much more accurately, since the atomic densities are better represented, particularly when semi-core states are present. This claim is further supported by Figure \ref{fig:rho_SAD_comp} where we compared the electrostatic energy field $\mathcal{E} = (\rho + b) \phi$, as calculated from the ($\rho$-based) ML model and the $\delta\rho$ ML model for a \ce{CrFeCoNi} system. Here $b$ denotes the  nuclear pseudo charge field and $\phi$ is the electrostatic potential that includes electron-electron, electron-nucleus and nucleus-nucleus interactions. The $\delta\rho$ ML model is seen to perform significantly better in terms of the error in the electrostatic energy field, particularly, near the nuclei.

%

\section{Discussion}
We have presented a machine learning (ML) framework that accurately predicts electron density for high entropy alloys at any composition. The model demonstrates strong generalization capabilities to various unseen configurations. It efficiently represents the chemical neighborhood, increasing modeling efficiency, and trains on an optimal set of the most informative compositions to reduce the amount of data required for training.
The electron density predicted by ML can be postprocessed to obtain energy and other physical properties of interest. 
{Currently, a generally accepted rule-of-thumb for quantum mechanical calculations is to aim for \emph{chemical accuracy}, i.e., a prediction error of $1.6$ mHa/atom ($1$ kcal/mole) or lesser, in the total energies \cite{suryanarayana2017nearsightedness, sauer2019ab, xu2018discrete, willand2013norm, science2016reproducibility}. This is often crucial for making realistic chemical predictions especially regarding thermochemical properties like ionization potentials and formation enthalpies. On average, for all the alloy systems studied here, our ML model demonstrated accuracies that met or were very close to achieving this threshold (see Figure \ref{fig:KeyResultsAccuracy}), thus making them accurate enough for the subsequent tasks they were applied to. Thus, the proposed ML model allows for the accelerated exploration of the complex composition space of high entropy alloys. Further improving the energy predictions of our model to enable routine calculations of quantities such as phonon spectra  which require more  accurate energies \cite{liu2023errors} remains the scope of future work. We also note that this appears to be an open area of research across a variety of ML based atomistic calculation models \cite{deng2025systematic}.}

The ML model employs a Bayesian neural network (BNN) to map atomic neighborhood descriptors of atomic configurations to electron densities. 
A key challenge for multi-element alloys is that the size of the descriptor vector increases rapidly with the number of alloying elements, necessitating more training data and larger ML models for accurate prediction.
To address this, we propose body-attached frame descriptors that  maintain approximately the same descriptor-vector size, regardless of the number of alloying elements. These proposed descriptors are a key enabler of our work. Moreover, they are easy to compute and inherently satisfy translational, rotational, and permutational invariances, eliminating the need for any handcrafting. Furthermore, obtaining the optimal number of descriptors required is simpler for these descriptors compared to the few proposed earlier in the literature. 

The composition space of multi-element alloys encompasses a vast number of compositions, demanding extensive \textit{ab initio} simulation data to develop an ML model that is accurate across the entire space. To address this challenge, we developed a Bayesian Active Learning approach to select a minimal number of training compositions sufficient for achieving high accuracy throughout the composition space. This approach leverages the uncertainty quantification (UQ) capability of a Bayesian Neural Network, generating data only at the compositions where the model has the greatest uncertainty, thereby minimizing the cost of data generation. 

We generate first principles data at various high temperatures, as thermalization helps produce data with a wide variety of atomic configurations for a given composition, enhancing the generalizability of the ML model beyond equilibrium configurations. Additionally, the Bayesian Neural Network enhances generalization through ensemble averaging of its stochastic parameters. 
The generalization capability of the ML model is demonstrated by its ability to accurately predict properties for systems not included in the training set, such as unseen alloy compositions, systems with localized defects, and systems with species segregation. The errors in energy for all test systems remain well within or close to chemical accuracy.

The proposed model demonstrates remarkable accuracy for  binary, ternary, and quaternary alloys, including \ce{SiGe}, \ce{SiGeSn} and \ce{CrFeCoNi}, all of which are of technical importance. However, the proposed framework can be applied to any alloys containing a large number of constituent elements. Although our examples involved bulk systems, the models extend also to low dimensional materials featuring chemical complexity and disorder. Furthermore, the model can be applied to predict other electronic fields. For the quaternary alloy, we develop a separate ML model to learn $\rho - \rho_{\text{SAD}}$ instead of $\rho$, enabling a more accurate representation of the density of semi-core states and significantly enhancing the overall accuracy of $\rho$ and energy predictions. 

Overall, the proposed model serves as a highly efficient tool for navigating the complex composition space of high entropy alloys and obtaining ground-state electron density at any composition. From this ground-state electron density, various physical properties of interest can be derived, making the model a powerful resource for identifying optimal material compositions tailored to specific target properties. Future work could focus on developing a universal ML framework that utilizes the proposed descriptors and functions accurately across diverse molecular structures and chemical spaces.

\section{Methods}\label{sec:Methods}

The methodology implemented in this work can be divided into six subsections: (1) the training data and test data generation; (2) the machine learning map for charge density prediction; (3) the atomic neighborhood descriptors; (4) the implemented Bayesian Neural Network; (5) Bayesian optimization and uncertainty quantification; (6) postprocessing and materials property analysis. In this following section, our methodology choices for each area are thoroughly discussed.

\subsection{Data Generation}

To generate the electron density data we use SPARC (\textbf{S}imulation \textbf{P}ackage for \textbf{A}b-initio \textbf{R}eal-space
 \textbf{C}alculations) which is an open-source finite difference based \textit{ab initio} simulation package \cite{xu2021sparc, zhang2024sparc, ghosh2017sparc, ghosh2017sparc_2}. We use the optimized norm-conserving Vanderbilt (ONCV) pseudo-potentials \cite{hamann2013optimized} for all the elements. For Si, Ge, and Sn pseudopotentials only the valence electrons are included, while for Cr, Fe, Ni, and Co semi-core states are also included. We use the Perdew-Burke-Ernzerhof (PBE) generalized gradient approximation (GGA) as the exchange-correlation functional \cite{perdew1996generalized}.
 
Real-space meshes of $0.4$ Bohr and $0.2$ Bohr were used for the \ce{SiGeSn} and CrFeCoNi systems respectively. These values were obtained after performing convergence testing on the bulk systems,  and guaranteed convergence of the total energy to within $10^{-4}$ Ha/atom. Periodic-Pulay mixing \cite{banerjee2016periodic} was employed for self-consistent field (SCF) convergence acceleration, and a tolerance of $10^{-6}$  was used. Only the gamma point in reciprocal space was sampled, as is common practice for large scale condensed matter systems. 
Fermi-Dirac smearing with an electronic temperature of $631.554$ Kelvin was used for all the simulations. 

{The atomic coordinate configurations that were fed into SPARC were obtained via sampling from high-temperature molecular dynamics trajectories --- either \emph{ab initio} molecular dynamics (AIMD) calculations or classical molecular dynamics (MD) using state-of-the-art machine learning interatomic potentials. To ensure comprehensive coverage of local atomic environments and to improve model generalizability, simulations were performed at elevated temperatures, consistent with our prior observations \cite{pathrudkar2024electronic}.  For each composition, atomic species labels were randomly assigned to lattice sites consistent with the target stoichiometry, and multiple distinct seeds (orderings)  were used as starting points for AIMD/MD trajectories. This procedure yields ensembles that, for the datasets used in this work, correspond to fully chemically disordered alloys. Additionally, we generated targeted ``hand-crafted'' configurations featuring, for example, species segregation and defects which were used in generalizability tests (described in detail in the Supplemental Materials).
}

For the \ce{SiGeSn} system, AIMD was performed, as per the methodology of our previous work \cite{pathrudkar2024electronic}. However, AIMD simulations can be time-consuming as one has to perform an electronic minimization at each MD step. The increased number of electrons required to model the \ce{CrFeCoNi} system motivated an alternative approach. In order to alleviate the computational burden of configurational sampling for the \ce{CrFeCoNi} system, we leveraged classical molecular dynamics (MD) instead of AIMD. The interatomic potential selected for the MD runs is the Materials 3-body Graph Network (M3GNet), a universal machine-learned potential implemented in the (Materials Graph Library) MatGL python package \cite{chen2021learning, chen2022universal}. The MD simulations are run through the Atomic Simulation Environment (ASE) interface built into MatGL. After extracting snapshots from the MD trajectory, a single electronic minimization step is performed to obtain the electron densities. MD with machine learned interatomic potentials is orders of magnitude cheaper compared to AIMD, and the subsequent electronic minimization tasks (for given system snapshots) can be conveniently parallelized. This approach facilitates rapid data generation for various configurations without any quality loss for the electron density training data. 

The compositions for which data was generated are shown in {Figure S1 
} for the \ce{SiGeSn} system and in {Figure S2 
} for the \ce{CrFeCoNi} system. For more details regarding the data generation, please refer to the Supplementary Material.

\subsection{Machine Learning Map for Charge Density Prediction}

Our ML model maps the atom coordinates $\{\textbf{R}_I\}_{ I = 1}^{N_\text{a}}$ and species (with atomic numbers $\{\textbf{Z}_I\}_{ I = 1}^{N_\text{a}}$) of the atoms, and a set of grid points $\{\textbf{r}_i\}_{i = 1}^{N_{\text{grid}}}$ in a computational domain, to the electron density values at those grid points. Here, $N_\text{a}$ and $N_{\text{grid}}$ refer to the number of atoms and the number of grid points, within the computational domain, respectively. We compute the aforementioned map in two steps. \emph{First}, given the atomic coordinates and species information,  we calculate atomic neighborhood descriptors for each grid point. \emph{Second}, a Bayesian Neural Network is used to map the descriptors to the electron density at each grid point. These two steps are discussed in more detail subsequently. 

\subsection{Atomic Neighborhood Descriptors}
One major challenge in predicting electron density for multi-element systems is the rapid increase in the number of descriptors as the number of species grows, which hampers both efficiency and accuracy. For example, the scalar product descriptors developed in \cite{pathrudkar2024electronic} increase rapidly with the number of species. 
Additionally, descriptors should be simple, easy to compute and optimize, and avoid manual adjustments like selecting basis functions. To address these issues, we propose a novel descriptor that utilizes position vectors to atoms represented in body-attached reference frames. The proposed descriptor overcome the scaling issue faced by the scalar product \cite{pathrudkar2024electronic}, tensor invariant based \cite{chandrasekaran2019solving} and SNAP \cite{thompson2015spectral} descriptors, since the number of position vectors needed depends only on the number of atoms in the atomic-neighborhood but are independent of the number of the species.   

We encode the local atomic neighborhood using descriptors $\mathcal{D}_i$. Descriptors are obtained for each gridpoint $\{\textbf{r}_i\}_{i = 1}^{N_{\text{grid}}}$ in the computational domain. Following the nearsightedness principle \cite{kohn1996density, prodan2005nearsightedness, zepeda2021deep}, we collect $M$ number of nearest atoms to the grid point $i$ to compute the descriptors for grid point $i$. This is analogous to setting a cutoff radius for obtaining the local atomic neighborhood. The descriptors for the grid point $i$ are denoted as $\mathcal{D}_i \in \mathbb{R}^{4M}$. For the $j$-th atom, descriptors are given as: 
{\begin{equation}
\mathcal(\mathcal{D}_i)^j = \bigg\{ \| \textbf{r} \|, \, \frac{r_{1}^0}{\| \textbf{r} \|}, \,  \frac{r_{2}^0}{\| \textbf{r} \|}, \, \frac{r_{3}^0}{\| \textbf{r} \|} \bigg\}
\end{equation} }
where {$(r_{1}^0, r_{2}^0, r_{3}^0)$} are {the} coordinates of the position vector {$\textbf{r}$} of atom $j$ with respect to a global reference frame at the grid point $i$. $j$ varies from  $1$ to $M$. The basis vectors for the global reference frame are denoted as $\mathbf{e}^{0}_{1}, \mathbf{e}^{0}_{2}, \mathbf{e}^{0}_{3}$. 

The above descriptors are not frame invariant and hence would change under rotation of the computational domain. Since the electron density is equivariant with respect to the given atomic arrangement, it is imperative to maintain equivariance. 
To address this issue, we propose to determine an unique local frame of reference for the atomic neighborhood and express these coordinates in that local reference frame. 
In previous works, such local frame of reference is constructed using two \cite{zeng2023deepmd} or three \cite{del2023deep} nearest atoms.  However, as mentioned in reference \cite{zeng2023deepmd}, these local frame descriptors exhibit non-smooth behavior when the order of nearest neighbors is altered or when there is a change in the nearest neighbors themselves. 
To address this issue, in this work, we obtain the local frame of reference using Principal Component Analysis (PCA) of an atomic neighborhood consisting of $M$ atoms. We apply PCA to position vectors  of these atoms and obtain principal directions, which yield an orthonormal basis set $\mathbf{e}_{1}, \mathbf{e}_{2}, \mathbf{e}_{3}$. We represent the components of the position vectors of the atoms with respect to this new basis set. Thus {the $p$-th component of} the position vector of atom $j$ with respect to a new reference frame at the grid point $i$ is given by {$r_p = \left(\mathbf{e}_{p}\cdot\mathbf{e}^{0}_{q}\right) \,r^0_q$. The Einstein summation convention is used; repeated indices have the range of 1, 2, 3.}  
The components of $\textbf{r}$ in the new reference frame are denoted by {$(r_{1},r_{2},r_{3})$} in the following.

In order to handle systems with multiple chemical species, species information needs to be encoded in the descriptors. One strategy proposed in previous work is to compute descriptors for individual species and concatenate the descriptors \cite{chandrasekaran2019solving}. Another strategy is to encode chemical species through one-hot vector \cite{zeng2023deepmd}. In this work, we encode the species information using the atomic number of the species. The atomic number of the $j$-th atom is denoted as  $Z_j$. Incorporating the species information, the updated descriptors are  $\mathcal{D}_i \in \mathbb{R}^{5M}$ are given as,
{\begin{equation}
    \mathcal{D}_i = \bigg\{ Z_j , \,  \| \textbf{r} \|, \, \frac{r_{1}}{\| \textbf{r} \|}, \,  \frac{r_{2}}{\| \textbf{r} \|}, \, \frac{r_{3}}{\| \textbf{r} \|} \bigg\}_{j=1,\cdots,M}
\end{equation}}
Therefore, the number of proposed descriptors does not increase with the number of species present in the alloy, for a fixed $M$. 

{The computational time required to calculate the proposed descriptors is about twice the time required by scalar product descriptors \cite{pathrudkar2022machine} and approximately the same as SNAP descriptors \cite{MalaGit, ellis2021accelerating}. }
\subsubsection{Selection of the Optimal Set of Descriptor} 
The nearsightedness principle \cite{kohn1996density, prodan2005nearsightedness} and screening effects \cite{ashcroft2022solid} imply that electron density at a given grid point is minimally influenced by atoms far away. 
This suggests that only descriptors from atoms  close to a grid point are necessary for the ML model. However, the optimal set of descriptors for accuracy are not known \textit{a priori} and can be computationally expensive to determine through a grid search  \cite{Barnard_RSC2023}.  

Using an excessive number of descriptors can increase the computational cost of descriptor calculation, model training, and inference. It can also lead to issues like the curse of dimensionality, reducing the model's prediction performance \cite{Hamer_Dupont2021JMRL, Guyon_Elisseeff2003JMRL, bishop2006pattern, yadav2021interpretable}, or may necessitate a larger neural network to learn effectively. Conversely, using too few descriptors results in an incomplete representation of atomic environments, leading to an inaccurate model.

Selection of optimal set of descriptors has been explored in prior works, particularly for Behler-Parinello symmetry functions \cite{gastegger2018wacsf, imbalzano2018automatic} or  widely used Smooth Overlap of Atomic Positions (SOAP) \cite{bartok2013representing} descriptors  \cite{Barnard_RSC2023}. These systematic procedures for descriptor selection eliminate the trial-and-error approach often used when finalizing a descriptor set. In \cite{imbalzano2018automatic}, the authors demonstrated that an optimized set of descriptors can enhance the efficiency of ML models.
Therefore, selecting an optimal set of descriptors for a given atomic system is crucial for balancing computational cost and prediction accuracy. Let $M$ ($M \leq N_\text{a}$) is a set of nearest neighboring atoms for grid points. We compute the descriptors for various $M$ and the corresponding errors in a ML model's prediction. The optimal value of \( M \) is the one that minimizes the error.
Figure \ref{fig:feature_convergence} shows the error in ML model's prediction for different values of $M$ for the \ce{SiGeSn} system, showing that the optimum value of $M$ is near 55.  
Computation of error in the ML model's prediction  for each $M$ involves descriptor computation, training of the neural network and testing, and therefore is computationally expensive. Given that a neural network needs to be trained for each selected $M$, descriptor optimization is challenging. In our previous work \cite{pathrudkar2024electronic}, we demonstrated descriptor convergence; it required training of 25 neural networks to obtain optimal number of descriptors for Aluminum. In this work, because of the proposed descriptors, descriptor convergence requires training of only 7 neural networks. 
Most existing approaches to descriptor convergence involve optimizing the cutoff radius (analogous to the number of nearest atoms) and the number of basis functions \cite{imbalzano2018automatic, chandrasekaran2019solving}. In contrast, the proposed descriptors in this work require optimization with respect to only one variable, $M$, the number of nearest atoms. This significantly reduces the time needed to identify the optimal set of descriptors. Once optimized, we used the same value of $M$ across binary, ternary and quaternary alloys. Our results show errors of similar magnitude across all these systems, giving us confidence in our choice.

\subsubsection{Equivariance through Invariant Descriptors}
The proposed descriptors are invariant to rotation and translation, as the position vectors are represented through a unique body-attached reference frame at the grid point. Additionally, invariance to the permutation of atomic indices is maintained, since the position vectors are sorted based on their distance from the origin. Given that the predicted electron density is a scalar-valued variable, the invariance of the input features is sufficient to ensure the equivariance of the predicted electron density under rotation, translation, and permutation of atomic indices, as noted in references \cite{koker2023higher,thomas2018tensor,pathrudkar2024electronic}.

\subsection{Bayesian Neural Network}

Bayesian Neural Networks are the stochastic counterparts of the traditional deterministic neural networks with advantages such as better generalization and robust uncertainty quantification.   
We train a Bayesian Neural Network (BNN) to predict the probability distribution, $P(\rho|\mathbf{x},{D})$, of the output electron density ($\rho$), given a set of training data, ${D} = \{\mathbf{x}_i,\rho_i\}_{i=1}^{N_d}$, and an input descriptor $\mathbf{x} \in \mathbb{R}^{N_\text{desc}}$. In BNNs, this is achieved by learning stochastic network parameters in contrast to the deterministic parameters learned in a traditional deep neural network. By assuming prior distribution $P(\mathbf{w})$ for the network parameters  $\mathbf{w} \in \Omega_{w}$, the posterior distribution  $P(\mathbf{w}|{D})$  is obtained from the Bayes' rule as $P(\mathbf{w}|{D}) = P({D}|\mathbf{w})P(\mathbf{w})/P({D})$. Here $\mathbf{w} \in \Omega_{w}$ is the set of parameters of the network and $ P({D}|\mathbf{w})$ is the likelihood of the data.

However, the term $P(D)$ -- known as the model evidence -- is intractable, since it involves a high dimensional integral which in turn results in an intractable posterior distribution $P(\mathbf{w}|{D})$. Therefore, the posterior distribution is approximated by variational inference \cite{hinton1993keeping,graves2011practical,blundell2015weight,thiagarajan2021explanation,thiagarajan2025jensen}. In variational inference, the intractable posterior  $P(\mathbf{w}|{D})$ is approximated by a tractable distribution, called the variational posterior ($q(\mathbf{w}|\boldsymbol{\theta})$), from a known family of distributions such as the Gaussian. The parameters ($\boldsymbol{\theta}$) of the distribution $q(\mathbf{w}|\boldsymbol{\theta})$ are optimized such that the statistical dissimilarity between the variational posterior and the true posterior is minimized. If the dissimilarity metric is taken as the KL divergence, we get the following optimization problem:
\begin{equation} \label{eq:kl_min}
\begin{split}
  \boldsymbol{\theta}^* &= \arg \min_{\boldsymbol{\theta}} \text{KL}\left[ q(\mathbf{w}|\boldsymbol{\theta})\:||\: P(\mathbf{w}|{D}) \right]\\
    &=  \arg \min_{\boldsymbol{\theta}} \int q(\mathbf{w}|\boldsymbol{\theta}) \log \left[\frac{q(\mathbf{w}|\boldsymbol{\theta})}{P(\mathbf{w})P({D}|\mathbf{w})} P({D})\right] d\mathbf{w}\,.   
\end{split}
\end{equation}
This leads to the following loss function for BNN that has to be minimized:

\begin{equation}\label{eq:elbo}
    \mathcal{F}_{KL}({D},\boldsymbol{\theta}) = \text{KL}\left[ q(\mathbf{w}|\boldsymbol{\theta})\:||\: P(\mathbf{w}) \right] - \mathbb{E}_{q(\mathbf{w}|\boldsymbol{\theta})} [\log P({D}|\mathbf{w})]\,.
\end{equation}

Once the posterior distribution of the parameters are approximated by variational inference, the probability distribution for the output can be evaluated by marginalizing over $\mathbf{w}$ as:
\begin{align}\label{eq:marg_rho}
    P(\rho|\mathbf{x},{D}) &= \int_{\Omega_{w}} P(\rho|\mathbf{x},\mathbf{w}) P(\mathbf{w}|{D}) d\mathbf{w} \\
                                &\approx  \int_{\Omega_{w}} P(\rho|\mathbf{x},\mathbf{w})q(\mathbf{w}|\boldsymbol{\theta}) d\mathbf{w}\,.
\end{align}
This marginalization helps in improving generalization, as it is equivalent to learning an ensemble of deterministic networks with different parameters $\mathbf{w}$. Furthermore, the variance of this distribution $P(\rho|\mathbf{x},{D})$ is a measure of model uncertainty in the predictions.

 \subsection{Uncertainty Quantification}
Bayesian Neural Networks provide a natural way to quantify uncertainties, since they predict a probability distribution for outputs. The uncertainties in the prediction can be classified as `aleatoric' and `epistemic' uncertainties. Aleatoric uncertainty stems from the natural variability in the system, such as noise in the training data. Whereas, epistemic uncertainties are a result of model uncertainties, such as the uncertainty in the parameters of the model.

Variance in the output distribution $P(\rho|\mathbf{x},{D})$ is a measure of uncertainty in the model prediction. The variance of this distribution is given as:
\begin{equation} \label{eq:var}
    \text{var}(\rho) = \sigma^2(\mathbf{x}) +\left[ \frac{1}{N_s} \sum_{j=1}^{N_s} \left(\widehat{\rho}_j\right)^2 - \left(\mathbb{E}(\widehat{\rho}_j)\right)^2 \right]\,.
\end{equation}

To evaluate this variance, a $j^{th}$ sample for each parameter is drawn following the learned posterior distributions $q(\mathbf{w}|\boldsymbol{\theta})$ for the parameters of the network. The network is then evaluated for this sample to predict the output, $\widehat{\rho}_j(x)$, for a given input. This process is repeated for a total of $N_s$ samples. This enables us to evaluate the epistemic uncertainty, which is the second term of Eq. \eqref{eq:var}. Next, $\sigma(\mathbf{x})$ -- which is a heterogeneous noise parameter representing the aleatoric uncertainty -- can be predicted by the network along with the output $\rho$.  For a Gaussian likelihood, the noise $\sigma (\mathbf{x})$ can be learned through the likelihood term of the loss function Eq.~\eqref{eq:elbo} following \cite{kendall2017uncertainties} as:
\begin{align}
\log P({D}|\mathbf{w}) = \sum_{i=1}^{N_d} -\frac{1}{2} \log \sigma_i^2 - \frac{1}{2\sigma_i^2}(f_N^\mathbf{w}(\mathbf{x}_i)-\rho_i)^2\,.
\end{align}
Here, $N_d$ is the size of the training data set. 

{In a well-calibrated model, the predictive distribution of the output closely resembles the empirical distribution of the data.  However, it is to be noted that the uncertainties presented in this work are uncalibrated. While calibration can provide better estimates of uncertainties, only the ordering of the uncertainty estimates among different compositions matters for the active learning framework employed here. Since calibration methods such as the ones presented in \cite{busk2021calibrated,gruich2023clarifying} do not affect this ordering, recalibration was not performed in this work.}

\subsection{Bayesian Active Learning}
The number of possible stoichiometric compositions in ternary and quaternary alloys is very large. Thus KS-DFT calculations on all of these compositions to create a ML model are quite expensive. There might be an optimal subset of compositions that contains sufficient information to train a ML model. However, such subsets are not known \textit{a priori}. We utilize the Active Learning technique to identify such an optimal subset of compositions to reduce the cost of data generation through KS-DFT.

Active learning is a machine learning algorithm that can query data points that need to be labeled to learn a surrogate model. Active learning is primarily used when the computational cost associated with generating the training labels is high. A schematic of Bayesian active learning is shown in Figure \ref{fig:active_learning}. In the first step, an initial set of training data is generated by \textit{ab initio} calculations and a Bayesian Neural Network model is trained on this initial training set. The second step involves optimizing an acquisition function. In active learning, an acquisition function explores the input space to find the next input point that is most informative to learn the input-output relationship. In this work, we hypothesize that the composition (or the set of compositions) with the highest epistemic uncertainty in the predictions contains the most information to learn the surrogate model. Therefore, the epistemic uncertainty in the predictions obtained by the Bayesian Neural Network as explained in the previous section is used as the acquisition function. Optimization of this acquisition function is achieved by evaluating the test compositions using the Bayesian network to obtain the ones with high uncertainties in their predictions.  As a third step of the active learning framework, \textit{ab initio} calculation needs to be performed for the compositions with high uncertainties found by optimizing the acquisition function. As a final step, this new data is appended to the training set, and the first, second, and third steps are repeated until a satisfactory model is learned. {In our present study, once a composition was identified for appending to the dataset, all the configuration snapshots (of varying atomic arrangements) associated with that composition were included in the next batch of training data.}

{To get a sense of the baseline errors while predicting across composition space, and to demonstrate the advantage of the Bayesian AL technique over the random selection of compositions, we have compared the errors from these two approaches in Figure~\ref{fig:baseline_model_comparison}. Both approaches used the same number ($20$) of compositions and the same amount of data. The advantage of the Bayesian AL technique is evident from the plot. Three different sets of randomly chosen compositions were used to develop three ML models and the error bars indicate the range of maximum NRMSE values observed across these models.}


\subsection{Tessellation-based Iterative Training} 
In Tessellation-based Iterative Training, we iteratively train the ML model on progressively larger subsets of compositions. We select the subsets by progressively refining the tessellation of the composition spaces. 
We tessellate the triangular and tetrahedral spaces of ternary and quaternary compositions using regular triangles and tetrahedrons. 
Successive levels of refinement are shown in Figure \ref{fig:T1toT4}.  The training compositions are chosen at the vertices of these triangles and tetrahedrons. The four triangular tessellations are denoted as T1, T2, T3, and T4, corresponding to 3, 6, 9, and 15 training points, respectively. However, the edge points of T3 do not include the edge points of T2. Therefore, we skip the T3 iteration and use T4 directly as the next iteration after T2 to ensure that no training data is discarded. 
For the quaternary system, tessellation iterations T1, T2, and T4, using regular tetrahedrons yields 4, 10, and 34 vertex points respectively. The second level of refinement includes all 10 compositions on the edges or vertices of the tetrahedrons and, therefore, does not have any composition that includes more than two elements. It has an octahedral space in the middle of the smaller tetrahedrons (see Figure 1 of \cite{gabbrielli2012families} and  Figure~\ref{fig:T1toT4}). We choose to use the midpoint of the octahedron as an additional training composition, leading to a total of 11 training compositions for the second level of refinement. 

\subsection{Postprocessing}
{Since much of the utility of predicting charge densities lies in the physical parameters that can be obtained from them, it is prudent to verify how well our model predicts downstream quantities. Here, we focus on computing the total ground state energy, as a postprocessing step to validate the predictions of our model. Further material properties of interest, e.g. defect formation energies, etc., can be calculated from these computed  energies. The postprocessing step is accomplished as follows. First, the predicted electron densities are rescaled by the total number of electrons:
\begin{align}
    \rho^{\text{scaled}}\left(\textbf{r}\right) = \rho^{\text{ML}}(\textbf{r})\frac{N_{\text{e}}}{\displaystyle\int_{\Omega}\rho^{\text{ML}}(\textbf{r})d\textbf{r}}\,.
    \label{eq:scaled_density}
\end{align}
\noindent where $\Omega$ is the periodic supercell used in the calculations, and $N_{\text{e}}$ is the number of electrons in the system. This step serves to ensure that the total system charge is accurately preserved by the ML predictions; this has been found to be important for obtaining high-quality predictions in the energy \cite{alred2018machine, pathrudkar2022machine}. Next, the scaled densities are input to the same real space electronic structure calculation framework, as used for data generation \cite{xu2020m, zhang2023version, xu2021sparc, zhang2024sparc, ghosh2017sparc, ghosh2017sparc_2}. The same calculation settings (e.g. real space mesh size, pseudopotentials, exchange-correlation functional, etc.) are chosen for the post-processing steps, which involves setting up of the Kohn-Sham Hamiltonian using the scaled electron density, diagonalization of the Hamiltonian and subsequent calculation of the Harris-Foulkes energy \cite{harris1985simplified, foulkes1989tight}: 
\begin{align}
\nonumber
{E}_{\text{Harris-Foulkes}} & = E_{\text{band}} + E_{\text{xc}} - E_{\text{Vxc}} \\ & + E_{\text{electrostatics}} + E_{\text{elec-entropy}}\,.
\end{align}
Here, the first term and the last term on the right hand side denote the electronic band-structure energy ($E_{\text{band}}$) and the electronic entropy contributions ($E_{\text{elec-entropy}}$), respectively.  These terms are directly dependent on the eigenstates of the Hamiltonian, while the remaining right hand terms are calculated readily from electron densities. The terms $E_{\text{xc}}$ and $E_{\text{Vxc}}$ denote contributions from the exchange correlation energy and its potential, respectively. The term $E_{\text{electrostatics}}$ arises from electrostatic interactions and includes electron-electron, electron-ion and ion-ion contributions, as well as corrections from pseudocharge self-interactions and overlaps \cite{ghosh2017sparc, ghosh2017sparc_2} . The specific forms of each of the terms on the right hand, as well as their implementation within the SPARC electronic structure code used in this work, are available in \cite{ghosh2017sparc, ghosh2017sparc_2}. Notably, the Harris-Foulkes energy is chosen since it is known to be less sensitive to self-consistency errors, and is therefore known to give a better estimate of the true Kohn-Sham ground-state energy \cite{foulkes1993accuracy}.}

\color{black}
The total energy errors for the systems considered in this work are summarized in {Figure \ref{fig:KeyResultsAccuracy}}. Additionally, {Figures S10-S11 
} in the Supplemental Material display the energy errors across the individual compositions considered. Performing this postprocessing step is an important component of the work, allowing us to observe the extent to which subtle errors in charge density predictions could propagate to downstream system properties.

\section*{Data Availability}\label{subsec:Data_availability}
\noindent Raw data were generated at Hoffman2 High-Performance Compute Cluster at UCLA's Institute for Digital Research and Education (IDRE) and National Energy Research Scientific Computing Center (NERSC). Derived data supporting the findings of this study are available from the corresponding author upon request.

\section*{Code Availability}\label{subsec:Code_availability}
Codes supporting the findings of this study are available from the corresponding author upon reasonable request.

\color{black}
\begin{acknowledgments}
This work was primarily supported by grant DE-SC0023432 funded by the U.S. Department of Energy, Office of Science. This research used resources of the National Energy Research Scientific Computing Center, a DOE Office of Science User Facility supported by the Office of Science of the U.S.~Department of Energy under Contract No.~DE-AC02-05CH11231, using NERSC awards  BES-ERCAP0033206, BES-ERCAP0025205, BES-ERCAP0025168, and BES-ERCAP0028072.  JM acknowledges support from the U.S.~Department of Energy under contracts DE-SC0018410 (FES) and DE-SC0020314 (BES). ASB and JM acknowledge funding through a UCLA SoHub seed grant. SP acknowledges the Doctoral Finishing Fellowship awarded by the Graduate School at MTU. The authors would like to thank UCLA's Institute for Digital Research and Education (IDRE), the Superior HPC facility at MTU, the MRI GPU cluster at MTU for making available some of the computing resources used in this work. The authors acknowledge the use of the GPT-4o (OpenAI) model to polish the language and edit grammatical errors in some sections of this manuscript. The authors subsequently inspected, validated and edited the text generated by the AI model, before incorporation.  
\end{acknowledgments}

\section*{Author contributions}
SP, PT and AK worked on developing the Bayesian-Active Learning framework, model descriptors and other machine learning (ML) aspects. ST, SA and AG worked on the KS-DFT data generation and post-processing calculations. ASB, SG and JM were involved in conceptualization, methodological design, supervision, and securing funding/resources. All authors contributed
to writing the manuscript.

\section*{Competing Interests}
\noindent The authors declare no competing interests.

\color{black}
\clearpage
\newpage


\begin{figure*}
    \centering
    \includegraphics[width=0.9\linewidth]{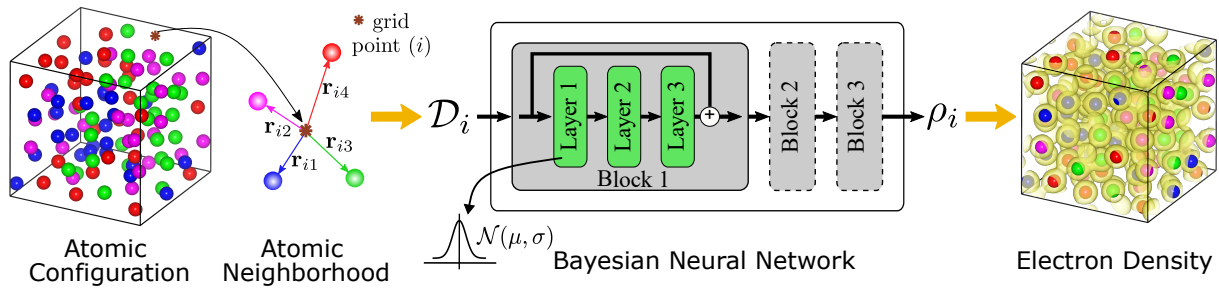}
    \caption{\textbf{Schematic representation of our Machine Learning model showing descriptor generation and mapping to electron density using Bayesian Neural Network.} The process begins with calculating atomic neighborhood descriptors $D(i)$ at each grid point, $i$, for the provided atomic configuration snapshot in the training data. A Bayesian Neural Network is trained to provide a probabilistic map from the atomic neighborhood descriptors $D(i)$  to the electronic charge density and corresponding uncertainty measure at grid point, $i$. Application of the trained model to generate charge density predictions for a given new query configuration requires: descriptor generation for the query configuration, forward propagation through the Bayesian Neural Network, and aggregation of the point-wise charge density predictions $\rho(i)$ and uncertainty values to obtain the charge density field $\rho$ and uncertainty field, respectively.}
    \label{fig:model_schematic}
\end{figure*}

\begin{figure}[htbp]
    \centering




\includegraphics[width=0.9\linewidth]{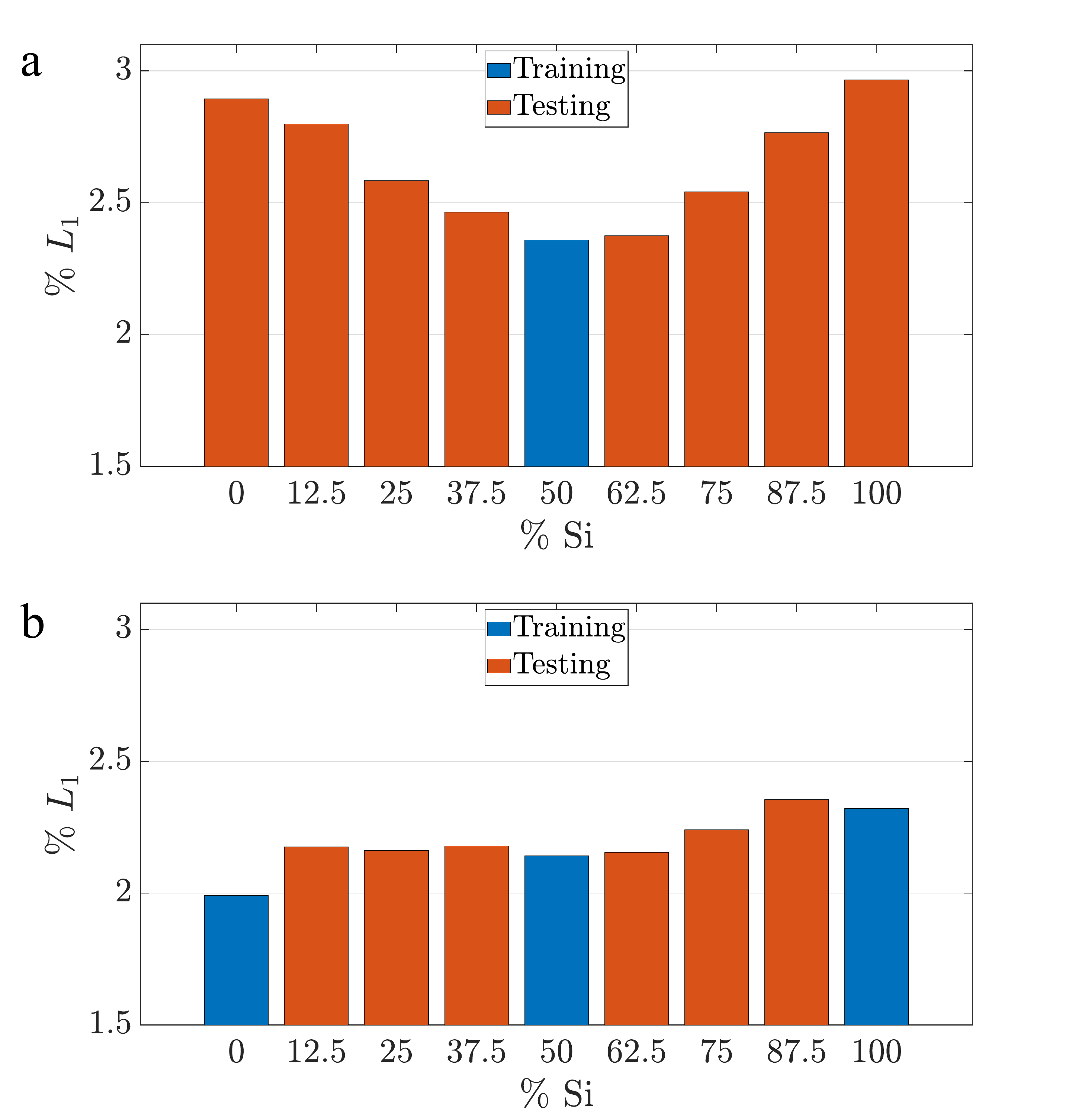}

    \caption{ \textbf{Iterative training for accurate prediction across composition space of binary alloy.} \textbf{(a)} Error in $\rho$ prediction for Si$_x$Ge$_{1-x}$, where the model was trained using only $x = 0.50$ and tested on all $x \neq 0.50$. \textbf{(b)} Error in $\rho$ prediction for Si$_x$Ge$_{1-x}$, where the model was trained using $x = 0,\, 0.50,\, 1.00$ and tested at other compositions. The error across entire composition space reduces significantly with the addition of only two extra training compositions. \mysquare{m0}: Training, \mysquare{m1}: Testing}
    \label{fig:sige_rel_l1}

\end{figure}

\begin{figure}[htbp]
    \centering
    \includegraphics[width=0.99\linewidth]{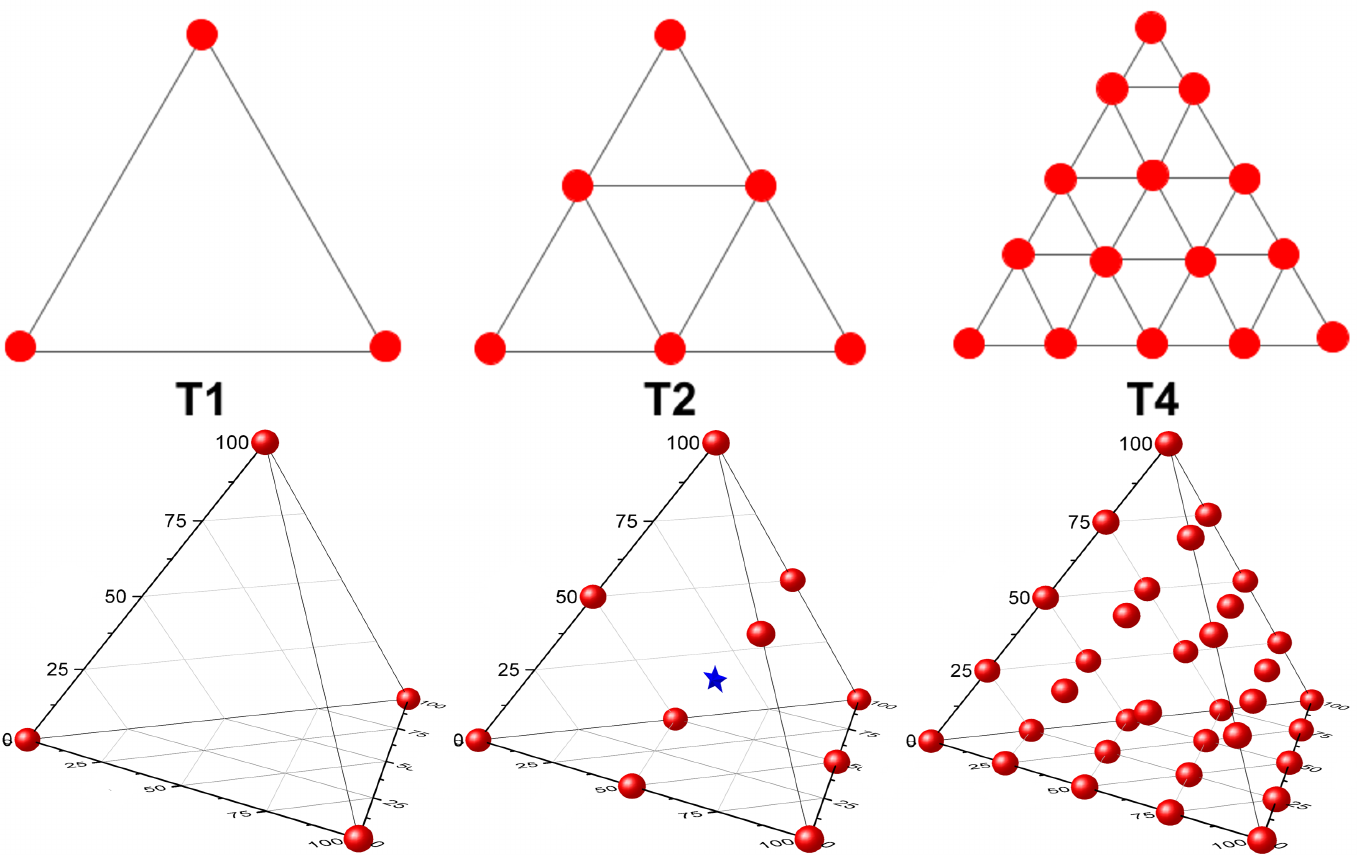}
    \caption{\textbf{Training compositions for three levels of tessellation (T1, T2 and T4).} The red dots show training compositions. The top row shows compositions for the ternary (SiGeSn) system and the bottom row shows compositions for the quaternary (CrFeCoNi) system. Note that we train the model T4 with the 4th iteration of tessellation, because the training compositions in the third iteration exclude available training compositions from the second iteration. The star depicts an additional point considered in the quaternary T2 model to capture information in the center, approximating the octahedron in the second tessellation of the tetrahedron.} 
    \label{fig:T1toT4}
\end{figure}

\begin{figure*}[htbp]
    \centering
    \includegraphics[width=0.8\linewidth]{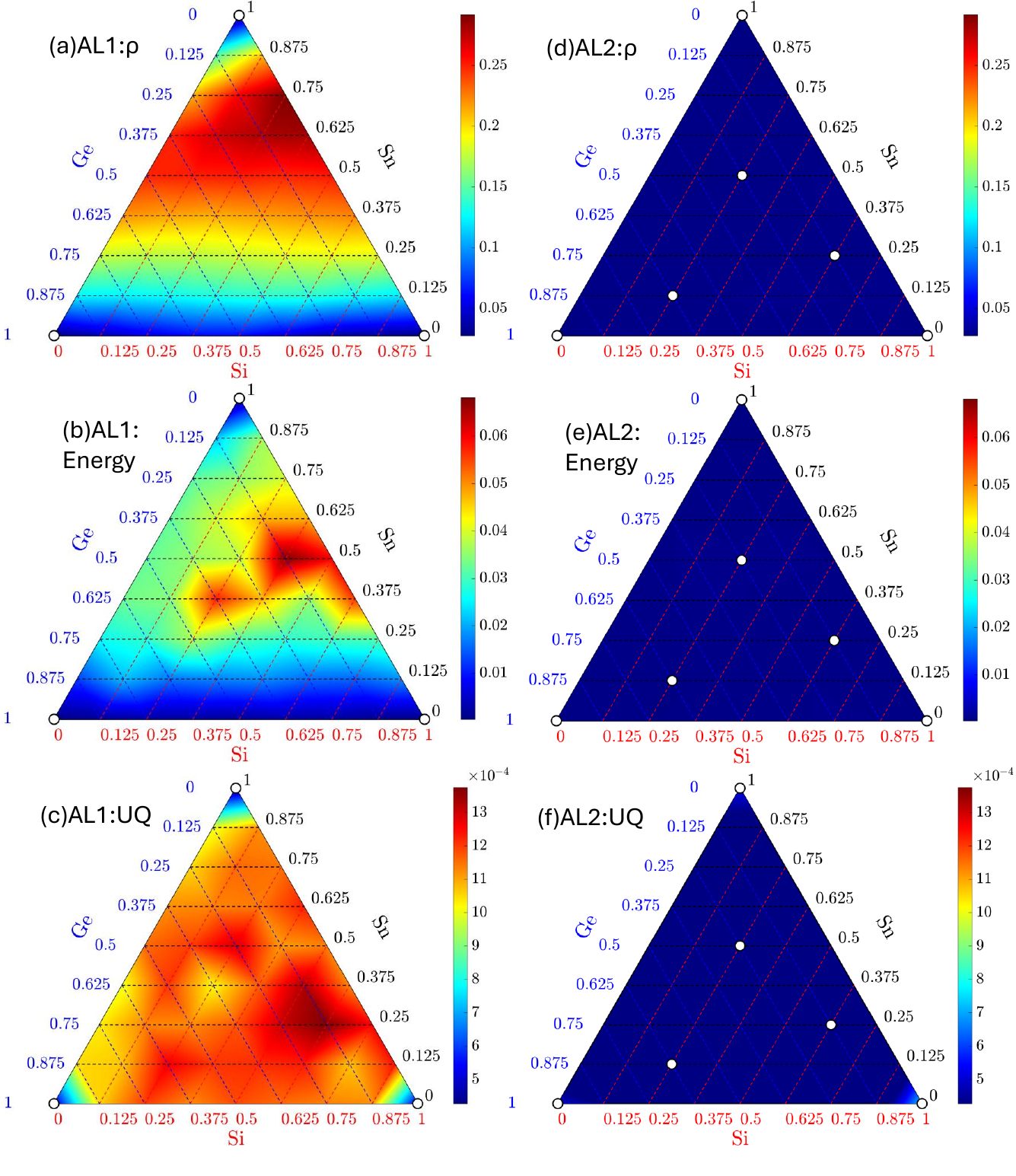}
    \caption{\textbf{Bayesian Active Learning to iteratively select training compositions to accurately predict across composition space of Ternary alloy.} (a) NRMSE across the composition space after 1st iteration of Active Learning, termed as AL1, trained using only 3 pure compositions shown using white circles. (b) Energy prediction error for model AL1 with 3 pure composition. (c) Epistemic Uncertainty in $\rho$ prediction across composition space after prediction with model AL1. Query points (additional training points) for the next iteration of Bayesian Active Learning are selected based on highest uncertainty regions shown in `f'. (d) NRMSE across the composition space after 2nd iteration of Active Learning. 3 additional training points are added as per the uncertainty contour in subfigure, `c'. This model is termed as AL2. We observe that the NRMSE is low and consistent across the composition space showing the effectiveness of query points selection through uncertainty. (e) Error in energy prediction across composition space. The unit of energy error is Ha/atom. The energy error is within chemical accuracy across the composition space. (f) Epistemic Uncertainty in $\rho$ prediction across composition space after prediction with model AL2. This figure uses same colorbars for AL1 and AL2 models. Refer to Figure S6 
    in the Supplemental Material for figure with distinct colorbars.}
    \label{fig:AL4figs}
\end{figure*}

\begin{figure*}[htbp]
    \centering

    \includegraphics[width=0.8\linewidth]{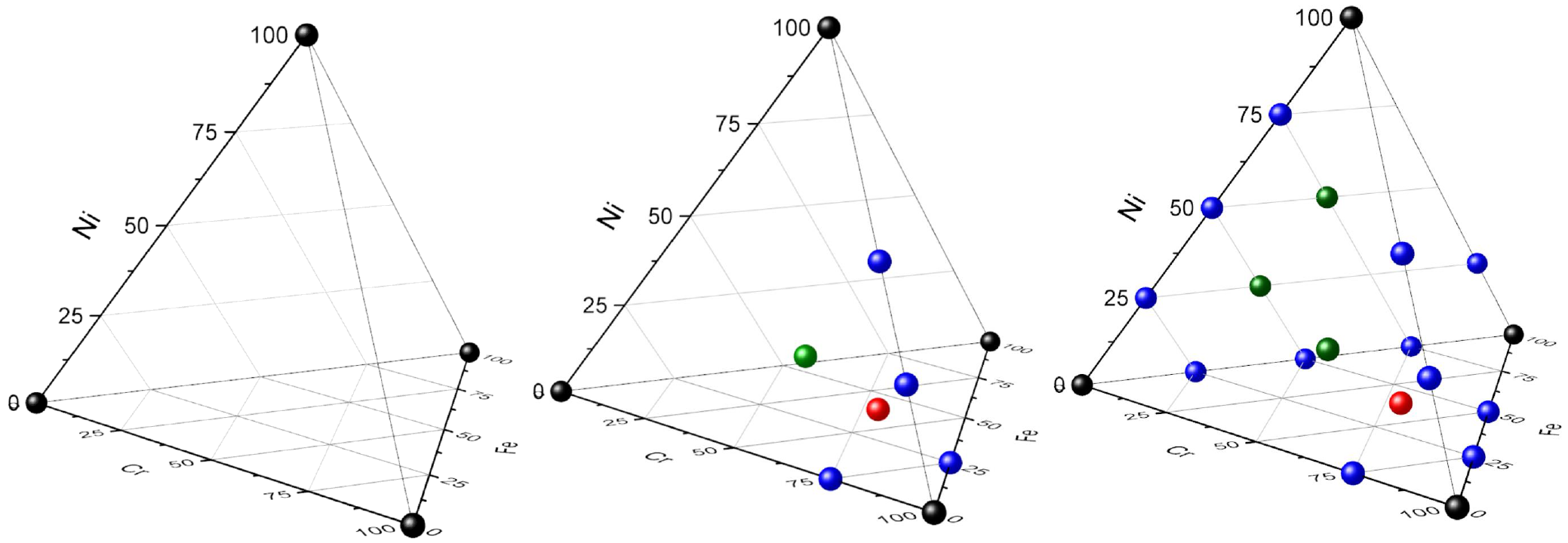}
 
    \caption{\textbf{Training compositions for Quaternary system for Active Learning models.} \textit{Left:} 4 training compositions used for model AL1, \textit{Middle:} 10 training compositions used for model AL2, \textit{Right:} 20 training compositions used for model AL3. Black spheres indicate compositions on vertex, blue spheres indicate compositions on edges, green spheres indicate compositions on faces and red spheres indicate compositions inside the tetrahedron.}
    \label{fig:AL_queries_quat}
\end{figure*}

\begin{figure*}
    \centering
    \includegraphics[width=0.9\linewidth]{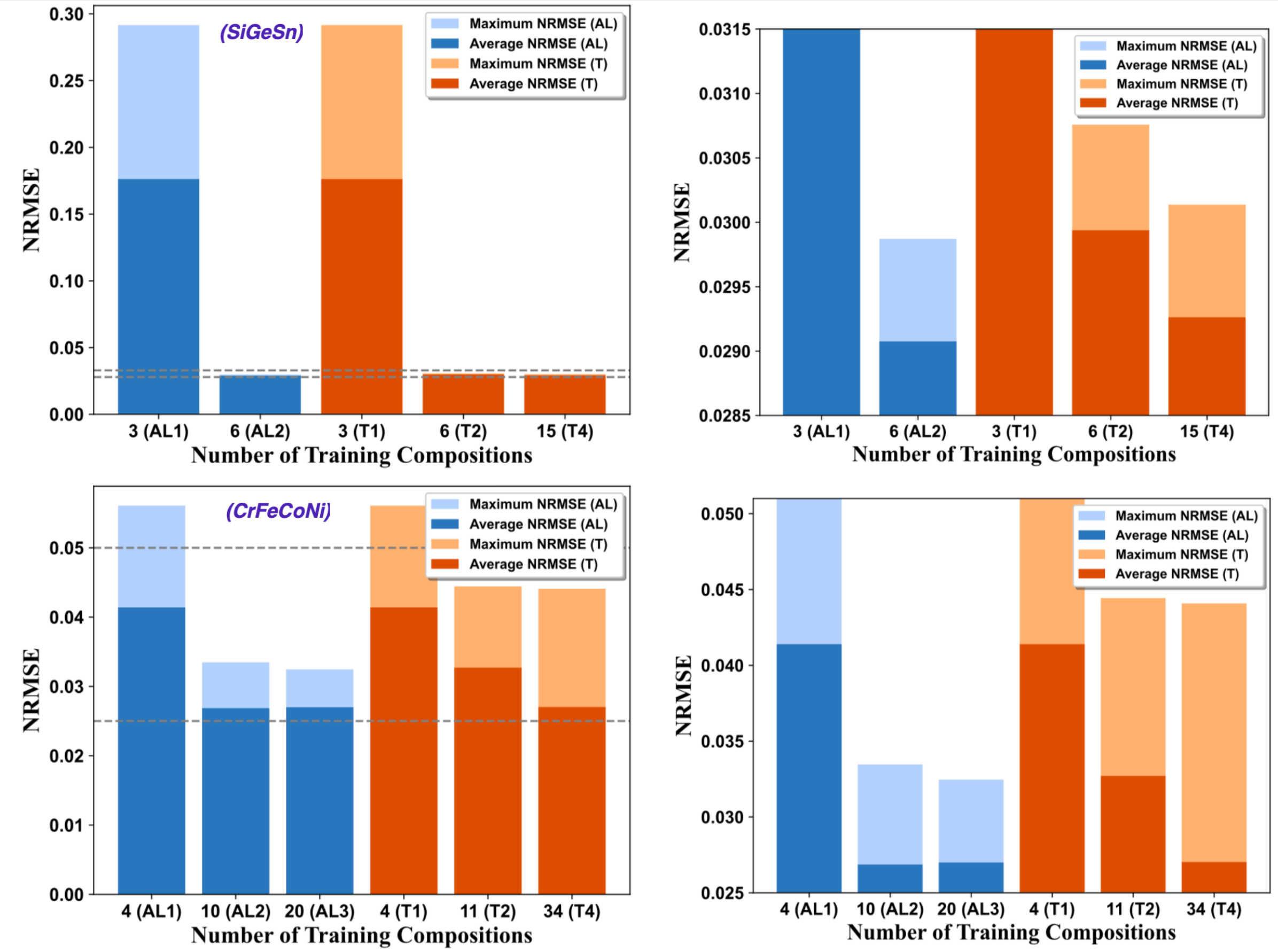}
    \caption{\textbf{Plots showing NRMSE across composition space with increasing number of training compositions for \ce{SiGeSn} (top) and \ce{CrFeCoNi} (bottom).} Right side plots are magnified version of the left side plots. The magnified region is indicated by black dashed line in the left plot. The training compositions for Tessellation models are shown in Figure \ref{fig:T1toT4}. The training compositions for Active Learning models of \ce{SiGeSn} are shown in Figure \ref{fig:AL4figs}. The training compositions for Active Learning models of \ce{CrFeCoNi} are shown in Figure \ref{fig:AL_queries_quat}. \mysquare{67fb}: Maximum NRMSE (AL), \mysquare{67db}: Average NRMSE (AL), \mysquare{67fo}: Maximum NRMSE (T), \mysquare{67do}: Average NRMSE (T)
    }
    \label{fig:max_nrmse_al_vs_ts}
\end{figure*}

\begin{figure*}[htbp]
    \centering
\includegraphics[width=0.9\linewidth]{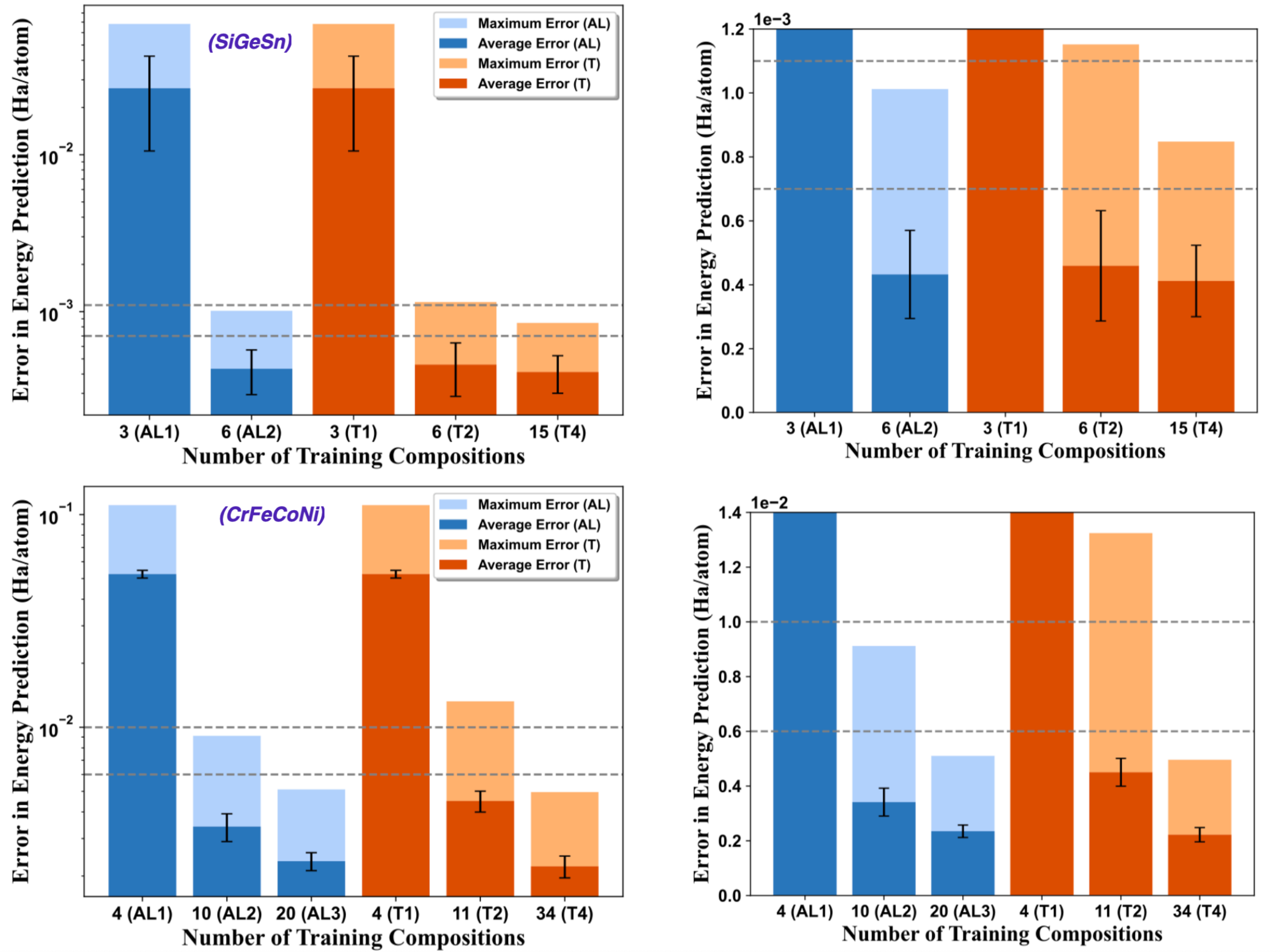}
    \caption{\textbf{Plots showing energy error in terms of Hartree/atom across composition space with increasing number of training compositions for \ce{SiGeSn} (top) and \ce{CrFeCoNi} (bottom)}. \textit{Top left:} Bulk 64-atom \ce{SiGeSn} results across composition space, logarithmic scale to emphasize the order of magnitude. \textit{Top right:} Magnified version of the \ce{SiGeSn} results, linear scale to emphasize the specific values. \textit{Bottom left:} Bulk 32-atom \ce{CrFeCoNi} results across composition space, logarithmic scale. \textit{Bottom right:} Magnified version of the \ce{CrFeCoNi} results, linear scale. The dashed lines are present to illustrate the magnification of the magnified plots. Standard deviation bars are shown in each of the plots. The training compositions for Tessellation models are shown in Figure \ref{fig:T1toT4}. The training compositions for Active Learning models of \ce{SiGeSn} are shown in Figure \ref{fig:AL4figs}. The training compositions for Active Learning models of \ce{CrFeCoNi} are shown in Figure \ref{fig:AL_queries_quat}.\mysquare{67fb}: Maximum Error (AL), \mysquare{67db}: Average Error (AL), \mysquare{67fo}: Maximum Error (T), \mysquare{67do}: Average Error (T)}
    \label{fig:ModelPlots}
\end{figure*}

\begin{figure*}[htbp]
    \centering

    \includegraphics[width=0.8\linewidth]{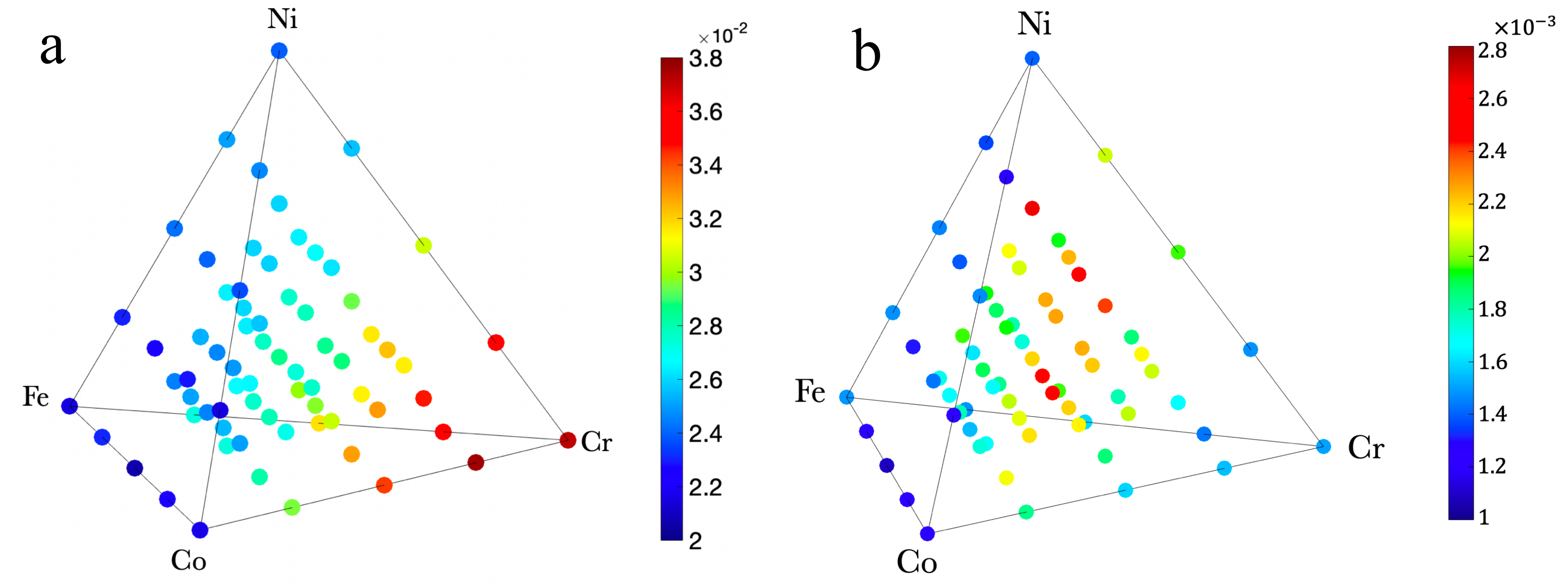}

    \caption{\textbf{Demonstration of accurate prediction of electron density and energy across composition space of Quaternary alloy.}
    \textbf{(a)} NRMSE in electron density for the pristine 32-atom \ce{CrFeCoNi} data set for AL2 model trained on $\delta\rho$. Note that the order of magnitude of the colorbar is $10^{-2}$.
    \textbf{(b)} Corresponding average error in energy at test compositions for the pristine 32-atom \ce{CrFeCoNi} data set, in terms of Ha/atom. Note that the order of magnitude of the colorbar is $10^{-3}$.}
    \label{fig:Quat_SAD_EnergyErrors}
\end{figure*}

\begin{figure}[htbp]
    \centering
    \includegraphics[width=0.85\linewidth]{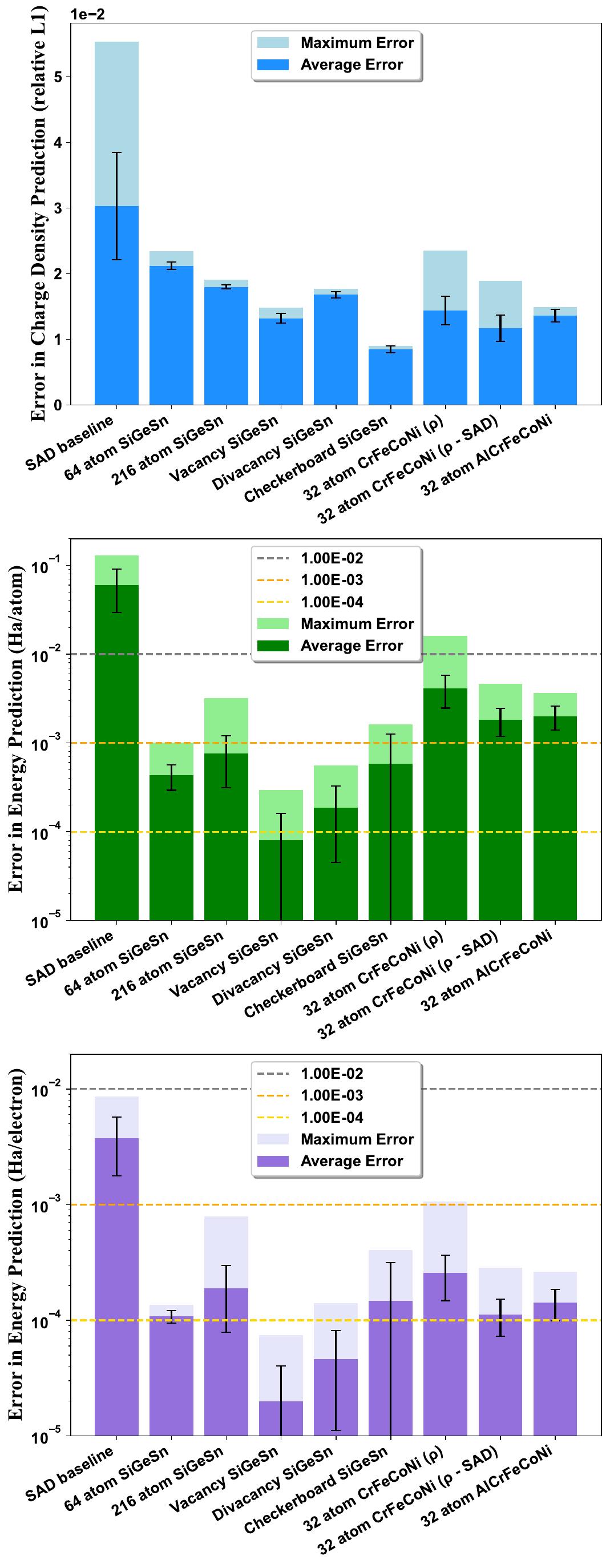}
    \caption{ \textbf{Plots showing highly accurate electron density and energy predictions for all systems assessed in this work. The comparative results shown here were obtained with the ternary AL2 model and the quaternary AL2 model respectively.} \textit{Top}: Accuracy in charge density predictions, in terms of relative L1. \textit{Middle}: Accuracy in energy predictions obtained from post-processing the charge densities, in terms of Hartree/atom. \textit{Bottom}: Accuracy in energy predictions, presented in terms of Hartree/electron. Note that Middle and Bottom plots have logarithmic scale. {The SAD baseline and AlCrFeCoNi system are discussed in Supplementary Material.}
    \textit{Top}: \mysquare{9tmax}: Maximum Error, \mysquare{9tavg}: Average Error. \textit{Middle}: \mysquare{9mmax}: Maximum Error, \mysquare{9mavg}: Average Error, \dashedgrey: $1\times 10^{-2}$, \dashedorange: $1 \times 10^{-3}$, \dashedyellow: $1\times 10^{-4}$.
    \textit{Bottom}: \mysquare{9bmax}: Maximum Error, \mysquare{9bavg}: Average Error, \dashedgrey: $1\times 10^{-2}$, \dashedorange: $1 \times 10^{-3}$, \dashedyellow: $1\times 10^{-4}$.
    }
    \label{fig:KeyResultsAccuracy}
    
\end{figure}

\begin{figure*}[t]
     \centering
     \includegraphics[width=0.8\linewidth]{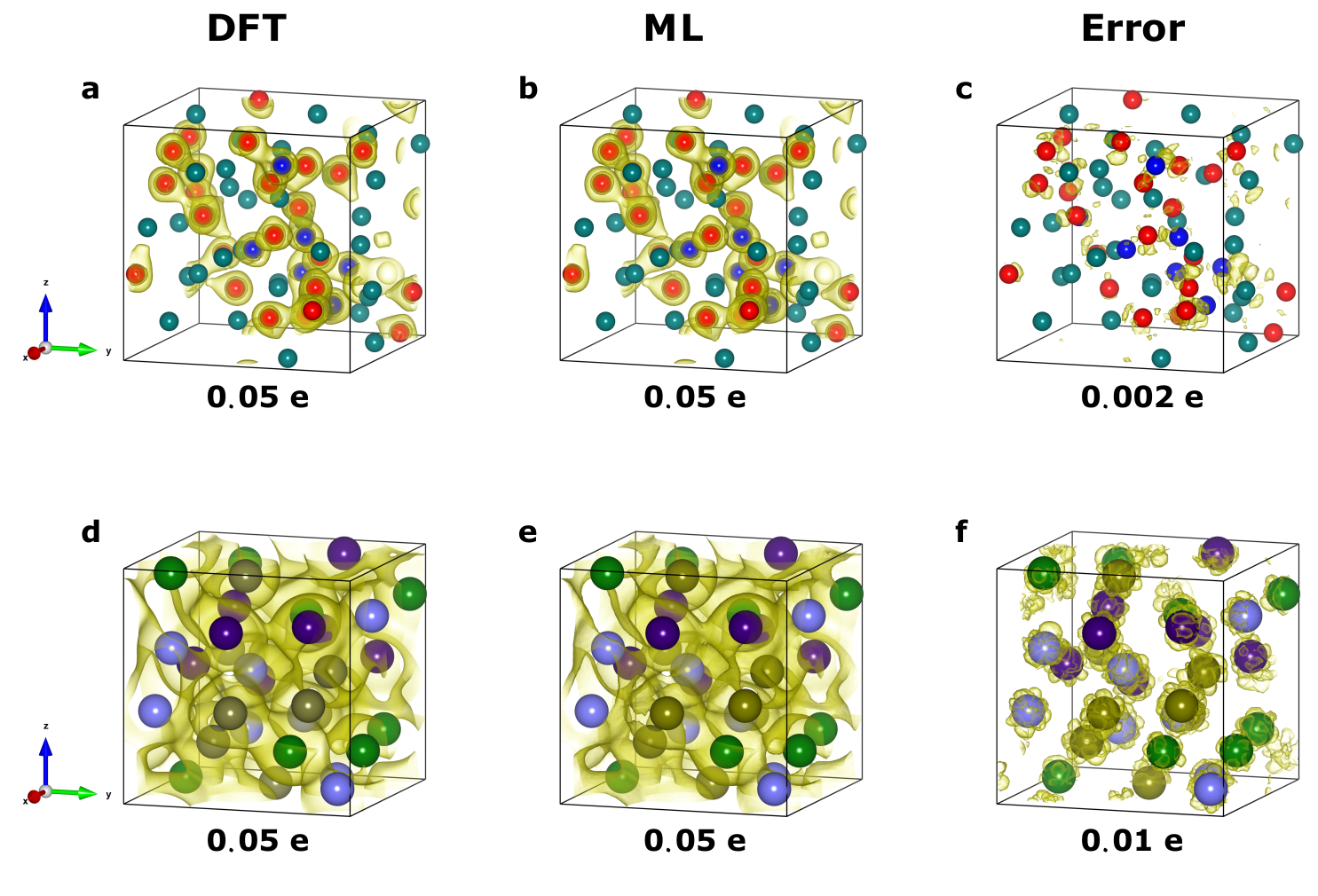}
     \caption{\textbf{Comparison of ML predicted and KS-DFT obtained electron density.} Electron densities (a, d) calculated by  KS-DFT and (b, e) predicted by ML, and the Error (absolute difference) between them (c, f) for
     \ce{SiGeSn} (a, b, c) and CrFeCoNi (d, e, f), using the AL2 model. Subplots (a, b, c) correspond to a 64-atom Si\textsubscript{12.5}Ge\textsubscript{37.5}Sn\textsubscript{50} simulation cell at 2400K.
     Subplots (d, e, f) are a 32-atom simulation cell at 5000K corresponding to  Cr\textsubscript{25}Fe\textsubscript{25}Ni\textsubscript{25}Co\textsubscript{25} $\delta\rho$ model respectively. The values below the snapshots refer to the iso-surface values. The visualization is done with the VESTA \cite{momma2008vesta} software.}
     \label{fig:cubeplots}
\end{figure*}

\begin{figure*}[t]
     \centering
     \includegraphics[width=0.8\linewidth]{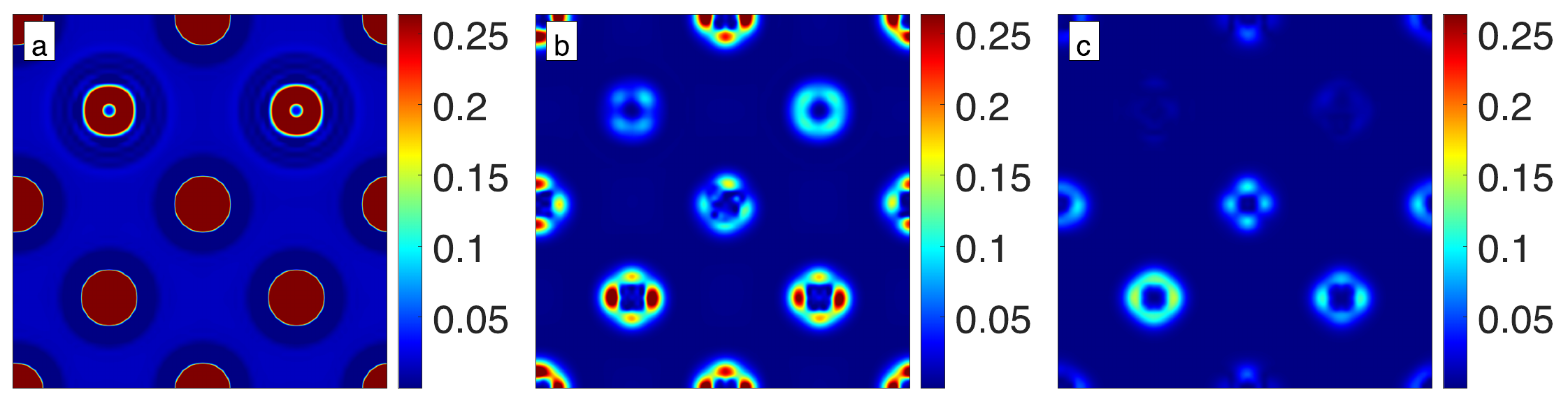}
     
     \caption{\textbf{Comparison of ML model and the $\delta\rho$ ML model by analysing errors in the electrostatic energy field, for the \ce{CrFeCoNi} system}. (a) Electrostatic energy field $\mathcal{E} = (\rho + b) \phi$ for the KS-DFT calculation. Here $\rho$ is the electron density, $b$ denotes the  nuclear pseudo charge field and $\phi$ is the electrostatic potential that includes electron-electron, electron-nucleus and nucleus-nucleus interactions. (b) The errors in the calculated electrostatic energy predicted field obtained through the ($\rho$-based) ML model. (c) The errors in the calculated electrostatic energy predicted field obtained through the $\delta\rho$ ML model. Most errors are seen to be concentrated around the atomic nuclei and are significantly reduced in case of the $\delta\rho$ ML model.  ML predictions are carried out using the AL2 model.}
     \label{fig:rho_SAD_comp}
\end{figure*}

\begin{figure}[htbp]
    \centering
   \includegraphics[width=0.99\linewidth]{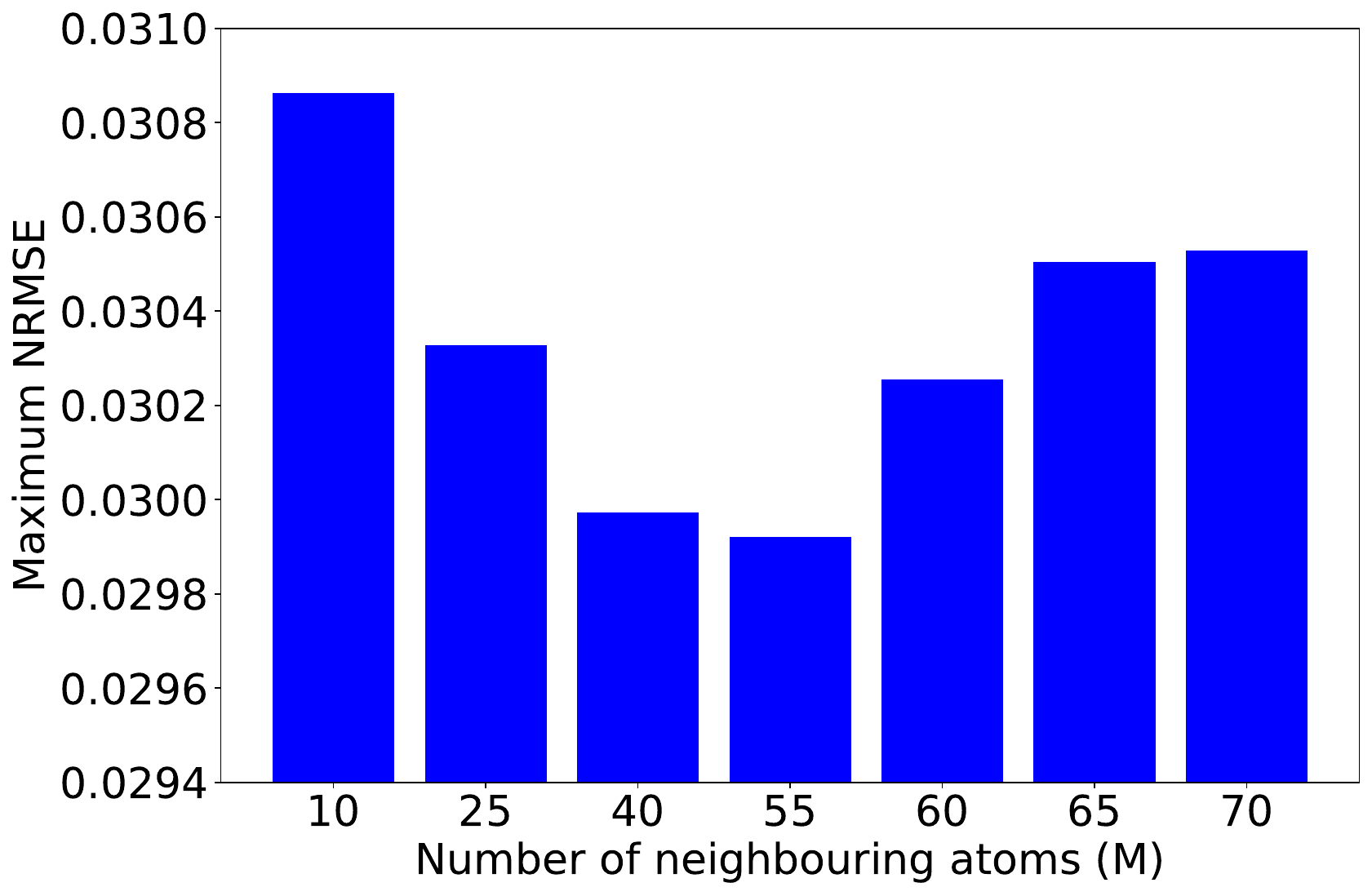}
    \caption{\textbf{Determination of the optimal set of descriptors}. For each ``M", we compute the descriptors for the training data, train the neural network and calculate the test NRMSE. The \ce{SiGeSn} system was used for this study.}
    \label{fig:feature_convergence}
\end{figure}

 \begin{figure}
     \centering
     \includegraphics[width=\linewidth]{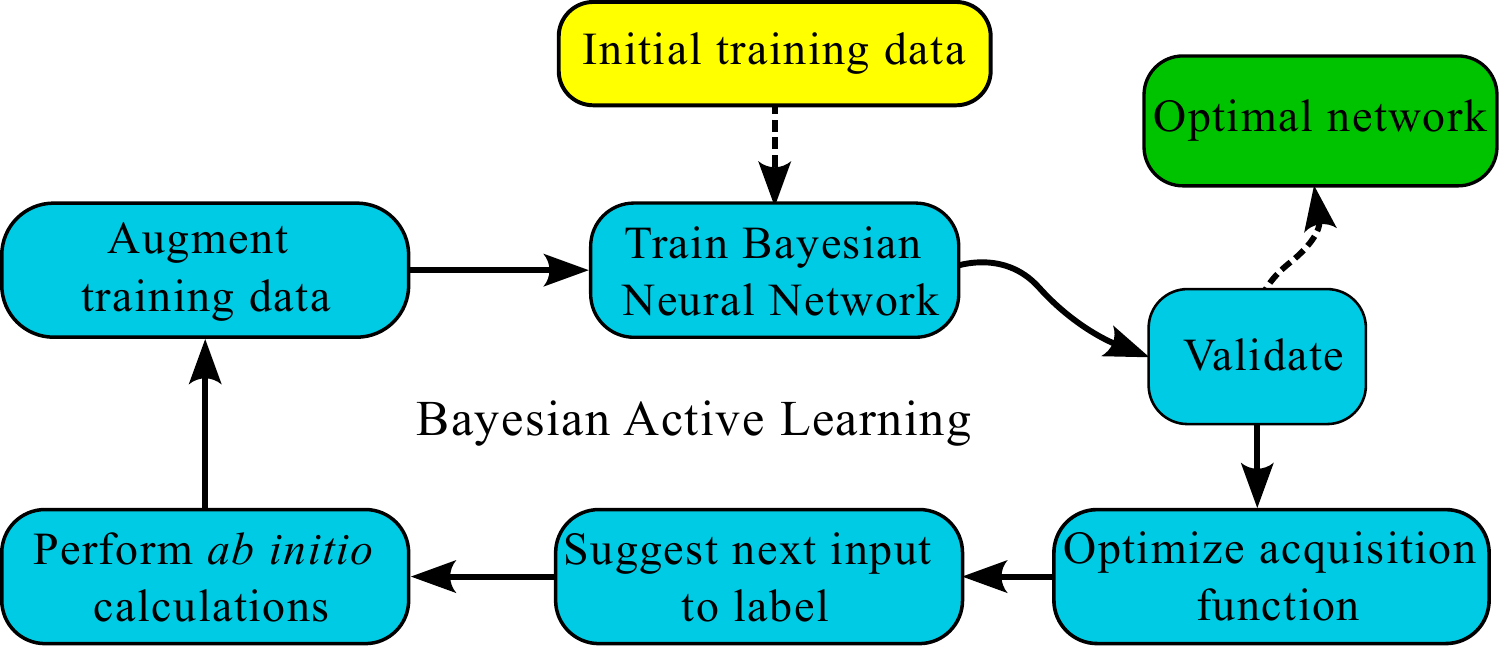}
     \caption{Schematic of the Bayesian active learning framework.}
     \label{fig:active_learning}
 \end{figure}

\begin{figure*}[htbp]
    \centering    
    \includegraphics[width=0.4\linewidth]{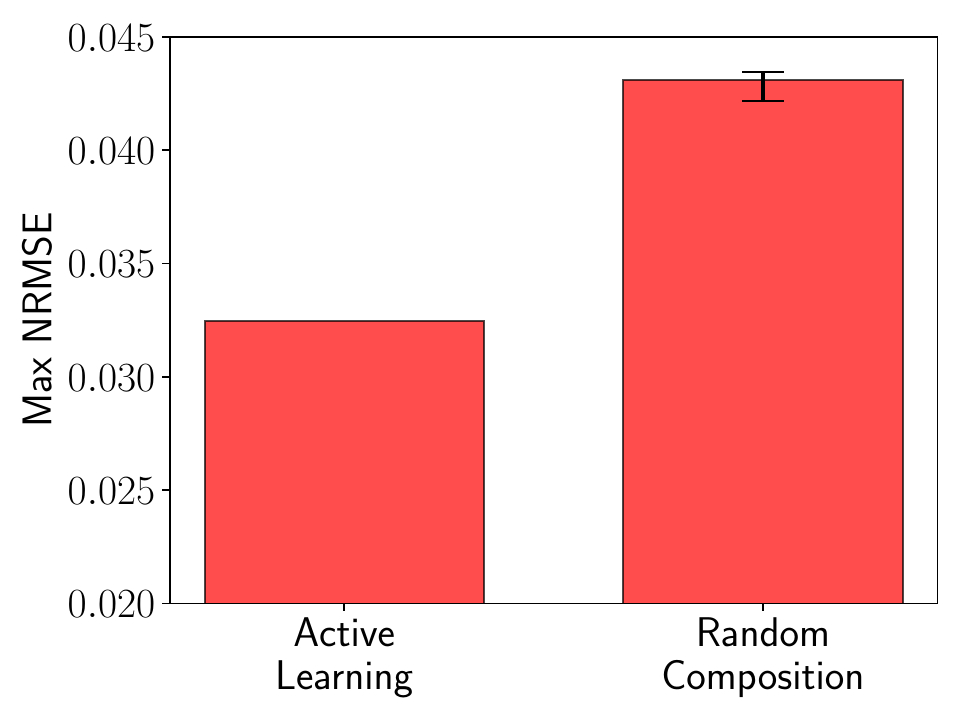}
    \caption{{\textbf{Advantage of Bayesian Active Learning over random selection of compositions.} This figure compares the maximum NRMSE across the composition space of the CrFeCoNi quaternary system using two different sampling strategies. The first bar shows the result from a model trained with 20 compositions selected via Bayesian Active Learning. The second bar corresponds to one of three models trained on 20 randomly selected compositions; the error bars indicate the range of maximum NRMSE values observed across the three models. All models were trained using the same number of data points, demonstrating the improved accuracy achieved through Bayesian Active Learning.}}
    \label{fig:baseline_model_comparison}
\end{figure*}

\clearpage
\newpage
\setcounter{page}{1}



\setcounter{subsection}{0}

\setcounter{figure}{0}
\renewcommand{\thefigure}{S\arabic{figure}}

\setcounter{table}{0}
\renewcommand{\thetable}{S\arabic{table}}

\begin{center}
\textbf{\large{Supplemental Material}}    
\end{center}

\subsection{Error Calculations}

The primary metrics used for error calculations in this work are:
\begin{equation*}\label{eq:nrmse}
    \text{NRMSE} = \frac{\sqrt{\frac{1}{N_{\text{grid}}} \sum_{i=1}^{N_{\text{grid}}} (\rho_i - \hat{\rho}_i)^2}}{\frac{1}{N_{\text{grid}}} \sum_{i=1}^{N_{\text{grid}}} |\rho_i|}
\end{equation*}

\begin{equation*}\label{eq:rel_l1}
    \text{Relative} \,L_\text{1} = \frac{{ \sum_{i=1}^{N_{\text{grid}}} |\rho_i - \hat{\rho}_i|}}{{ \sum_{i=1}^{N_{\text{grid}}}|\rho_{i}| }}
\end{equation*}
where, $\rho$ is the ground truth electron density and $\hat{\rho}$ is the ML-predicted electron density.

The error in energy predictions presented in the paper are absolute difference in Kohn-Sham Density Functional Theory (KS-DFT) obtained energy and energy obtained by postprocessing ML predicted electron density field, and is reported in Hartree/atom. 

\subsection{Compositions Used in the \textit{ab initio} Calculations}

Training and testing compositions for \ce{SiGeSn} are given in Figure \ref{fig:compSiGeSn}, while for \ce{CrFeCoNi} are given in Figure \ref{fig:quatcomp}. 

The complete list of training and testing compositions for the 64 and 216 atom \ce{SiGeSn} data are given in Tables \ref{tab:SiGeSn_64atom_complist} and \ref{tab:SiGeSn_216atom_complist}, respectively. Finally, Table \ref{tab:CrFeCoNi_32atom_complist} contains the complete list of training and testing compositions in the 32 atom \ce{CrFeCoNi} data.

\twocolumngrid

\begin{table}[htbp]
    \centering
    \begin{tabular}{|c|c|c|c|c|c|c|c|}\hline
    \shortstack{System \\ Index} & \% Si & \% Ge & \% Sn & \shortstack{System \\ Index} & \% Si & \% Ge & \% Sn \\\hline
    t-1  & 0    & 0     & 100   & t-24 & 25   & 75    & 0   \\\hline
    t-2  & 0    & 12.5  & 87.5  & t-25 & 37.5 & 0     & 62.5  \\\hline
    t-3  & 0    & 25    & 75    & t-26 & 37.5 & 12.5  & 50  \\\hline
    t-4  & 0    & 37.5  & 62.5  & t-27 & 37.5 & 25    & 37.5  \\\hline
    t-5  & 0    & 50    & 50    & t-28 & 37.5 & 37.5  & 25  \\\hline
    t-6  & 0    & 62.5  & 37.5  & t-29 & 37.5 & 50    & 12.5  \\\hline
    t-7  & 0    & 75    & 25    & t-30 & 37.5 & 62.5  & 0   \\\hline
    t-8  & 0    & 87.5  & 12.5  & t-31 & 50   & 0     & 50  \\\hline
    t-9  & 0    & 100   & 0     & t-32 & 50   & 12.5  & 37.5  \\\hline
    t-10 & 12.5 & 0     & 87.5  & t-33 & 50   & 25    & 25  \\\hline
    t-11 & 12.5 & 12.5  & 75    & t-34 & 50   & 37.5  & 12.5  \\\hline
    t-12 & 12.5 & 25    & 62.5  & t-35 & 50   & 50    & 0   \\\hline
    t-13 & 12.5 & 37.5  & 50    & t-36 & 62.5 & 0     & 37.5  \\\hline
    t-14 & 12.5 & 50    & 37.5  & t-37 & 62.5 & 12.5  & 25  \\\hline
    t-15 & 12.5 & 62.5  & 25    & t-38 & 62.5 & 25    & 12.5  \\\hline
    t-16 & 12.5 & 75    & 12.5  & t-39 & 62.5 & 37.5  & 0   \\\hline
    t-17 & 12.5 & 87.5  & 0     & t-40 & 75   & 0     & 25  \\\hline
    t-18 & 25   & 0     & 75    & t-41 & 75   & 12.5  & 12.5  \\\hline
    t-19 & 25   & 12.5  & 62.5  & t-42 & 75   & 25    & 0   \\\hline
    t-20 & 25   & 25    & 50    & t-43 & 87.5 & 0     & 12.5  \\\hline
    t-21 & 25   & 37.5  & 37.5  & t-44 & 87.5 & 12.5  & 0   \\\hline
    t-22 & 25   & 50    & 25    & t-45 & 100  & 0     & 0   \\\hline
    t-23 & 25   & 62.5  & 12.5  &      &      &       &     \\\hline
    \end{tabular}
    \caption{\textbf{List of training and testing compositions in the 64 atom \ce{SiGeSn} data set.} The error in electron density and energy prediction for these composition is given in \ref{fig:SiGeSn_DensEnergyError} and Main Text Figure \ref{fig:AL4figs}.}
    \label{tab:SiGeSn_64atom_complist}
\end{table}

    \begin{table}[htbp]
    \centering
    \begin{tabular}{|c|c|c|c|}\hline
    \shortstack{System \\ Index} & \% Si & \% Ge & \% Sn \\\hline
    T-1 & 9.259 & 29.63 & 61.111  \\\hline
    T-2 & 9.259 & 64.815 & 25.926  \\\hline
    T-3 & 29.63 & 60.185 & 10.185  \\\hline
    T-4 & 64.815 & 25 & 10.185  \\\hline
    T-5 & 60.185 & 9.259 & 30.556  \\\hline
    T-6 & 30.093 & 30.093 & 39.815  \\\hline
    T-7 & 30.093 & 39.815 & 30.093  \\\hline
    T-8 & 25 & 9.259 & 65.741  \\\hline
    T-9 & 39.815 & 30.093 & 30.093  \\\hline
    T-10 & 30.556 & 14.815 & 54.63  \\\hline
    T-11 & 50 & 14.815 & 35.185  \\\hline
    T-12 & 54.63 & 30.556 & 14.815  \\\hline
    T-13 & 0 & 50 & 50   \\\hline
    T-14 & 50 & 0 & 50  \\\hline
    T-15 & 50 & 50 & 0  \\\hline
    T-16 & 100 & 0 & 0  \\\hline
    T-17 & 0 & 100 & 0  \\\hline
    T-18 & 0 & 0 & 100  \\\hline
    T-19 & 35.185 & 50 & 14.815  \\\hline
    T-20 & 14.815 & 54.63 & 30.556 \\\hline
    T-21 & 14.815 & 35.185 & 50 \\\hline

    \end{tabular}	
			
    \caption{\textbf{List of compositions in the 216-atom \ce{SiGeSn} data set. These systems are used for testing purposes only.}  The error in electron density and energy prediction for these composition is given in Figure \ref{fig:errorSiGeSn_216atom}.}
    \label{tab:SiGeSn_216atom_complist}
    \end{table}


\begin{table}[htbp]
    \centering
    \scriptsize 
    \begin{adjustbox}{totalheight=10.5cm} 
    \begin{tabular}{|c|c|c|c|c|c|c|c|c|c|}\hline
    \shortstack{System \\ Index} & \% Cr & \% Fe & \% Co & \% Ni & \shortstack{System \\ Index} & \% Cr & \% Fe & \% Co & \% Ni \\\hline
    q-1  & 25    & 25    & 25    & 25    & q-36  & 12.5  & 12.5  & 12.5  & 62.5  \\\hline
    q-2  & 100   & 0     & 0     & 0     & q-37  & 12.5  & 12.5  & 25    & 50    \\\hline
    q-3  & 0     & 100   & 0     & 0     & q-38  & 12.5  & 12.5  & 37.5  & 37.5  \\\hline
    q-4  & 0     & 0     & 100   & 0     & q-39  & 12.5  & 12.5  & 50    & 25    \\\hline
    q-5  & 0     & 0     & 0     & 100   & q-40  & 12.5  & 12.5  & 62.5  & 12.5  \\\hline
    q-6  & 75    & 25    & 0     & 0     & q-41  & 12.5  & 25    & 12.5  & 50    \\\hline
    q-7  & 25    & 75    & 0     & 0     & q-42  & 12.5  & 25    & 25    & 37.5  \\\hline
    q-8  & 75    & 0     & 0     & 25    & q-43  & 12.5  & 25    & 37.5  & 25    \\\hline
    q-9  & 25    & 0     & 0     & 75    & q-44  & 12.5  & 25    & 50    & 12.5  \\\hline
    q-10 & 0     & 0     & 75    & 25    & q-45  & 12.5  & 37.5  & 12.5  & 37.5  \\\hline
    q-11 & 0     & 0     & 25    & 75    & q-46  & 12.5  & 37.5  & 25    & 25    \\\hline
    q-12 & 0     & 75    & 25    & 0     & q-47  & 12.5  & 37.5  & 37.5  & 12.5  \\\hline
    q-13 & 0     & 25    & 75    & 0     & q-48  & 12.5  & 50    & 12.5  & 25    \\\hline
    q-14 & 0     & 75    & 0     & 25    & q-49  & 12.5  & 50    & 25    & 12.5  \\\hline
    q-15 & 0     & 25    & 0     & 75    & q-50  & 12.5  & 62.5  & 12.5  & 12.5  \\\hline
    q-16 & 75    & 0     & 25    & 0     & q-51  & 25    & 12.5  & 12.5  & 50    \\\hline
    q-17 & 25    & 0     & 75    & 0     & q-52  & 25    & 12.5  & 25    & 37.5  \\\hline
    q-18 & 50    & 50    & 0     & 0     & q-53  & 25    & 12.5  & 37.5  & 25    \\\hline
    q-19 & 50    & 0     & 50    & 0     & q-54  & 25    & 12.5  & 50    & 12.5  \\\hline
    q-20 & 50    & 0     & 0     & 50    & q-55  & 25    & 25    & 12.5  & 37.5  \\\hline
    q-21 & 0     & 50    & 50    & 0     & q-56  & 25    & 25    & 37.5  & 12.5  \\\hline
    q-22 & 0     & 50    & 0     & 50    & q-57  & 25    & 37.5  & 12.5  & 25    \\\hline
    q-23 & 0     & 0     & 50    & 50    & q-58  & 25    & 37.5  & 25    & 12.5  \\\hline
    q-24 & 50    & 25    & 25    & 0     & q-59  & 25    & 50    & 12.5  & 12.5  \\\hline
    q-25 & 50    & 25    & 0     & 25    & q-60  & 37.5  & 12.5  & 12.5  & 37.5  \\\hline
    q-26 & 50    & 0     & 25    & 25    & q-61  & 37.5  & 12.5  & 25    & 25    \\\hline
    q-27 & 25    & 50    & 25    & 0     & q-62  & 37.5  & 12.5  & 37.5  & 12.5  \\\hline
    q-28 & 25    & 50    & 0     & 25    & q-63  & 37.5  & 25    & 12.5  & 25    \\\hline
    q-29 & 0     & 50    & 25    & 25    & q-64  & 37.5  & 25    & 25    & 12.5  \\\hline
    q-30 & 25    & 25    & 50    & 0     & q-65  & 37.5  & 37.5  & 12.5  & 12.5  \\\hline
    q-31 & 25    & 0     & 50    & 25    & q-66  & 50    & 12.5  & 12.5  & 25    \\\hline
    q-32 & 0     & 25    & 50    & 25    & q-67  & 50    & 12.5  & 25    & 12.5  \\\hline
    q-33 & 25    & 25    & 0     & 50    & q-68  & 50    & 25    & 12.5  & 12.5  \\\hline
    q-34 & 25    & 0     & 25    & 50    & q-69  & 62.5  & 12.5  & 12.5  & 12.5  \\\hline
    q-35 & 0     & 25    & 25    & 50    & & & & & \\\hline
    \end{tabular}
    \end{adjustbox}
    \caption{\textbf{List of training and testing compositions in the 32-atom \ce{CrFeCoNi} data set}. Note that not all compositions are used for training. The error in electron density and energy prediction for these composition is given in Figures  \ref{fig:DensEnergyError_CrFeCoNi}.}
    \label{tab:CrFeCoNi_32atom_complist}
\end{table}

\newpage

\subsection{Details of Data Generation Methodology:}

In the context of this work, the `data' that was used for both testing and training the model consisted of a large batch of snapshots. Each snapshot represents an atomic arrangement in the simulation cell. For each of the snapshots, a grid in real space is considered and the corresponding electron density value at each grid point is obtained. The process of obtaining these `snapshot' files was as follows:

\begin{enumerate}
    \item An atomic configuration was selected.
    \item A Kohn-Sham density functional theory (KS-DFT) calculation was performed to obtain the ground state electron densities associated with that atomic configuration.
    \item Text-processing was performed on the electron density output to format it for further calculations. 
\end{enumerate}

\noindent For each input atomic configuration, one snapshot file would be obtained, which could then be used for either training or testing. 

To produce the atomic configurations, one of three options was leveraged: sampling from the trajectory of an \textit{ab initio} molecular dynamics (AIMD) simulation, sampling from the trajectory of a classical molecular dynamics (MD) simulation, or handcrafting more unique systems for the purpose of test data (e.g. defects, checkerboard boundaries). 

The first option, AIMD, was used for all of the \ce{SiGeSn} system data (ternary, binary, and unary derivatives). This option was selected due to the ease of implementation; we obtained coordinates and electron densities, and both of these were calculated with high fidelity at each step of the AIMD trajectory. We leveraged this approach in our previous work \cite{pathrudkar2024electronic}, and it was straightforward to extend the method to the ternary \ce{SiGeSn} system. This streamlined many aspects of the data generation. However, the downside of this approach is its high computational expense; since most of the steps in the trajectory are not included in the final data pool (they are simply intermediate steps), a great deal of computational resources go into electronic structure calculations that are --- in the context of data generation --- unused.

In an effort to leverage a significantly cheaper alternative to AIMD, the second option, classical molecular dynamics (MD), was used generating the atomic configurations for all of the \ce{CrFeCoNi} system data (quaternary, ternary, binary, and unary derivatives). Since a quaternary system has a higher degree of freedom with respect to the compositions, a greater number of simulations were required to obtain an adequate pool of data across composition space. Furthermore, each electronic structure calculation was itself more expensive, due to the inclusion of semi-core electrons for the atomic species Cr, Fe, Co, and Ni. Leveraging MD meant introducing an additional step into the data generation pipeline, but it vastly increased the efficiency at which atomic configurations and their corresponding electron densities could be obtained.

{\textbf{Lattice constant calculations:}} One parameter needed as input in both our MatGL-enabled MD simulations and SPARC-enabled AIMD and static KS-DFT calculations is the lattice constant. For a flexible approach that allowed for the simulation of any composition choice within our alloy systems, we opted for implementing the rule-of-mixtures, Vegard's Law \cite{jacob2007vegard}. First, we obtained the lattice constant predicted for the pure elements Si, Ge, Sn, Cr, Fe, Co, and Ni by each respective pseudopotential, imposing the specified lattice geometry. Due to their impact on the rest of the data generation, these calculations were done at a higher level of precision. Then, the lattice constant for any alloy composition was obtained by taking a weighted average of the lattice constants from the pure elements, in proportion to the elemental composition of that alloy. The pure lattice constants are shown in Table \ref{tab:latticeconstants}. While this is a simplifying assumption, it was sufficient for our purposes. It is encouraging to note that when the lattice parameter is relaxed with MatGL, the optimized lattice parameter obtained is not significantly different, as shown in Table \ref{tab:m3gnetlc}. As a consequence the averaged electron density also approximately follows Vegard's law. This makes sense because the average electron density is contingent on the volume, and the volume of our cubic systems is simply the lattice constant cubed.

    \begin{table}[t]
    \centering
    \begin{tabular}{|c|c|c|c|c|}\hline
    Element & \shortstack{Lattice \\ Geometry} & \shortstack{Lattice \\ Constant \\ Obtained} & \shortstack{Materials \\ Project \\ Reference} & \shortstack{Experimental \\ Reference \cite{Barrett1966}} \\\hline
    Si & diamond & \textbf{5.47} & \textit{5.44} & \textit{5.42} \\\hline
    Ge & diamond & \textbf{5.76} & \textit{5.67} & \textit{5.65} \\\hline
    Sn & diamond & \textbf{6.63} & \textit{6.57} & \textit{6.46} \\\hline
    Cr & FCC & \textbf{3.65} & \textit{3.58} & \textit{-}\\\hline
    Cr & BCC & \textit{2.86} & \textit{2.97} & \textit{2.88}\\\hline
    Fe & FCC & \textbf{3.45} & \textit{3.66} & \textit{3.56} \\\hline
    Fe & BCC & \textit{2.76} & \textit{2.86} & \textit{2.93} \\\hline
    Co & FCC & \textbf{3.45} & \textit{3.51} & \textit{3.55} \\\hline
    Ni & FCC & \textbf{3.51} & \textit{3.48} & \textit{3.52} \\\hline

    \end{tabular}

    \caption{Lattice constants obtained from the pseudopotentials employed in this study, in units of Angstrom, for the 8-atom diamond unit cell, the 4-atom face-centered cubic (FCC) unit cell, and the 2-atom body-centered cubic (BCC) unit cell. Values in \textbf{bold} were leveraged for the rule-of-mixtures appraoch. Values in \textit{italics} were not leveraged, but are provided here to for the purpose of comparison. Note that while Fe and Co both exhibit FCC phases at higher temperature, Cr does not have an experimentally observed FCC phase in nature. The lattice constant for FCC Cr was used to match the FCC lattice imposed upon our \ce{CrFeCoNi} alloy systems.}
    \label{tab:latticeconstants}
    \end{table}

    \begin{table}[htbp]
    \centering
    \begin{tabular}{|c|c|c|}\hline
    \shortstack{Alloy \\ Composition} & \shortstack{M3GNet-\\Optimized \\ Lattice \\Constant} & \shortstack{Vegard's Law \\ Lattice\\ Constant} \\\hline

    Cr & 3.61 & 3.65 \\\hline
    Fe & 3.46 & 3.45 \\\hline
    Ni & 3.50 & 3.51 \\\hline
    Co & 3.52 & 3.45 \\\hline
    Cr$_{0.5}$Fe$_{0.5}$ & 3.57 & 3.55 \\\hline
    Fe$_{0.5}$Ni$_{0.5}$ & 3.56 & 3.48 \\\hline
    Ni$_{0.5}$Co$_{0.5}$ & 3.50 & 3.48 \\\hline
    Cr$_{0.5}$Ni$_{0.5}$ & 3.53 & 3.58 \\\hline
    Cr$_{0.25}$Fe$_{0.5}$Ni$_{0.25}$ & 3.55 & 3.51 \\\hline
    Cr$_{0.5}$Fe$_{0.25}$Co$_{0.25}$ & 3.55 & 3.55 \\\hline
    Cr$_{0.5}$Co$_{0.25}$Ni$_{0.25}$ & 3.54 & 3.57 \\\hline
    Fe$_{0.25}$Co$_{0.25}$Ni$_{0.5}$ & 3.53 & 3.51 \\\hline
    Cr$_{0.25}$Fe$_{0.25}$Co$_{0.25}$Ni$_{0.25}$ & 3.54 & 3.52 \\\hline
    Cr$_{0.125}$Fe$_{0.25}$Co$_{0.375}$Ni$_{0.25}$ & 3.53 & 3.49 \\\hline
    Cr$_{0.25}$Fe$_{0.125}$Co$_{0.25}$Ni$_{0.375}$ & 3.52 & 3.52 \\\hline
    Cr$_{0.25}$Fe$_{0.375}$Co$_{0.25}$Ni$_{0.125}$ & 3.55 & 3.51 \\\hline
    \end{tabular}

    \caption{Comparison between unit cell lattice parameter obtained from volume relaxation with MatGL and the unit cell lattice parameter obtained from the rule-of-mixtures weighted average approach described in the text. These MatGL simulations were done with 6x6x6 supercells of 864 atoms in FCC lattice geometry, and the lattice constants shown are scaled down by a factor of six to allow for facile comparison with Table \ref{tab:latticeconstants}. Units in Angstrom.} 
    \label{tab:m3gnetlc}
    \end{table}

\textbf{{Data generation of \ce{SiGeSn} systems:}} 
For \ce{SiGeSn}, the fractional coordinates for an 8-atom unit cell of the diamond lattice structure were scaled up to produce 64-atom and 216-atom supercells. 
Atom labels were randomly appended to these coordinates, in accordance with the desired alloy composition. These initial atomic configurations were then converted to the SPARC-required format of a .ion file, and the fractional coordinates were scaled by the lattice constant obtained via Vegard's Law. 
For each .ion file, a corresponding .inpt file containing the AIMD settings for SPARC was generated. These files were then fed into SPARC and allowed to run, generating an atomic trajectory. After a short equilibration period to allow the temperature and energy fluctuations to stabilize, snapshots were extracted at fixed intervals and the corresponding electron densities associated with those snapshots were collected. 
For the 64-atom case, there were 45 unique compositions, 4 random initial atomic configurations for each composition, 1 AIMD temperature (2400K), and 6 snapshots collected from each AIMD run. This yielded a total of 900 data points. For the 216-atom case, there were 21 unique compositions, 1 random initial atomic configuration, 1 AIMD temperature (2400K), and 3 snapshots collected. This yielded a total of 63 data points. The 216-atom \ce{SiGeSn} data was used for testing, while the 64-atom data was used for both training and testing. See {Figure \ref{fig:compSiGeSn}} (and \textbf{Tables \ref{tab:SiGeSn_64atom_complist}} and \textbf{\ref{tab:SiGeSn_216atom_complist}}) for the \ce{SiGeSn} compositions.

\textbf{{Data generation of \ce{CrFeCoNi} systems:}} For \ce{CrFeCoNi}, the fractional coordinates for a 4-atom unit cell of the face-centered cubic (FCC) structure were scaled up to produce 32-atom supercells. Randomly selected atom labels were assigned to the coordinates, in proportion to the alloy composition. These initial atomic configurations were converted to a data structure that MatGL could read, a Python file containing the molecular dynamics settings was generated, and the molecular dynamics (MD) simulation was run.  Snapshots were collected from the trajectory at fixed intervals after a short equilibration time. 
SPARC was then run, thus producing the electron densities associated with the extracted snapshots. 
For the 32-atom case, there were 69 unique compositions, 4 random initial atomic configurations per composition, 4 MD temperatures (4000K, 5000K, 6000K, 7000K), and 7 snapshots collected. This yielded a total of 7728 data points. The 32-atom data was used for both training and testing. See {Figure \ref{fig:quatcomp}} (and \textbf{Tables \ref{tab:CrFeCoNi_32atom_complist}} for the \ce{CrFeCoNi} compositions.

\textbf{{Data generation of special \ce{SiGeSn} systems:}} Three forms of additional \ce{SiGeSn} test data were produced for model generalizability studies: monovacancy-containing, divacancy-containing, and handcrafted `checkerboard' systems (with species segregation). For convenience, the \textit{atomsk} package \cite{hirel2015atomsk} was used in generating some of these system configurations. For the monovacancy and divacancy data, atom labels were randomly assigned to the coordinate sites of a pristine 64-atom diamond cubic lattice cell such that the final composition matched one of twelve pre-selected compositions. To generate a monovacancy-containing system, a randomly selected atom that matched the desired atom type would be removed from the cell. By choosing to remove either a Si, Ge, or Sn atom, each snapshot produced three derivative data points. To generate a divacancy-containing system, a randomly selected atom, and its neighbor would be removed as a pair, producing one of six possible scenarios: a missing SiSi, GeGe, SnSn, \ce{SiGe}, GeSn, or SiSn pair.  Since it was also of interest to see how well the trained model could make predictions for intersections of bulk elemental regions, handcrafted `checkerboard' systems were produced. Cubic simulation cells of 64 and 216 atoms occupying diamond lattice sites were divided up into smaller cubic sub-regions, i.e. either 8 bins (2x2x2) for the 64-atom and 216-atom cells, or 27 bins (3x3x3) for the 216-atom cell. Elemental labels were then assigned to each bin, such that no two neighboring bins contained the atoms of the same element, with periodic boundaries taken into consideration as well. In the 8-bin case, three compositions were considered: $\text{Si}_{0.25}\text{Ge}_{0.375}\text{Sn}_{0.375}$, $\text{Si}_{0.375}\text{Ge}_{0.25}\text{Sn}_{0.375}$, and $\text{Si}_{0.375}\text{Ge}_{0.375}\text{Sn}_{0.25}$. In the 27-bin case, just the equiatomic \ce{SiGeSn} case was considered (e.g. $\text{Si}_{0.33}\text{Ge}_{0.33}\text{Sn}_{0.33}$).

In total, the data generated for this study consisted of:
\begin{itemize}
    \item 1080 snapshots for the 64-atom \ce{SiGeSn} system
    \item 63 snapshots for the 216-atom \ce{SiGeSn} system
    \item 7728 snapshots for the 32-atom \ce{CrFeCoNi} system
    \item 36 snapshots for the vacancy-containing \ce{SiGeSn} systems
    \item 72 snapshots for the divacancy-containing \ce{SiGeSn} systems
    \item 7 snapshots for the handcrafted, `checkerboard' \ce{SiGeSn} systems.
\end{itemize}

It is noted that being able to obtain more data more quickly with the augmented MD method enhances the feasibility of further studies with large system sizes and greater number of alloying elements.\\

\newpage

\subsection{Additional results} 

Additional supporting plots are presented in this section. Figure \ref{fig:Quat_SAD_EnergyErrors}\,b in the main text shows a plot of error in energy for the AL2 model trained on $\delta\rho$; for the sake of comparison, Figure \ref{fig:Quat_EnergyErrors} shown here presents error in energy for the AL2 model trained on $\rho$.

Figure \ref{fig:errorSiGeSn_216atom} shows density and energy prediction errors for the larger 216-atom \ce{SiGeSn} system; recall that only the 64-atom \ce{SiGeSn} data was used in the training set.

In Figure \ref{fig:AL4figs} in the main text, \ce{SiGeSn} prediction results for the Bayesian Active Learning approach are shown. Figure \ref{fig:Fig26_HEAML_Ternary_Collage_colorbar} shows the analogous results for the Tessellation approach. Also, Figure \ref{fig:AL4figs} in the main text has the same colorbar scales for AL1 and AL2 model. This is to illustrate the decrease in errors obtained from the AL2 model vs. the AL1 model. However, presenting the results in that fashion inhibits the readability of the AL2 values. To circumvent the readability issue, Figure \ref{fig:AL4figs_true_scale} is presented below, with separate colorbar scales. 

Figures \ref{fig:defect_slices}, \ref{fig:cubeplots_2} and \ref{fig:cubeplots_3} demonstrate the difference between `ground-truth' KS-DFT-obtained electron densities and ML-obtained electron densities for a sample of systems. Specifically, Figure \ref{fig:defect_slices} shows a \ce{SiGeSn} vacancy-containing snapshot and Figure \ref{fig:cubeplots_2} shows a thermalized 216-atom \ce{SiGeSn} snapshot and a thermalized 32-atom CrFeCoNi snapshot.

Figure \ref{fig:KeyResultsAccuracy} in the main text shows key results for the systems considered in this work, but the results shown there are aggregated over all compositions. Since the composition-dependent values may also be of interest, Figures \ref{fig:SiGeSn_DensEnergyError} and \ref{fig:DensEnergyError_CrFeCoNi} plot the error in electron density and error in energy by composition. Figure \ref{fig:SiGeSn_DensEnergyError} shows the results for the 45 compositions present in the 64-atom \ce{SiGeSn} test data. Figure \ref{fig:DensEnergyError_CrFeCoNi} shows the results for the 69 compositions present in the 32-atom CrFeCoNi test data. Note that the \ce{SiGeSn} results in Figure \ref{fig:SiGeSn_DensEnergyError} come from a model trained on $\rho$ (the charge density field), while the CrFeCoNi results in Figure \ref{fig:DensEnergyError_CrFeCoNi} come from a model trained on $\delta\rho$ (the difference between the charge density field and atomic densities). In both cases, though, the density errors displayed are for the charge density itself.

\clearpage
\subsection{Extension to Quinary System: AlCrFeCoNi}
To further illustrate that our methodology works well for typical high entropy alloys, we trained an additional model based on the quinary AlCrFeCoNi system, and focused on near-equiatomic compositions.

The details of this model are as follows. Data was generated in the same fashion as the quaternary \ce{CrFeCoNi} system; a crystalline 32-atom face-centered cubic supercell was assigned atom labels corresponding to different composition percentages. The twenty near-equi-atomic compositions that were selected for data generation are shown in Table \ref{tab:AlCrFeCoNi_complist}. 
The crystalline system was set as the initial configuration for a classical molecular dynamics simulation performed at 4000K with MatGL \cite{chen2021learning, chen2022universal}. After an equilibration period of 1000 timesteps (each timestep is one femtosecond), snapshots were extracted from the trajectory, at 100-timestep intervals. Data generation was minimal; only one trajectory per alloy composition was employed, yielding a total of 10 configuration snapshots for each composition (the initial crystalline configuration and nine thermalized configurations drawn from the molecular dynamics trajectory). Twenty compositions with ten configurations each yielded a total data pool of 200 snapshots for this model. 

We constrained our quinary \ce{AlCrFeCoNi} model to the face-centered cubic (FCC) system only, neglecting any potential phase transitions to other lattices. The possibility of exploration of the utility of our ML model across crystal systems is left as future work.

The architecture of the \ce{AlCrFeCoNi} neural network was identical to the other models produced in this study. 
The model was trained on data from 16 of the compositions, and the remaining 4 compositions were leveraged for testing. 
Since the $\rho-$SAD approach worked well at reducing the error metrics in the quaternary system, we trained the \ce{AlCrFeCoNi} model using $\rho-$SAD instead of just $\rho$. Model training took only 14 CPU-hours for this system. Figure 9 in the main text shows the error in electron density and energy prediction, as obtained from the 4 near-equiatomic test compositions.
The values obtained are just as good in accuracy as those obtained for the quaternary system, suggesting the successful application of the ML model to typical 5-element high entropy alloy systems.

    \begin{table}[htbp]
    \centering
    \begin{tabular}{|c|c|c|c|c|c|}\hline
    \shortstack{System \\ Index} & \% Al & \% Cr & \% Fe & \% Co & \% Ni \\\hline
    R-1 & 0 & 0 & 0 & 0 & 100  \\\hline
    R-2 & 0 & 0 & 0 & 100 & 0  \\\hline
    R-3 & 0 & 0 & 100 & 0 & 0  \\\hline
    R-4 & 0 & 100 & 0 & 0 & 0 \\\hline
    R-5 & 12.5 & 12.5 & 25 & 25 & 25  \\\hline
    R-6 & 12.5 & 25 & 12.5 & 25 & 25  \\\hline
    R-7 & 12.5 & 25 & 25 & 12.5 & 25  \\\hline
    R-8 & 12.5 & 25 & 25 & 25 & 12.5  \\\hline
    R-9 & 18.75 & 18.75 & 18.75 & 18.75 & 25  \\\hline
    R-10 & 18.75 & 18.75 & 18.75 & 25 & 18.75  \\\hline
    R-11 & 18.75 & 18.75 & 25 & 18.75 & 18.75  \\\hline
    R-12 & 18.75 & 25 & 18.75 & 18.75 & 18.75  \\\hline
    R-13 & 25 & 12.5 & 12.5 & 25 & 25  \\\hline
    R-14 & 25 & 12.5 & 25 & 12.5 & 25  \\\hline
    R-15 & 25 & 12.5 & 25 & 25 & 12.5  \\\hline
    R-16 & 25 & 18.75 & 18.75 & 18.75 & 18.75  \\\hline
    R-17 & 25 & 25 & 12.5 & 12.5 & 25   \\\hline
    R-18 & 25 & 25 & 12.5 & 25 & 12.5 \\\hline
    R-19 & 25 & 25 & 25 & 12.5 & 12.5  \\\hline
    R-20 & 100 & 0 & 0 & 0 & 0 \\\hline

    \end{tabular}	
			
    \caption{{\textbf{List of compositions in the 32-atom \ce{AlCrFeCoNi} data set.} Compositions R-8, R-9, R-10, and R-11 were used for testing and the other compositions were included in the training dataset. The four testing compositions were selected at random.}}
    \label{tab:AlCrFeCoNi_complist}
    \end{table}

\color{black}

\newpage
\clearpage
\subsection{Efficiency Comparison}
In the following, we elaborate on how much faster the ML approach is, for compositional space exploration compared to the conventional Kohn-Sham Density Functional Theory (KS-DFT) approach. A key feature of the analysis presented below is the inclusion of ``offline'' or training data generation costs, that is often left out in other similar studies \cite{teh2021machine}. We consider the following four systems presented in the paper:
\begin{itemize}
    \item 64-atom SiGeSn system
    \item 216-atom SiGeSn system
    \item 32-atom CrFeCoNi system
    \item 32-atom AlCrFeCoNi system
\end{itemize}
For the above four systems we consider the cost of exploring the composition space. For the purpose of illustration, we presume that we wish to  predict the electron density for $1000$ compositions, with $10$ multiple distinct configurations each (disordered alloy properties for a given composition are often required to be obtained as averages over several configurations). Given that, the cost of exploration using KS-DFT will be:
\begin{equation}
    \text{Cost}_{DFT} = N_C \times N_S \times C_{DFT}
    \label{eqn:dftcost}
\end{equation}
where, $N_C$ is the number of compositions, $N_S$ is the number of configurations per composition and $C_{DFT}$ is the computational cost of a single electron density calculation via KS-DFT.

The cost of exploration using ML approach will be:
\begin{equation}
    \text{Cost}_{ML} = \text{Cost}_{\text{Data}} + \text{Cost}_{\text{Training}} +  \text{Cost}_{\text{Inference}}
\end{equation}
where, $\text{Cost}_{\text{Data}}$ is the training data generation cost, $\text{Cost}_{\text{Training}}$ is the cost of training the ML model and $\text{Cost}_{\text{Inference}}$ is the cost of prediction from the ML model. The cost for training data generation can be calculated using Equation \ref{eqn:dftcost} based on the number of compositions and configurations used for training only. The cost of prediction however, will involve prediction at all compositions and all configurations. 

To illustrate our claim of improved efficiency in more quantitative terms, {Table \ref{tab:costDemonstration}} compares the cost of the KS-DFT approach vs. the ML approach, as determined from our experiences with data generation and model training in this study. (Note that the ML model trained using 64-atom SiGeSn data is used for 216-atom SiGeSn system as well.) 
It is evident that the `total cost' values shown in Table \ref{tab:costDemonstration} are dependent on the number of system snapshots to be processed (where the number of snapshots is equal to `number of compositions to obtain electron densities for' times `number of configurations at each composition'). At lower snapshot quantities, the KS-DFT approach is more efficient, because the cost of training a model presents a higher upfront cost. However, the cost of using the ML model to predict the density is much cheaper than the cost of performing a KS-DFT calculation. Thus, even though the KS-DFT approach is cheaper at smaller snapshot quantities, as the number of snapshots for which to obtain electron densities increases, the efficiency of the ML approach will outperform that of KS-DFT. Also note that for the sake of a fair comparison, all computations relevant to obtaining these costs were carried out  on CPUs (for both KS-DFT and ML approaches). In realistic deployment scenarios, ML training, testing and inference would be carried out GPUs, thus making the large performance gains of the ML based approach even more likely. 

In line with the above discussion, for each system, we observe a \textit{crossover point} where the ML approach becomes less computationally costly than the KS-DFT approach.
Figure \ref{fig:efficiencyGain} displays these crossover points for the four models that we trained. The location of the crossover point depends primarily on how expensive one KS-DFT calculation is for that system, and also on how expensive the ML model is to train. Since the 64-atom SiGeSn, owing to soft pseudopotentials and coarser KS-DFT calculation meshes, has a very low computational cost for each KS-DFT calculation (as listed in Table \ref{tab:costDemonstration}), it takes a larger number of snapshots before the ML approach overtakes the KS-DFT approach in terms of computational efficiency. Additionally, Figure \ref{fig:consolidatedPlot} consolidates the subplots of Figure \ref{fig:efficiencyGain} into a single plot, to enable a direct comparison of efficiency gains from scalability.
We emphasize that the values provided in Table \ref{tab:costDemonstration}, Figure \ref{fig:efficiencyGain}, and Figure \ref{fig:consolidatedPlot} reflect the specific models that we trained, and these values could change depending on pseudopotential choices, KS-DFT calculation parameters, model training choices, and so forth. Nevertheless, we believe that these examples are illustrative of the efficiency gains from the ML approach.

Based on the above discussion, and upon looking over the crossover points shown in Figure \ref{fig:efficiencyGain}, we are led to the fact that the relative efficiency of the KS-DFT and ML approaches for compositional exploration is ultimately dependent on how many compositions are required for exploration of the composition space for a given alloy system. Table \ref{tab:countingUniqueCompositions} addresses this question. It is clear that the number of alloy compositions scales rapidly with the number of species, as well as fineness with which the composition space is sampled (i.e., the  percentage increments in each elemental concentration), leading to a combinatorial explosion that renders exhaustive first-principles exploration computationally prohibitive. We also remark that since our KS-DFT calculations do not use any kind of statistical averaging over atomic configurations,  the simulation supercell size directly constrains the achievable percentage increments. Smaller increments require larger supercells, which in turn increase computational cost due to the cubic scaling of KS-DFT with system size. As a result, the ML-based exploration approaches presented here become even more attractive as the number of species grows and/or the desired resolution in compositional space increases.

\newpage

\color{black}
\onecolumngrid

    \begin{table}[htbp]
    \centering
    \begin{tabular}{|c|c|c|c|c|c|}\hline
    \shortstack{ \textcolor{white}{place} \\ \textcolor{white}{holder}} & \shortstack{Comparison \\ Category} & \shortstack{\\32-atom \\ CrFeCoNi \\ Example} & \shortstack{\\64-atom \\ SiGeSn \\ Example} & \shortstack{\\216-atom \\ SiGeSn \\ Example} & \shortstack{\\32-atom \\ AlCrFeCoNi \\ Example} \\\hline
    \shortstack{Number of compositions to \\ obtain electron densities for}& KS-DFT & 1000 & 1000 & 1000 & 1000 \\\hline
    \shortstack{Number of configurations \\ at each composition} & KS-DFT & 10 & 10 & 10 & 10 \\\hline
    \shortstack{Cost of a single electron \\ density calculation, via KS-DFT \\ (in CPU hours)} & KS-DFT & 0.025 & 0.005 & 0.12 & 0.03 \\\hline
    \shortstack{\textbf{Total cost of KS-DFT approach} \\ \textbf{(in CPU hours)}} & \textbf{KS-DFT} & \textbf{250} & \textbf{50} & \textbf{1200} & \textbf{300} \\\hline
    \shortstack{Number of compositions to \\ obtain training data for}& ML & 10 & 6 & 6 & 16 \\\hline
    \shortstack{Number of configurations \\ at each training composition} & ML & 112 & 24 & 24 & 11 \\\hline
    \shortstack{Cost of a single electron \\ density calculation, via KS-DFT \\ (in CPU hours)} & ML & 0.025 & 0.005 & 0.12 & 0.03 \\\hline
    \shortstack{\textit{Total cost of training} \\ \textit{data generation (in CPU hours)}} & ML & \textit{28} & \textit{0.72} & \textit{0.72} & \textit{5.28} \\\hline
    \shortstack{Number of compositions to  \\ predict electron densities for} & ML & 1000 & 1000 & 1000 & 1000 \\\hline
    \shortstack{Number of configurations at \\ each prediction composition} & ML & 10 & 10 & 10 & 10 \\\hline
    \shortstack{Cost of a single electron density \\ prediction (in CPU hours) } & ML & 0.0025 & 0.0028 & 0.0039 & 0.0014 \\\hline
    \shortstack{\textit{Total cost of prediction} \\ \textit{(in CPU hours)}} & ML & \textit{25} & \textit{28} & \textit{39} & \textit{14} \\\hline
    \shortstack{Cost of model training \\ (in CPU hours)}& ML & 25 & 29 & 29 & 14 \\\hline
    \shortstack{\textbf{Total cost of ML approach} \\ \textbf{(in CPU hours)}} & \textbf{ML} & \textbf{78} & \textbf{57.72} & \textbf{68.72} & \textbf{33.28} \\\hline

    \end{tabular}	
    \caption{{Comparison of the computational cost between exploring the composition space of different alloy systems via the KS-DFT and ML approaches. For the sake of a fair comparison, all computations relevant to obtaining these costs were carried out  on CPUs (for both  KS-DFT and ML approaches).}}
    \label{tab:costDemonstration}
    \end{table}

    \begin{table}[htbp]
    \centering
    \begin{tabular}{|c|c|c|c|c|c|}\hline
    \shortstack{ \textcolor{white}{place} \\ \textcolor{white}{holder}} & \shortstack{\textbf{3-element} \\ \textit{(e.g. SiGeSn)}} & \shortstack{\textbf{4-element} \\ \textit{(e.g. CrFeCoNi)}} & \shortstack{\textbf{5-element} \\ \textit{(e.g. AlCrFeCoNi)}}  \\\hline
    Number of compositions for increment of 20\% & 21& 56& 126 \\\hline
    Number of compositions for increment of 10\% & 66 & 286 & 1,001 \\\hline
    Number of compositions for increment of 5\% & 231 & 1,771 & 10,626 \\\hline
    Number of compositions for increment of 1\% & 5,151 & 17,6851 & 4,598,126 \\\hline
    Number of compositions for increment of 0.1\% & 501,501 & 167,668,501 & 42,084,793,751 \\\hline
    \end{tabular}	
    \caption{{Number of unique alloy compositions needed to fully map the composition space for different values of increment of the alloying element concentration. This count includes the sub-systems where one or more elements have $0\%$ concentration. Multiply these numbers by number of unique configurational snapshots in order to obtain the number of snapshots shown in the x-axis of Figure \ref{fig:efficiencyGain}.}}
    \label{tab:countingUniqueCompositions}
    \end{table}

\color{black}
\twocolumngrid

\newpage
\clearpage
\subsection{SAD Baseline Details}
In order to appreciate the accuracy of the model predictions obtained in this study, it is helpful to compare results with suitable baselines. Indeed, it is quite common in the ML literature to compare new models against community-accepted baselines, so as to standardize demonstrated performance improvements.
However, at present, there is no community-accepted baseline for ML models that predict the electron density. Additionally, we are also not aware of other ML models that predict electron density across composition space for the ternary and quaternary alloys considered here; this work is novel in that regard. Thus, given that our work is proposing an entire framework, rather than simply improving upon an existing approach, obvious baselines are not readily available. To address this issue, we have selected the superposition of atomic densities (SAD) \cite{jorgensen2022equivariant, li2024image} to serve as a baseline model for electron density prediction. Additionally, since our ML models across composition space are developed using Bayesian Active Learning, they are compared against models developed by random selection of compositions. Each of these are discussed further below.

Our reason for choosing SAD as the baseline model for electron density prediction is that it has long been recognized \cite{read1989tests, foulkes1993accuracy, bellchambers2011approximate} to capture a good fraction of the actual electron density in various systems (a recent study \cite{li2024image} estimates it to be $\sim$85\% accurate in getting the electron density of molecular systems). Furthermore, SAD are inexpensive to compute (no KS-DFT calculations are needed). Since our ML models are trained on KS-DFT data, which \emph{do} contain atomic bonding information, while the SAD \emph{do not}, the SAD make for a convenient baseline. Notably, as mentioned in the main text, we already utilize SAD in our work; we provided an example of a CrFeCoNi model trained on the difference between electron density and SAD (i.e. $\rho-$SAD) and found that this approach was effective for error reduction.

To obtain our own assessment of the ``baseline'' error that results from just using the SAD, we compute this quantity as a field over the grid points (which we already had to do when using our $\rho-$SAD model training approach).  Thereafter, we treat it as if the values were  predicted by an ML model, and calculate the density errors and post-process the field to obtain energy errors. We performed this analysis for $455$ snapshots across the $69$ compositions that comprised the quaternary CrFeCoNi dataset. The results are displayed in Fig.~9 of the main text. The performance improvements of our ML models compared to the SAD baseline are evident.

To get a sense of the baseline errors while predicting across composition space, and to demonstrate the advantage of the Bayesian AL technique over the random selection of compositions, we have compared the errors from these two approaches in Fig.~14 of the main text. In this comparison, both approaches used the same number of compositions and the same amount of data. The advantage of the Bayesian AL technique is evident from the error plot. Three different sets of 20 randomly chosen compositions were used to develop three ML models and their errors are shown in the Fig.~14 of the main text (error bars indicate the range of maximum NRMSE values observed across the three models).
\color{black}

\newpage
\clearpage
\bibliography{main}

\begin{thebibliography}{153}%
\makeatletter
\providecommand \@ifxundefined [1]{%
 \@ifx{#1\undefined}
}%
\providecommand \@ifnum [1]{%
 \ifnum #1\expandafter \@firstoftwo
 \else \expandafter \@secondoftwo
 \fi
}%
\providecommand \@ifx [1]{%
 \ifx #1\expandafter \@firstoftwo
 \else \expandafter \@secondoftwo
 \fi
}%
\providecommand \natexlab [1]{#1}%
\providecommand \enquote  [1]{``#1''}%
\providecommand \bibnamefont  [1]{#1}%
\providecommand \bibfnamefont [1]{#1}%
\providecommand \citenamefont [1]{#1}%
\providecommand \href@noop [0]{\@secondoftwo}%
\providecommand \href [0]{\begingroup \@sanitize@url \@href}%
\providecommand \@href[1]{\@@startlink{#1}\@@href}%
\providecommand \@@href[1]{\endgroup#1\@@endlink}%
\providecommand \@sanitize@url [0]{\catcode `\\12\catcode `\$12\catcode `\&12\catcode `\#12\catcode `\^12\catcode `\_12\catcode `\%12\relax}%
\providecommand \@@startlink[1]{}%
\providecommand \@@endlink[0]{}%
\providecommand \url  [0]{\begingroup\@sanitize@url \@url }%
\providecommand \@url [1]{\endgroup\@href {#1}{\urlprefix }}%
\providecommand \urlprefix  [0]{URL }%
\providecommand \Eprint [0]{\href }%
\providecommand \doibase [0]{https://doi.org/}%
\providecommand \selectlanguage [0]{\@gobble}%
\providecommand \bibinfo  [0]{\@secondoftwo}%
\providecommand \bibfield  [0]{\@secondoftwo}%
\providecommand \translation [1]{[#1]}%
\providecommand \BibitemOpen [0]{}%
\providecommand \bibitemStop [0]{}%
\providecommand \bibitemNoStop [0]{.\EOS\space}%
\providecommand \EOS [0]{\spacefactor3000\relax}%
\providecommand \BibitemShut  [1]{\csname bibitem#1\endcsname}%
\let\auto@bib@innerbib\@empty
\bibitem [{\citenamefont {Hohenberg}\ and\ \citenamefont {Kohn}(1964)}]{hohenberg1964inhomogeneous}%
  \BibitemOpen
  \bibfield  {author} {\bibinfo {author} {\bibfnamefont {P.}~\bibnamefont {Hohenberg}}\ and\ \bibinfo {author} {\bibfnamefont {W.}~\bibnamefont {Kohn}},\ }\href@noop {} {\bibfield  {journal} {\bibinfo  {journal} {Physical review}\ }\textbf {\bibinfo {volume} {136}},\ \bibinfo {pages} {B864} (\bibinfo {year} {1964})}\BibitemShut {NoStop}%
\bibitem [{\citenamefont {Kohn}\ and\ \citenamefont {Sham}(1965)}]{kohn1965self}%
  \BibitemOpen
  \bibfield  {author} {\bibinfo {author} {\bibfnamefont {W.}~\bibnamefont {Kohn}}\ and\ \bibinfo {author} {\bibfnamefont {L.~J.}\ \bibnamefont {Sham}},\ }\href@noop {} {\bibfield  {journal} {\bibinfo  {journal} {Physical review}\ }\textbf {\bibinfo {volume} {140}},\ \bibinfo {pages} {A1133} (\bibinfo {year} {1965})}\BibitemShut {NoStop}%
\bibitem [{\citenamefont {Martin}(2004)}]{Martin_ES}%
  \BibitemOpen
  \bibfield  {author} {\bibinfo {author} {\bibfnamefont {R.~M.}\ \bibnamefont {Martin}},\ }\href@noop {} {\emph {\bibinfo {title} {Electronic Structure: Basic Theory and Practical Methods}}},\ \bibinfo {edition} {1st}\ ed.\ (\bibinfo  {publisher} {Cambridge University Press},\ \bibinfo {year} {2004})\BibitemShut {NoStop}%
\bibitem [{\citenamefont {Hafner}\ \emph {et~al.}(2006)\citenamefont {Hafner}, \citenamefont {Wolverton},\ and\ \citenamefont {Ceder}}]{hafner2006toward}%
  \BibitemOpen
  \bibfield  {author} {\bibinfo {author} {\bibfnamefont {J.}~\bibnamefont {Hafner}}, \bibinfo {author} {\bibfnamefont {C.}~\bibnamefont {Wolverton}},\ and\ \bibinfo {author} {\bibfnamefont {G.}~\bibnamefont {Ceder}},\ }\href@noop {} {\bibfield  {journal} {\bibinfo  {journal} {MRS bulletin}\ }\textbf {\bibinfo {volume} {31}},\ \bibinfo {pages} {659} (\bibinfo {year} {2006})}\BibitemShut {NoStop}%
\bibitem [{\citenamefont {Saal}\ \emph {et~al.}(2013)\citenamefont {Saal}, \citenamefont {Kirklin}, \citenamefont {Aykol}, \citenamefont {Meredig},\ and\ \citenamefont {Wolverton}}]{saal2013materials}%
  \BibitemOpen
  \bibfield  {author} {\bibinfo {author} {\bibfnamefont {J.~E.}\ \bibnamefont {Saal}}, \bibinfo {author} {\bibfnamefont {S.}~\bibnamefont {Kirklin}}, \bibinfo {author} {\bibfnamefont {M.}~\bibnamefont {Aykol}}, \bibinfo {author} {\bibfnamefont {B.}~\bibnamefont {Meredig}},\ and\ \bibinfo {author} {\bibfnamefont {C.}~\bibnamefont {Wolverton}},\ }\href@noop {} {\bibfield  {journal} {\bibinfo  {journal} {Jom}\ }\textbf {\bibinfo {volume} {65}},\ \bibinfo {pages} {1501} (\bibinfo {year} {2013})}\BibitemShut {NoStop}%
\bibitem [{\citenamefont {Emery}\ and\ \citenamefont {Wolverton}(2017)}]{emery2017high}%
  \BibitemOpen
  \bibfield  {author} {\bibinfo {author} {\bibfnamefont {A.~A.}\ \bibnamefont {Emery}}\ and\ \bibinfo {author} {\bibfnamefont {C.}~\bibnamefont {Wolverton}},\ }\href@noop {} {\bibfield  {journal} {\bibinfo  {journal} {Scientific data}\ }\textbf {\bibinfo {volume} {4}},\ \bibinfo {pages} {1} (\bibinfo {year} {2017})}\BibitemShut {NoStop}%
\bibitem [{\citenamefont {Choudhary}\ \emph {et~al.}(2020)\citenamefont {Choudhary}, \citenamefont {Garrity}, \citenamefont {Sharma}, \citenamefont {Biacchi}, \citenamefont {Hight~Walker},\ and\ \citenamefont {Tavazza}}]{choudhary2020high}%
  \BibitemOpen
  \bibfield  {author} {\bibinfo {author} {\bibfnamefont {K.}~\bibnamefont {Choudhary}}, \bibinfo {author} {\bibfnamefont {K.~F.}\ \bibnamefont {Garrity}}, \bibinfo {author} {\bibfnamefont {V.}~\bibnamefont {Sharma}}, \bibinfo {author} {\bibfnamefont {A.~J.}\ \bibnamefont {Biacchi}}, \bibinfo {author} {\bibfnamefont {A.~R.}\ \bibnamefont {Hight~Walker}},\ and\ \bibinfo {author} {\bibfnamefont {F.}~\bibnamefont {Tavazza}},\ }\href@noop {} {\bibfield  {journal} {\bibinfo  {journal} {npj computational materials}\ }\textbf {\bibinfo {volume} {6}},\ \bibinfo {pages} {64} (\bibinfo {year} {2020})}\BibitemShut {NoStop}%
\bibitem [{\citenamefont {Jain}\ \emph {et~al.}(2013)\citenamefont {Jain}, \citenamefont {Ong}, \citenamefont {Hautier}, \citenamefont {Chen}, \citenamefont {Richards}, \citenamefont {Dacek}, \citenamefont {Cholia}, \citenamefont {Gunter}, \citenamefont {Skinner}, \citenamefont {Ceder} \emph {et~al.}}]{jain2013commentary}%
  \BibitemOpen
  \bibfield  {author} {\bibinfo {author} {\bibfnamefont {A.}~\bibnamefont {Jain}}, \bibinfo {author} {\bibfnamefont {S.~P.}\ \bibnamefont {Ong}}, \bibinfo {author} {\bibfnamefont {G.}~\bibnamefont {Hautier}}, \bibinfo {author} {\bibfnamefont {W.}~\bibnamefont {Chen}}, \bibinfo {author} {\bibfnamefont {W.~D.}\ \bibnamefont {Richards}}, \bibinfo {author} {\bibfnamefont {S.}~\bibnamefont {Dacek}}, \bibinfo {author} {\bibfnamefont {S.}~\bibnamefont {Cholia}}, \bibinfo {author} {\bibfnamefont {D.}~\bibnamefont {Gunter}}, \bibinfo {author} {\bibfnamefont {D.}~\bibnamefont {Skinner}}, \bibinfo {author} {\bibfnamefont {G.}~\bibnamefont {Ceder}}, \emph {et~al.},\ }\href@noop {} {\bibfield  {journal} {\bibinfo  {journal} {APL materials}\ }\textbf {\bibinfo {volume} {1}} (\bibinfo {year} {2013})}\BibitemShut {NoStop}%
\bibitem [{\citenamefont {Jain}\ \emph {et~al.}(2020)\citenamefont {Jain}, \citenamefont {Montoya}, \citenamefont {Dwaraknath}, \citenamefont {Zimmermann}, \citenamefont {Dagdelen}, \citenamefont {Horton}, \citenamefont {Huck}, \citenamefont {Winston}, \citenamefont {Cholia}, \citenamefont {Ong} \emph {et~al.}}]{jain2020materials}%
  \BibitemOpen
  \bibfield  {author} {\bibinfo {author} {\bibfnamefont {A.}~\bibnamefont {Jain}}, \bibinfo {author} {\bibfnamefont {J.}~\bibnamefont {Montoya}}, \bibinfo {author} {\bibfnamefont {S.}~\bibnamefont {Dwaraknath}}, \bibinfo {author} {\bibfnamefont {N.~E.}\ \bibnamefont {Zimmermann}}, \bibinfo {author} {\bibfnamefont {J.}~\bibnamefont {Dagdelen}}, \bibinfo {author} {\bibfnamefont {M.}~\bibnamefont {Horton}}, \bibinfo {author} {\bibfnamefont {P.}~\bibnamefont {Huck}}, \bibinfo {author} {\bibfnamefont {D.}~\bibnamefont {Winston}}, \bibinfo {author} {\bibfnamefont {S.}~\bibnamefont {Cholia}}, \bibinfo {author} {\bibfnamefont {S.~P.}\ \bibnamefont {Ong}}, \emph {et~al.},\ }\href@noop {} {\bibfield  {journal} {\bibinfo  {journal} {Handbook of Materials Modeling: Methods: Theory and Modeling}\ ,\ \bibinfo {pages} {1751}} (\bibinfo {year} {2020})}\BibitemShut {NoStop}%
\bibitem [{\citenamefont {Gavini}\ \emph {et~al.}(2023)\citenamefont {Gavini}, \citenamefont {Baroni}, \citenamefont {Blum}, \citenamefont {Bowler}, \citenamefont {Buccheri}, \citenamefont {Chelikowsky}, \citenamefont {Das}, \citenamefont {Dawson}, \citenamefont {Delugas}, \citenamefont {Dogan} \emph {et~al.}}]{gavini2023roadmap}%
  \BibitemOpen
  \bibfield  {author} {\bibinfo {author} {\bibfnamefont {V.}~\bibnamefont {Gavini}}, \bibinfo {author} {\bibfnamefont {S.}~\bibnamefont {Baroni}}, \bibinfo {author} {\bibfnamefont {V.}~\bibnamefont {Blum}}, \bibinfo {author} {\bibfnamefont {D.~R.}\ \bibnamefont {Bowler}}, \bibinfo {author} {\bibfnamefont {A.}~\bibnamefont {Buccheri}}, \bibinfo {author} {\bibfnamefont {J.~R.}\ \bibnamefont {Chelikowsky}}, \bibinfo {author} {\bibfnamefont {S.}~\bibnamefont {Das}}, \bibinfo {author} {\bibfnamefont {W.}~\bibnamefont {Dawson}}, \bibinfo {author} {\bibfnamefont {P.}~\bibnamefont {Delugas}}, \bibinfo {author} {\bibfnamefont {M.}~\bibnamefont {Dogan}}, \emph {et~al.},\ }\href@noop {} {\bibfield  {journal} {\bibinfo  {journal} {Modelling and Simulation in Materials Science and Engineering}\ }\textbf {\bibinfo {volume} {31}},\ \bibinfo {pages} {063301} (\bibinfo {year} {2023})}\BibitemShut {NoStop}%
\bibitem [{\citenamefont {Goedecker}(1999)}]{goedecker1999linear}%
  \BibitemOpen
  \bibfield  {author} {\bibinfo {author} {\bibfnamefont {S.}~\bibnamefont {Goedecker}},\ }\href@noop {} {\bibfield  {journal} {\bibinfo  {journal} {Reviews of Modern Physics}\ }\textbf {\bibinfo {volume} {71}},\ \bibinfo {pages} {1085} (\bibinfo {year} {1999})}\BibitemShut {NoStop}%
\bibitem [{\citenamefont {Banerjee}\ \emph {et~al.}(2016{\natexlab{a}})\citenamefont {Banerjee}, \citenamefont {Lin}, \citenamefont {Hu}, \citenamefont {Yang},\ and\ \citenamefont {Pask}}]{banerjee2016chebyshev}%
  \BibitemOpen
  \bibfield  {author} {\bibinfo {author} {\bibfnamefont {A.~S.}\ \bibnamefont {Banerjee}}, \bibinfo {author} {\bibfnamefont {L.}~\bibnamefont {Lin}}, \bibinfo {author} {\bibfnamefont {W.}~\bibnamefont {Hu}}, \bibinfo {author} {\bibfnamefont {C.}~\bibnamefont {Yang}},\ and\ \bibinfo {author} {\bibfnamefont {J.~E.}\ \bibnamefont {Pask}},\ }\href@noop {} {\bibfield  {journal} {\bibinfo  {journal} {Journal of Chemical Physics}\ }\textbf {\bibinfo {volume} {145}},\ \bibinfo {pages} {154101} (\bibinfo {year} {2016}{\natexlab{a}})}\BibitemShut {NoStop}%
\bibitem [{\citenamefont {Banerjee}\ \emph {et~al.}(2018)\citenamefont {Banerjee}, \citenamefont {Lin}, \citenamefont {Suryanarayana}, \citenamefont {Yang},\ and\ \citenamefont {Pask}}]{banerjee2018two}%
  \BibitemOpen
  \bibfield  {author} {\bibinfo {author} {\bibfnamefont {A.~S.}\ \bibnamefont {Banerjee}}, \bibinfo {author} {\bibfnamefont {L.}~\bibnamefont {Lin}}, \bibinfo {author} {\bibfnamefont {P.}~\bibnamefont {Suryanarayana}}, \bibinfo {author} {\bibfnamefont {C.}~\bibnamefont {Yang}},\ and\ \bibinfo {author} {\bibfnamefont {J.~E.}\ \bibnamefont {Pask}},\ }\href@noop {} {\bibfield  {journal} {\bibinfo  {journal} {Journal of Chemical Theory and Computation}\ }\textbf {\bibinfo {volume} {14}},\ \bibinfo {pages} {2930} (\bibinfo {year} {2018})}\BibitemShut {NoStop}%
\bibitem [{\citenamefont {Motamarri}\ and\ \citenamefont {Gavini}(2014)}]{motamarri2014subquadratic}%
  \BibitemOpen
  \bibfield  {author} {\bibinfo {author} {\bibfnamefont {P.}~\bibnamefont {Motamarri}}\ and\ \bibinfo {author} {\bibfnamefont {V.}~\bibnamefont {Gavini}},\ }\href@noop {} {\bibfield  {journal} {\bibinfo  {journal} {Physical Review B}\ }\textbf {\bibinfo {volume} {90}},\ \bibinfo {pages} {115127} (\bibinfo {year} {2014})}\BibitemShut {NoStop}%
\bibitem [{\citenamefont {Lin}\ \emph {et~al.}(2014)\citenamefont {Lin}, \citenamefont {Garc{\'\i}a}, \citenamefont {Huhs},\ and\ \citenamefont {Yang}}]{lin2014siesta}%
  \BibitemOpen
  \bibfield  {author} {\bibinfo {author} {\bibfnamefont {L.}~\bibnamefont {Lin}}, \bibinfo {author} {\bibfnamefont {A.}~\bibnamefont {Garc{\'\i}a}}, \bibinfo {author} {\bibfnamefont {G.}~\bibnamefont {Huhs}},\ and\ \bibinfo {author} {\bibfnamefont {C.}~\bibnamefont {Yang}},\ }\href@noop {} {\bibfield  {journal} {\bibinfo  {journal} {Journal of Physics: Condensed Matter}\ }\textbf {\bibinfo {volume} {26}},\ \bibinfo {pages} {305503} (\bibinfo {year} {2014})}\BibitemShut {NoStop}%
\bibitem [{\citenamefont {Dogan}\ \emph {et~al.}(2023)\citenamefont {Dogan}, \citenamefont {Liou},\ and\ \citenamefont {Chelikowsky}}]{dogan2023real}%
  \BibitemOpen
  \bibfield  {author} {\bibinfo {author} {\bibfnamefont {M.}~\bibnamefont {Dogan}}, \bibinfo {author} {\bibfnamefont {K.-H.}\ \bibnamefont {Liou}},\ and\ \bibinfo {author} {\bibfnamefont {J.~R.}\ \bibnamefont {Chelikowsky}},\ }\href@noop {} {\bibfield  {journal} {\bibinfo  {journal} {The Journal of Chemical Physics}\ }\textbf {\bibinfo {volume} {158}} (\bibinfo {year} {2023})}\BibitemShut {NoStop}%
\bibitem [{\citenamefont {Gavini}\ \emph {et~al.}(2007)\citenamefont {Gavini}, \citenamefont {Bhattacharya},\ and\ \citenamefont {Ortiz}}]{gavini2007vacancy}%
  \BibitemOpen
  \bibfield  {author} {\bibinfo {author} {\bibfnamefont {V.}~\bibnamefont {Gavini}}, \bibinfo {author} {\bibfnamefont {K.}~\bibnamefont {Bhattacharya}},\ and\ \bibinfo {author} {\bibfnamefont {M.}~\bibnamefont {Ortiz}},\ }\href@noop {} {\bibfield  {journal} {\bibinfo  {journal} {Physical Review B}\ }\textbf {\bibinfo {volume} {76}},\ \bibinfo {pages} {180101} (\bibinfo {year} {2007})}\BibitemShut {NoStop}%
\bibitem [{\citenamefont {Carr}\ \emph {et~al.}(2020)\citenamefont {Carr}, \citenamefont {Fang},\ and\ \citenamefont {Kaxiras}}]{carr2020electronic}%
  \BibitemOpen
  \bibfield  {author} {\bibinfo {author} {\bibfnamefont {S.}~\bibnamefont {Carr}}, \bibinfo {author} {\bibfnamefont {S.}~\bibnamefont {Fang}},\ and\ \bibinfo {author} {\bibfnamefont {E.}~\bibnamefont {Kaxiras}},\ }\href@noop {} {\bibfield  {journal} {\bibinfo  {journal} {Nature Reviews Materials}\ }\textbf {\bibinfo {volume} {5}},\ \bibinfo {pages} {748} (\bibinfo {year} {2020})}\BibitemShut {NoStop}%
\bibitem [{\citenamefont {Jaros}(1985)}]{jaros1985electronic}%
  \BibitemOpen
  \bibfield  {author} {\bibinfo {author} {\bibfnamefont {M.}~\bibnamefont {Jaros}},\ }\href@noop {} {\bibfield  {journal} {\bibinfo  {journal} {Reports on Progress in Physics}\ }\textbf {\bibinfo {volume} {48}},\ \bibinfo {pages} {1091} (\bibinfo {year} {1985})}\BibitemShut {NoStop}%
\bibitem [{\citenamefont {Wei}\ \emph {et~al.}(1990)\citenamefont {Wei}, \citenamefont {Ferreira}, \citenamefont {Bernard},\ and\ \citenamefont {Zunger}}]{wei1990electronic}%
  \BibitemOpen
  \bibfield  {author} {\bibinfo {author} {\bibfnamefont {S.-H.}\ \bibnamefont {Wei}}, \bibinfo {author} {\bibfnamefont {L.}~\bibnamefont {Ferreira}}, \bibinfo {author} {\bibfnamefont {J.~E.}\ \bibnamefont {Bernard}},\ and\ \bibinfo {author} {\bibfnamefont {A.}~\bibnamefont {Zunger}},\ }\href@noop {} {\bibfield  {journal} {\bibinfo  {journal} {Physical Review B}\ }\textbf {\bibinfo {volume} {42}},\ \bibinfo {pages} {9622} (\bibinfo {year} {1990})}\BibitemShut {NoStop}%
\bibitem [{\citenamefont {George}\ \emph {et~al.}(2019)\citenamefont {George}, \citenamefont {Raabe},\ and\ \citenamefont {Ritchie}}]{george2019high}%
  \BibitemOpen
  \bibfield  {author} {\bibinfo {author} {\bibfnamefont {E.~P.}\ \bibnamefont {George}}, \bibinfo {author} {\bibfnamefont {D.}~\bibnamefont {Raabe}},\ and\ \bibinfo {author} {\bibfnamefont {R.~O.}\ \bibnamefont {Ritchie}},\ }\href@noop {} {\bibfield  {journal} {\bibinfo  {journal} {Nature reviews materials}\ }\textbf {\bibinfo {volume} {4}},\ \bibinfo {pages} {515} (\bibinfo {year} {2019})}\BibitemShut {NoStop}%
\bibitem [{\citenamefont {Wang}\ \emph {et~al.}(2021)\citenamefont {Wang}, \citenamefont {Xiong}, \citenamefont {Li}, \citenamefont {Zeng}, \citenamefont {Xiong},\ and\ \citenamefont {Chai}}]{wang2021comparison}%
  \BibitemOpen
  \bibfield  {author} {\bibinfo {author} {\bibfnamefont {S.}~\bibnamefont {Wang}}, \bibinfo {author} {\bibfnamefont {J.}~\bibnamefont {Xiong}}, \bibinfo {author} {\bibfnamefont {D.}~\bibnamefont {Li}}, \bibinfo {author} {\bibfnamefont {Q.}~\bibnamefont {Zeng}}, \bibinfo {author} {\bibfnamefont {M.}~\bibnamefont {Xiong}},\ and\ \bibinfo {author} {\bibfnamefont {X.}~\bibnamefont {Chai}},\ }\href@noop {} {\bibfield  {journal} {\bibinfo  {journal} {Materials Letters}\ }\textbf {\bibinfo {volume} {282}},\ \bibinfo {pages} {128754} (\bibinfo {year} {2021})}\BibitemShut {NoStop}%
\bibitem [{\citenamefont {Tian}\ \emph {et~al.}(2016)\citenamefont {Tian}, \citenamefont {Varga}, \citenamefont {Shen},\ and\ \citenamefont {Vitos}}]{tian2016calculating}%
  \BibitemOpen
  \bibfield  {author} {\bibinfo {author} {\bibfnamefont {F.}~\bibnamefont {Tian}}, \bibinfo {author} {\bibfnamefont {L.~K.}\ \bibnamefont {Varga}}, \bibinfo {author} {\bibfnamefont {J.}~\bibnamefont {Shen}},\ and\ \bibinfo {author} {\bibfnamefont {L.}~\bibnamefont {Vitos}},\ }\href@noop {} {\bibfield  {journal} {\bibinfo  {journal} {Computational materials science}\ }\textbf {\bibinfo {volume} {111}},\ \bibinfo {pages} {350} (\bibinfo {year} {2016})}\BibitemShut {NoStop}%
\bibitem [{\citenamefont {Karabin}\ \emph {et~al.}(2022)\citenamefont {Karabin}, \citenamefont {Mondal}, \citenamefont {{\"O}stlin}, \citenamefont {Ho}, \citenamefont {Dobrosavljevic}, \citenamefont {Tam}, \citenamefont {Terletska}, \citenamefont {Chioncel}, \citenamefont {Wang},\ and\ \citenamefont {Eisenbach}}]{karabin2022ab}%
  \BibitemOpen
  \bibfield  {author} {\bibinfo {author} {\bibfnamefont {M.}~\bibnamefont {Karabin}}, \bibinfo {author} {\bibfnamefont {W.~R.}\ \bibnamefont {Mondal}}, \bibinfo {author} {\bibfnamefont {A.}~\bibnamefont {{\"O}stlin}}, \bibinfo {author} {\bibfnamefont {W.-G.~D.}\ \bibnamefont {Ho}}, \bibinfo {author} {\bibfnamefont {V.}~\bibnamefont {Dobrosavljevic}}, \bibinfo {author} {\bibfnamefont {K.-M.}\ \bibnamefont {Tam}}, \bibinfo {author} {\bibfnamefont {H.}~\bibnamefont {Terletska}}, \bibinfo {author} {\bibfnamefont {L.}~\bibnamefont {Chioncel}}, \bibinfo {author} {\bibfnamefont {Y.}~\bibnamefont {Wang}},\ and\ \bibinfo {author} {\bibfnamefont {M.}~\bibnamefont {Eisenbach}},\ }\href@noop {} {\bibfield  {journal} {\bibinfo  {journal} {Journal of Materials Science}\ }\textbf {\bibinfo {volume} {57}},\ \bibinfo {pages} {10677} (\bibinfo {year} {2022})}\BibitemShut {NoStop}%
\bibitem [{\citenamefont {Gao}\ \emph {et~al.}(2016)\citenamefont {Gao}, \citenamefont {Niu}, \citenamefont {Jiang},\ and\ \citenamefont {Irving}}]{gao2016applications}%
  \BibitemOpen
  \bibfield  {author} {\bibinfo {author} {\bibfnamefont {M.~C.}\ \bibnamefont {Gao}}, \bibinfo {author} {\bibfnamefont {C.}~\bibnamefont {Niu}}, \bibinfo {author} {\bibfnamefont {C.}~\bibnamefont {Jiang}},\ and\ \bibinfo {author} {\bibfnamefont {D.~L.}\ \bibnamefont {Irving}},\ }\href@noop {} {\bibfield  {journal} {\bibinfo  {journal} {High-entropy alloys: Fundamentals and applications}\ ,\ \bibinfo {pages} {333}} (\bibinfo {year} {2016})}\BibitemShut {NoStop}%
\bibitem [{\citenamefont {Lewis}\ \emph {et~al.}(2021)\citenamefont {Lewis}, \citenamefont {Grisafi}, \citenamefont {Ceriotti},\ and\ \citenamefont {Rossi}}]{lewis2021learning}%
  \BibitemOpen
  \bibfield  {author} {\bibinfo {author} {\bibfnamefont {A.~M.}\ \bibnamefont {Lewis}}, \bibinfo {author} {\bibfnamefont {A.}~\bibnamefont {Grisafi}}, \bibinfo {author} {\bibfnamefont {M.}~\bibnamefont {Ceriotti}},\ and\ \bibinfo {author} {\bibfnamefont {M.}~\bibnamefont {Rossi}},\ }\href@noop {} {\bibfield  {journal} {\bibinfo  {journal} {Journal of Chemical Theory and Computation}\ }\textbf {\bibinfo {volume} {17}},\ \bibinfo {pages} {7203} (\bibinfo {year} {2021})}\BibitemShut {NoStop}%
\bibitem [{\citenamefont {J{\o}rgensen}\ and\ \citenamefont {Bhowmik}(2022)}]{jorgensen2022equivariant}%
  \BibitemOpen
  \bibfield  {author} {\bibinfo {author} {\bibfnamefont {P.~B.}\ \bibnamefont {J{\o}rgensen}}\ and\ \bibinfo {author} {\bibfnamefont {A.}~\bibnamefont {Bhowmik}},\ }\href@noop {} {\bibfield  {journal} {\bibinfo  {journal} {npj Computational Materials}\ }\textbf {\bibinfo {volume} {8}},\ \bibinfo {pages} {183} (\bibinfo {year} {2022})}\BibitemShut {NoStop}%
\bibitem [{\citenamefont {Zepeda-N{\'u}{\~n}ez}\ \emph {et~al.}(2021)\citenamefont {Zepeda-N{\'u}{\~n}ez}, \citenamefont {Chen}, \citenamefont {Zhang}, \citenamefont {Jia}, \citenamefont {Zhang},\ and\ \citenamefont {Lin}}]{zepeda2021deep}%
  \BibitemOpen
  \bibfield  {author} {\bibinfo {author} {\bibfnamefont {L.}~\bibnamefont {Zepeda-N{\'u}{\~n}ez}}, \bibinfo {author} {\bibfnamefont {Y.}~\bibnamefont {Chen}}, \bibinfo {author} {\bibfnamefont {J.}~\bibnamefont {Zhang}}, \bibinfo {author} {\bibfnamefont {W.}~\bibnamefont {Jia}}, \bibinfo {author} {\bibfnamefont {L.}~\bibnamefont {Zhang}},\ and\ \bibinfo {author} {\bibfnamefont {L.}~\bibnamefont {Lin}},\ }\href@noop {} {\bibfield  {journal} {\bibinfo  {journal} {Journal of Computational Physics}\ }\textbf {\bibinfo {volume} {443}},\ \bibinfo {pages} {110523} (\bibinfo {year} {2021})}\BibitemShut {NoStop}%
\bibitem [{\citenamefont {Chandrasekaran}\ \emph {et~al.}(2019)\citenamefont {Chandrasekaran}, \citenamefont {Kamal}, \citenamefont {Batra}, \citenamefont {Kim}, \citenamefont {Chen},\ and\ \citenamefont {Ramprasad}}]{chandrasekaran2019solving}%
  \BibitemOpen
  \bibfield  {author} {\bibinfo {author} {\bibfnamefont {A.}~\bibnamefont {Chandrasekaran}}, \bibinfo {author} {\bibfnamefont {D.}~\bibnamefont {Kamal}}, \bibinfo {author} {\bibfnamefont {R.}~\bibnamefont {Batra}}, \bibinfo {author} {\bibfnamefont {C.}~\bibnamefont {Kim}}, \bibinfo {author} {\bibfnamefont {L.}~\bibnamefont {Chen}},\ and\ \bibinfo {author} {\bibfnamefont {R.}~\bibnamefont {Ramprasad}},\ }\href@noop {} {\bibfield  {journal} {\bibinfo  {journal} {npj Computational Materials}\ }\textbf {\bibinfo {volume} {5}},\ \bibinfo {pages} {22} (\bibinfo {year} {2019})}\BibitemShut {NoStop}%
\bibitem [{\citenamefont {Fiedler}\ \emph {et~al.}(2023)\citenamefont {Fiedler}, \citenamefont {Modine}, \citenamefont {Schmerler}, \citenamefont {Vogel}, \citenamefont {Popoola}, \citenamefont {Thompson}, \citenamefont {Rajamanickam},\ and\ \citenamefont {Cangi}}]{fiedler2023predicting}%
  \BibitemOpen
  \bibfield  {author} {\bibinfo {author} {\bibfnamefont {L.}~\bibnamefont {Fiedler}}, \bibinfo {author} {\bibfnamefont {N.~A.}\ \bibnamefont {Modine}}, \bibinfo {author} {\bibfnamefont {S.}~\bibnamefont {Schmerler}}, \bibinfo {author} {\bibfnamefont {D.~J.}\ \bibnamefont {Vogel}}, \bibinfo {author} {\bibfnamefont {G.~A.}\ \bibnamefont {Popoola}}, \bibinfo {author} {\bibfnamefont {A.~P.}\ \bibnamefont {Thompson}}, \bibinfo {author} {\bibfnamefont {S.}~\bibnamefont {Rajamanickam}},\ and\ \bibinfo {author} {\bibfnamefont {A.}~\bibnamefont {Cangi}},\ }\href@noop {} {\bibfield  {journal} {\bibinfo  {journal} {npj Computational Materials}\ }\textbf {\bibinfo {volume} {9}},\ \bibinfo {pages} {115} (\bibinfo {year} {2023})}\BibitemShut {NoStop}%
\bibitem [{\citenamefont {Brockherde}\ \emph {et~al.}(2017)\citenamefont {Brockherde}, \citenamefont {Vogt}, \citenamefont {Li}, \citenamefont {Tuckerman}, \citenamefont {Burke},\ and\ \citenamefont {M{\"u}ller}}]{brockherde2017bypassing}%
  \BibitemOpen
  \bibfield  {author} {\bibinfo {author} {\bibfnamefont {F.}~\bibnamefont {Brockherde}}, \bibinfo {author} {\bibfnamefont {L.}~\bibnamefont {Vogt}}, \bibinfo {author} {\bibfnamefont {L.}~\bibnamefont {Li}}, \bibinfo {author} {\bibfnamefont {M.~E.}\ \bibnamefont {Tuckerman}}, \bibinfo {author} {\bibfnamefont {K.}~\bibnamefont {Burke}},\ and\ \bibinfo {author} {\bibfnamefont {K.-R.}\ \bibnamefont {M{\"u}ller}},\ }\href@noop {} {\bibfield  {journal} {\bibinfo  {journal} {Nature communications}\ }\textbf {\bibinfo {volume} {8}},\ \bibinfo {pages} {872} (\bibinfo {year} {2017})}\BibitemShut {NoStop}%
\bibitem [{\citenamefont {del Rio}\ \emph {et~al.}(2023)\citenamefont {del Rio}, \citenamefont {Phan},\ and\ \citenamefont {Ramprasad}}]{del2023deep}%
  \BibitemOpen
  \bibfield  {author} {\bibinfo {author} {\bibfnamefont {B.~G.}\ \bibnamefont {del Rio}}, \bibinfo {author} {\bibfnamefont {B.}~\bibnamefont {Phan}},\ and\ \bibinfo {author} {\bibfnamefont {R.}~\bibnamefont {Ramprasad}},\ }\href@noop {} {\bibfield  {journal} {\bibinfo  {journal} {npj Computational Materials}\ }\textbf {\bibinfo {volume} {9}},\ \bibinfo {pages} {158} (\bibinfo {year} {2023})}\BibitemShut {NoStop}%
\bibitem [{\citenamefont {Tang}\ \emph {et~al.}(2024)\citenamefont {Tang}, \citenamefont {Zou}, \citenamefont {Li}, \citenamefont {Wang}, \citenamefont {Yuan}, \citenamefont {Tao}, \citenamefont {Li}, \citenamefont {Chen}, \citenamefont {Zhao}, \citenamefont {Sun} \emph {et~al.}}]{tang2024improving}%
  \BibitemOpen
  \bibfield  {author} {\bibinfo {author} {\bibfnamefont {Z.}~\bibnamefont {Tang}}, \bibinfo {author} {\bibfnamefont {N.}~\bibnamefont {Zou}}, \bibinfo {author} {\bibfnamefont {H.}~\bibnamefont {Li}}, \bibinfo {author} {\bibfnamefont {Y.}~\bibnamefont {Wang}}, \bibinfo {author} {\bibfnamefont {Z.}~\bibnamefont {Yuan}}, \bibinfo {author} {\bibfnamefont {H.}~\bibnamefont {Tao}}, \bibinfo {author} {\bibfnamefont {Y.}~\bibnamefont {Li}}, \bibinfo {author} {\bibfnamefont {Z.}~\bibnamefont {Chen}}, \bibinfo {author} {\bibfnamefont {B.}~\bibnamefont {Zhao}}, \bibinfo {author} {\bibfnamefont {M.}~\bibnamefont {Sun}}, \emph {et~al.},\ }\href@noop {} {\bibfield  {journal} {\bibinfo  {journal} {arXiv preprint arXiv:2406.17561}\ } (\bibinfo {year} {2024})}\BibitemShut {NoStop}%
\bibitem [{\citenamefont {Shao}\ \emph {et~al.}(2023)\citenamefont {Shao}, \citenamefont {Paetow}, \citenamefont {Tuckerman},\ and\ \citenamefont {Pavanello}}]{shao2023machine}%
  \BibitemOpen
  \bibfield  {author} {\bibinfo {author} {\bibfnamefont {X.}~\bibnamefont {Shao}}, \bibinfo {author} {\bibfnamefont {L.}~\bibnamefont {Paetow}}, \bibinfo {author} {\bibfnamefont {M.~E.}\ \bibnamefont {Tuckerman}},\ and\ \bibinfo {author} {\bibfnamefont {M.}~\bibnamefont {Pavanello}},\ }\href@noop {} {\bibfield  {journal} {\bibinfo  {journal} {Nature communications}\ }\textbf {\bibinfo {volume} {14}},\ \bibinfo {pages} {6281} (\bibinfo {year} {2023})}\BibitemShut {NoStop}%
\bibitem [{\citenamefont {Hazra}\ \emph {et~al.}(2024)\citenamefont {Hazra}, \citenamefont {Patil},\ and\ \citenamefont {Sanvito}}]{hazra2024predicting}%
  \BibitemOpen
  \bibfield  {author} {\bibinfo {author} {\bibfnamefont {S.}~\bibnamefont {Hazra}}, \bibinfo {author} {\bibfnamefont {U.}~\bibnamefont {Patil}},\ and\ \bibinfo {author} {\bibfnamefont {S.}~\bibnamefont {Sanvito}},\ }\href@noop {} {\bibfield  {journal} {\bibinfo  {journal} {Journal of Chemical Theory and Computation}\ } (\bibinfo {year} {2024})}\BibitemShut {NoStop}%
\bibitem [{\citenamefont {Sager-Smith}\ and\ \citenamefont {Mazziotti}(2022)}]{sager2022reducing}%
  \BibitemOpen
  \bibfield  {author} {\bibinfo {author} {\bibfnamefont {L.~M.}\ \bibnamefont {Sager-Smith}}\ and\ \bibinfo {author} {\bibfnamefont {D.~A.}\ \bibnamefont {Mazziotti}},\ }\href@noop {} {\bibfield  {journal} {\bibinfo  {journal} {Journal of the American Chemical Society}\ }\textbf {\bibinfo {volume} {144}},\ \bibinfo {pages} {18959} (\bibinfo {year} {2022})}\BibitemShut {NoStop}%
\bibitem [{\citenamefont {Teh}\ \emph {et~al.}(2021)\citenamefont {Teh}, \citenamefont {Ghosh},\ and\ \citenamefont {Bhattacharya}}]{teh2021machine}%
  \BibitemOpen
  \bibfield  {author} {\bibinfo {author} {\bibfnamefont {Y.~S.}\ \bibnamefont {Teh}}, \bibinfo {author} {\bibfnamefont {S.}~\bibnamefont {Ghosh}},\ and\ \bibinfo {author} {\bibfnamefont {K.}~\bibnamefont {Bhattacharya}},\ }\href@noop {} {\bibfield  {journal} {\bibinfo  {journal} {Mechanics of Materials}\ }\textbf {\bibinfo {volume} {163}},\ \bibinfo {pages} {104070} (\bibinfo {year} {2021})}\BibitemShut {NoStop}%
\bibitem [{\citenamefont {Pathrudkar}\ \emph {et~al.}(2022)\citenamefont {Pathrudkar}, \citenamefont {Yu}, \citenamefont {Ghosh},\ and\ \citenamefont {Banerjee}}]{pathrudkar2022machine}%
  \BibitemOpen
  \bibfield  {author} {\bibinfo {author} {\bibfnamefont {S.}~\bibnamefont {Pathrudkar}}, \bibinfo {author} {\bibfnamefont {H.~M.}\ \bibnamefont {Yu}}, \bibinfo {author} {\bibfnamefont {S.}~\bibnamefont {Ghosh}},\ and\ \bibinfo {author} {\bibfnamefont {A.~S.}\ \bibnamefont {Banerjee}},\ }\href@noop {} {\bibfield  {journal} {\bibinfo  {journal} {Physical Review B}\ }\textbf {\bibinfo {volume} {105}},\ \bibinfo {pages} {195141} (\bibinfo {year} {2022})}\BibitemShut {NoStop}%
\bibitem [{\citenamefont {Arora}\ \emph {et~al.}(2022)\citenamefont {Arora}, \citenamefont {Manzoor},\ and\ \citenamefont {Aidhy}}]{arora2022charge}%
  \BibitemOpen
  \bibfield  {author} {\bibinfo {author} {\bibfnamefont {G.}~\bibnamefont {Arora}}, \bibinfo {author} {\bibfnamefont {A.}~\bibnamefont {Manzoor}},\ and\ \bibinfo {author} {\bibfnamefont {D.~S.}\ \bibnamefont {Aidhy}},\ }\href@noop {} {\bibfield  {journal} {\bibinfo  {journal} {Journal of Applied Physics}\ }\textbf {\bibinfo {volume} {132}} (\bibinfo {year} {2022})}\BibitemShut {NoStop}%
\bibitem [{\citenamefont {Banerjee}(2021)}]{banerjee2021ab}%
  \BibitemOpen
  \bibfield  {author} {\bibinfo {author} {\bibfnamefont {A.~S.}\ \bibnamefont {Banerjee}},\ }\href@noop {} {\bibfield  {journal} {\bibinfo  {journal} {Journal of the Mechanics and Physics of Solids}\ }\textbf {\bibinfo {volume} {154}},\ \bibinfo {pages} {104515} (\bibinfo {year} {2021})}\BibitemShut {NoStop}%
\bibitem [{\citenamefont {Yu}\ and\ \citenamefont {Banerjee}(2022)}]{yu2022density}%
  \BibitemOpen
  \bibfield  {author} {\bibinfo {author} {\bibfnamefont {H.~M.}\ \bibnamefont {Yu}}\ and\ \bibinfo {author} {\bibfnamefont {A.~S.}\ \bibnamefont {Banerjee}},\ }\href@noop {} {\bibfield  {journal} {\bibinfo  {journal} {Journal of Computational Physics}\ }\textbf {\bibinfo {volume} {456}},\ \bibinfo {pages} {111023} (\bibinfo {year} {2022})}\BibitemShut {NoStop}%
\bibitem [{\citenamefont {Ghosh}\ \emph {et~al.}(2019)\citenamefont {Ghosh}, \citenamefont {Banerjee},\ and\ \citenamefont {Suryanarayana}}]{ghosh2019symmetry}%
  \BibitemOpen
  \bibfield  {author} {\bibinfo {author} {\bibfnamefont {S.}~\bibnamefont {Ghosh}}, \bibinfo {author} {\bibfnamefont {A.~S.}\ \bibnamefont {Banerjee}},\ and\ \bibinfo {author} {\bibfnamefont {P.}~\bibnamefont {Suryanarayana}},\ }\href@noop {} {\bibfield  {journal} {\bibinfo  {journal} {Physical Review B}\ }\textbf {\bibinfo {volume} {100}},\ \bibinfo {pages} {125143} (\bibinfo {year} {2019})}\BibitemShut {NoStop}%
\bibitem [{\citenamefont {Grisafi}\ \emph {et~al.}(2018)\citenamefont {Grisafi}, \citenamefont {Fabrizio}, \citenamefont {Meyer}, \citenamefont {Wilkins}, \citenamefont {Corminboeuf},\ and\ \citenamefont {Ceriotti}}]{grisafi2018transferable}%
  \BibitemOpen
  \bibfield  {author} {\bibinfo {author} {\bibfnamefont {A.}~\bibnamefont {Grisafi}}, \bibinfo {author} {\bibfnamefont {A.}~\bibnamefont {Fabrizio}}, \bibinfo {author} {\bibfnamefont {B.}~\bibnamefont {Meyer}}, \bibinfo {author} {\bibfnamefont {D.~M.}\ \bibnamefont {Wilkins}}, \bibinfo {author} {\bibfnamefont {C.}~\bibnamefont {Corminboeuf}},\ and\ \bibinfo {author} {\bibfnamefont {M.}~\bibnamefont {Ceriotti}},\ }\href@noop {} {\bibfield  {journal} {\bibinfo  {journal} {ACS central science}\ }\textbf {\bibinfo {volume} {5}},\ \bibinfo {pages} {57} (\bibinfo {year} {2018})}\BibitemShut {NoStop}%
\bibitem [{\citenamefont {Fabrizio}\ \emph {et~al.}(2019)\citenamefont {Fabrizio}, \citenamefont {Grisafi}, \citenamefont {Meyer}, \citenamefont {Ceriotti},\ and\ \citenamefont {Corminboeuf}}]{fabrizio2019electron}%
  \BibitemOpen
  \bibfield  {author} {\bibinfo {author} {\bibfnamefont {A.}~\bibnamefont {Fabrizio}}, \bibinfo {author} {\bibfnamefont {A.}~\bibnamefont {Grisafi}}, \bibinfo {author} {\bibfnamefont {B.}~\bibnamefont {Meyer}}, \bibinfo {author} {\bibfnamefont {M.}~\bibnamefont {Ceriotti}},\ and\ \bibinfo {author} {\bibfnamefont {C.}~\bibnamefont {Corminboeuf}},\ }\href@noop {} {\bibfield  {journal} {\bibinfo  {journal} {Chemical science}\ }\textbf {\bibinfo {volume} {10}},\ \bibinfo {pages} {9424} (\bibinfo {year} {2019})}\BibitemShut {NoStop}%
\bibitem [{\citenamefont {Fu}\ \emph {et~al.}(2024)\citenamefont {Fu}, \citenamefont {Rosen}, \citenamefont {Bystrom}, \citenamefont {Wang}, \citenamefont {Musaelian}, \citenamefont {Kozinsky}, \citenamefont {Smidt},\ and\ \citenamefont {Jaakkola}}]{fu2024recipe}%
  \BibitemOpen
  \bibfield  {author} {\bibinfo {author} {\bibfnamefont {X.}~\bibnamefont {Fu}}, \bibinfo {author} {\bibfnamefont {A.}~\bibnamefont {Rosen}}, \bibinfo {author} {\bibfnamefont {K.}~\bibnamefont {Bystrom}}, \bibinfo {author} {\bibfnamefont {R.}~\bibnamefont {Wang}}, \bibinfo {author} {\bibfnamefont {A.}~\bibnamefont {Musaelian}}, \bibinfo {author} {\bibfnamefont {B.}~\bibnamefont {Kozinsky}}, \bibinfo {author} {\bibfnamefont {T.}~\bibnamefont {Smidt}},\ and\ \bibinfo {author} {\bibfnamefont {T.}~\bibnamefont {Jaakkola}},\ }\href@noop {} {\bibfield  {journal} {\bibinfo  {journal} {arXiv preprint arXiv:2405.19276}\ } (\bibinfo {year} {2024})}\BibitemShut {NoStop}%
\bibitem [{\citenamefont {Qiao}\ \emph {et~al.}(2022)\citenamefont {Qiao}, \citenamefont {Christensen}, \citenamefont {Welborn}, \citenamefont {Manby}, \citenamefont {Anandkumar},\ and\ \citenamefont {Miller~III}}]{qiao2022informing}%
  \BibitemOpen
  \bibfield  {author} {\bibinfo {author} {\bibfnamefont {Z.}~\bibnamefont {Qiao}}, \bibinfo {author} {\bibfnamefont {A.~S.}\ \bibnamefont {Christensen}}, \bibinfo {author} {\bibfnamefont {M.}~\bibnamefont {Welborn}}, \bibinfo {author} {\bibfnamefont {F.~R.}\ \bibnamefont {Manby}}, \bibinfo {author} {\bibfnamefont {A.}~\bibnamefont {Anandkumar}},\ and\ \bibinfo {author} {\bibfnamefont {T.~F.}\ \bibnamefont {Miller~III}},\ }\href@noop {} {\bibfield  {journal} {\bibinfo  {journal} {Proceedings of the National Academy of Sciences}\ }\textbf {\bibinfo {volume} {119}},\ \bibinfo {pages} {e2205221119} (\bibinfo {year} {2022})}\BibitemShut {NoStop}%
\bibitem [{\citenamefont {Pathrudkar}\ \emph {et~al.}(2024)\citenamefont {Pathrudkar}, \citenamefont {Thiagarajan}, \citenamefont {Agarwal}, \citenamefont {Banerjee},\ and\ \citenamefont {Ghosh}}]{pathrudkar2024electronic}%
  \BibitemOpen
  \bibfield  {author} {\bibinfo {author} {\bibfnamefont {S.}~\bibnamefont {Pathrudkar}}, \bibinfo {author} {\bibfnamefont {P.}~\bibnamefont {Thiagarajan}}, \bibinfo {author} {\bibfnamefont {S.}~\bibnamefont {Agarwal}}, \bibinfo {author} {\bibfnamefont {A.~S.}\ \bibnamefont {Banerjee}},\ and\ \bibinfo {author} {\bibfnamefont {S.}~\bibnamefont {Ghosh}},\ }\href@noop {} {\bibfield  {journal} {\bibinfo  {journal} {npj Computational Materials}\ }\textbf {\bibinfo {volume} {10}},\ \bibinfo {pages} {175} (\bibinfo {year} {2024})}\BibitemShut {NoStop}%
\bibitem [{\citenamefont {Gong}\ \emph {et~al.}(2019)\citenamefont {Gong}, \citenamefont {Xie}, \citenamefont {Zhu}, \citenamefont {Wang}, \citenamefont {Fadel}, \citenamefont {Li},\ and\ \citenamefont {Grossman}}]{gong2019predicting}%
  \BibitemOpen
  \bibfield  {author} {\bibinfo {author} {\bibfnamefont {S.}~\bibnamefont {Gong}}, \bibinfo {author} {\bibfnamefont {T.}~\bibnamefont {Xie}}, \bibinfo {author} {\bibfnamefont {T.}~\bibnamefont {Zhu}}, \bibinfo {author} {\bibfnamefont {S.}~\bibnamefont {Wang}}, \bibinfo {author} {\bibfnamefont {E.~R.}\ \bibnamefont {Fadel}}, \bibinfo {author} {\bibfnamefont {Y.}~\bibnamefont {Li}},\ and\ \bibinfo {author} {\bibfnamefont {J.~C.}\ \bibnamefont {Grossman}},\ }\href@noop {} {\bibfield  {journal} {\bibinfo  {journal} {Physical Review B}\ }\textbf {\bibinfo {volume} {100}},\ \bibinfo {pages} {184103} (\bibinfo {year} {2019})}\BibitemShut {NoStop}%
\bibitem [{\citenamefont {Pope}\ and\ \citenamefont {Jacobs}(2024)}]{pope2024towards}%
  \BibitemOpen
  \bibfield  {author} {\bibinfo {author} {\bibfnamefont {P.}~\bibnamefont {Pope}}\ and\ \bibinfo {author} {\bibfnamefont {D.}~\bibnamefont {Jacobs}},\ }\href@noop {} {\bibfield  {journal} {\bibinfo  {journal} {Advances in Neural Information Processing Systems}\ }\textbf {\bibinfo {volume} {36}} (\bibinfo {year} {2024})}\BibitemShut {NoStop}%
\bibitem [{\citenamefont {Zhang}\ \emph {et~al.}(2023{\natexlab{a}})\citenamefont {Zhang}, \citenamefont {Wang}, \citenamefont {Huang}, \citenamefont {Xiang}, \citenamefont {Xiong}, \citenamefont {Xu}, \citenamefont {Ma}, \citenamefont {Fu}, \citenamefont {Kai}, \citenamefont {Kang} \emph {et~al.}}]{zhang2023design}%
  \BibitemOpen
  \bibfield  {author} {\bibinfo {author} {\bibfnamefont {J.}~\bibnamefont {Zhang}}, \bibinfo {author} {\bibfnamefont {C.}~\bibnamefont {Wang}}, \bibinfo {author} {\bibfnamefont {S.}~\bibnamefont {Huang}}, \bibinfo {author} {\bibfnamefont {X.}~\bibnamefont {Xiang}}, \bibinfo {author} {\bibfnamefont {Y.}~\bibnamefont {Xiong}}, \bibinfo {author} {\bibfnamefont {B.}~\bibnamefont {Xu}}, \bibinfo {author} {\bibfnamefont {S.}~\bibnamefont {Ma}}, \bibinfo {author} {\bibfnamefont {H.}~\bibnamefont {Fu}}, \bibinfo {author} {\bibfnamefont {J.}~\bibnamefont {Kai}}, \bibinfo {author} {\bibfnamefont {X.}~\bibnamefont {Kang}}, \emph {et~al.},\ }\href@noop {} {\bibfield  {journal} {\bibinfo  {journal} {Joule}\ }\textbf {\bibinfo {volume} {7}},\ \bibinfo {pages} {1832} (\bibinfo {year} {2023}{\natexlab{a}})}\BibitemShut {NoStop}%
\bibitem [{\citenamefont {Zhang}\ \emph {et~al.}(2023{\natexlab{b}})\citenamefont {Zhang}, \citenamefont {Xiang}, \citenamefont {Xu}, \citenamefont {Huang}, \citenamefont {Xiong}, \citenamefont {Ma}, \citenamefont {Fu}, \citenamefont {Ma}, \citenamefont {Chen}, \citenamefont {Wu} \emph {et~al.}}]{zhang2023rational}%
  \BibitemOpen
  \bibfield  {author} {\bibinfo {author} {\bibfnamefont {J.}~\bibnamefont {Zhang}}, \bibinfo {author} {\bibfnamefont {X.}~\bibnamefont {Xiang}}, \bibinfo {author} {\bibfnamefont {B.}~\bibnamefont {Xu}}, \bibinfo {author} {\bibfnamefont {S.}~\bibnamefont {Huang}}, \bibinfo {author} {\bibfnamefont {Y.}~\bibnamefont {Xiong}}, \bibinfo {author} {\bibfnamefont {S.}~\bibnamefont {Ma}}, \bibinfo {author} {\bibfnamefont {H.}~\bibnamefont {Fu}}, \bibinfo {author} {\bibfnamefont {Y.}~\bibnamefont {Ma}}, \bibinfo {author} {\bibfnamefont {H.}~\bibnamefont {Chen}}, \bibinfo {author} {\bibfnamefont {Z.}~\bibnamefont {Wu}}, \emph {et~al.},\ }\href@noop {} {\bibfield  {journal} {\bibinfo  {journal} {Current Opinion in Solid State and Materials Science}\ }\textbf {\bibinfo {volume} {27}},\ \bibinfo {pages} {101057} (\bibinfo {year} {2023}{\natexlab{b}})}\BibitemShut {NoStop}%
\bibitem [{\citenamefont {Li}\ \emph {et~al.}(2025)\citenamefont {Li}, \citenamefont {Sharir}, \citenamefont {Yuan},\ and\ \citenamefont {Chan}}]{li2024image}%
  \BibitemOpen
  \bibfield  {author} {\bibinfo {author} {\bibfnamefont {C.}~\bibnamefont {Li}}, \bibinfo {author} {\bibfnamefont {O.}~\bibnamefont {Sharir}}, \bibinfo {author} {\bibfnamefont {S.}~\bibnamefont {Yuan}},\ and\ \bibinfo {author} {\bibfnamefont {G.~K.-L.}\ \bibnamefont {Chan}},\ }\href@noop {} {\bibfield  {journal} {\bibinfo  {journal} {Nature Communications}\ }\textbf {\bibinfo {volume} {16}},\ \bibinfo {pages} {4811} (\bibinfo {year} {2025})}\BibitemShut {NoStop}%
\bibitem [{\citenamefont {Koker}\ \emph {et~al.}(2023)\citenamefont {Koker}, \citenamefont {Quigley}, \citenamefont {Taw}, \citenamefont {Tibbetts},\ and\ \citenamefont {Li}}]{koker2023higher}%
  \BibitemOpen
  \bibfield  {author} {\bibinfo {author} {\bibfnamefont {T.}~\bibnamefont {Koker}}, \bibinfo {author} {\bibfnamefont {K.}~\bibnamefont {Quigley}}, \bibinfo {author} {\bibfnamefont {E.}~\bibnamefont {Taw}}, \bibinfo {author} {\bibfnamefont {K.}~\bibnamefont {Tibbetts}},\ and\ \bibinfo {author} {\bibfnamefont {L.}~\bibnamefont {Li}},\ }\href@noop {} {\bibfield  {journal} {\bibinfo  {journal} {arXiv preprint arXiv:2312.05388}\ } (\bibinfo {year} {2023})}\BibitemShut {NoStop}%
\bibitem [{\citenamefont {Okabe}\ \emph {et~al.}(2024)\citenamefont {Okabe}, \citenamefont {Chotrattanapituk}, \citenamefont {Boonkird}, \citenamefont {Andrejevic}, \citenamefont {Fu}, \citenamefont {Jaakkola}, \citenamefont {Song}, \citenamefont {Nguyen}, \citenamefont {Drucker}, \citenamefont {Mu} \emph {et~al.}}]{okabe2024virtual}%
  \BibitemOpen
  \bibfield  {author} {\bibinfo {author} {\bibfnamefont {R.}~\bibnamefont {Okabe}}, \bibinfo {author} {\bibfnamefont {A.}~\bibnamefont {Chotrattanapituk}}, \bibinfo {author} {\bibfnamefont {A.}~\bibnamefont {Boonkird}}, \bibinfo {author} {\bibfnamefont {N.}~\bibnamefont {Andrejevic}}, \bibinfo {author} {\bibfnamefont {X.}~\bibnamefont {Fu}}, \bibinfo {author} {\bibfnamefont {T.~S.}\ \bibnamefont {Jaakkola}}, \bibinfo {author} {\bibfnamefont {Q.}~\bibnamefont {Song}}, \bibinfo {author} {\bibfnamefont {T.}~\bibnamefont {Nguyen}}, \bibinfo {author} {\bibfnamefont {N.}~\bibnamefont {Drucker}}, \bibinfo {author} {\bibfnamefont {S.}~\bibnamefont {Mu}}, \emph {et~al.},\ }\href@noop {} {\bibfield  {journal} {\bibinfo  {journal} {Nature Computational Science}\ }\textbf {\bibinfo {volume} {4}},\ \bibinfo {pages} {522} (\bibinfo {year} {2024})}\BibitemShut {NoStop}%
\bibitem [{\citenamefont {Fung}\ \emph {et~al.}(2021)\citenamefont {Fung}, \citenamefont {Zhang}, \citenamefont {Juarez},\ and\ \citenamefont {Sumpter}}]{fung2021benchmarking}%
  \BibitemOpen
  \bibfield  {author} {\bibinfo {author} {\bibfnamefont {V.}~\bibnamefont {Fung}}, \bibinfo {author} {\bibfnamefont {J.}~\bibnamefont {Zhang}}, \bibinfo {author} {\bibfnamefont {E.}~\bibnamefont {Juarez}},\ and\ \bibinfo {author} {\bibfnamefont {B.~G.}\ \bibnamefont {Sumpter}},\ }\href@noop {} {\bibfield  {journal} {\bibinfo  {journal} {npj Computational Materials}\ }\textbf {\bibinfo {volume} {7}},\ \bibinfo {pages} {84} (\bibinfo {year} {2021})}\BibitemShut {NoStop}%
\bibitem [{\citenamefont {Jiang}\ \emph {et~al.}(2021)\citenamefont {Jiang}, \citenamefont {Wu}, \citenamefont {Hsieh}, \citenamefont {Chen}, \citenamefont {Liao}, \citenamefont {Wang}, \citenamefont {Shen}, \citenamefont {Cao}, \citenamefont {Wu},\ and\ \citenamefont {Hou}}]{jiang2021could}%
  \BibitemOpen
  \bibfield  {author} {\bibinfo {author} {\bibfnamefont {D.}~\bibnamefont {Jiang}}, \bibinfo {author} {\bibfnamefont {Z.}~\bibnamefont {Wu}}, \bibinfo {author} {\bibfnamefont {C.-Y.}\ \bibnamefont {Hsieh}}, \bibinfo {author} {\bibfnamefont {G.}~\bibnamefont {Chen}}, \bibinfo {author} {\bibfnamefont {B.}~\bibnamefont {Liao}}, \bibinfo {author} {\bibfnamefont {Z.}~\bibnamefont {Wang}}, \bibinfo {author} {\bibfnamefont {C.}~\bibnamefont {Shen}}, \bibinfo {author} {\bibfnamefont {D.}~\bibnamefont {Cao}}, \bibinfo {author} {\bibfnamefont {J.}~\bibnamefont {Wu}},\ and\ \bibinfo {author} {\bibfnamefont {T.}~\bibnamefont {Hou}},\ }\href@noop {} {\bibfield  {journal} {\bibinfo  {journal} {Journal of cheminformatics}\ }\textbf {\bibinfo {volume} {13}},\ \bibinfo {pages} {12} (\bibinfo {year} {2021})}\BibitemShut {NoStop}%
\bibitem [{\citenamefont {Pasini1}\ \emph {et~al.}(2020)\citenamefont {Pasini1}, \citenamefont {Li}, \citenamefont {Yin}, \citenamefont {Zhang}, \citenamefont {Barros},\ and\ \citenamefont {Eisenbach}}]{eisenbach2020solidsolution}%
  \BibitemOpen
  \bibfield  {author} {\bibinfo {author} {\bibfnamefont {M.}~\bibnamefont {Pasini1}}, \bibinfo {author} {\bibfnamefont {Y.}~\bibnamefont {Li}}, \bibinfo {author} {\bibfnamefont {J.}~\bibnamefont {Yin}}, \bibinfo {author} {\bibfnamefont {J.}~\bibnamefont {Zhang}}, \bibinfo {author} {\bibfnamefont {K.}~\bibnamefont {Barros}},\ and\ \bibinfo {author} {\bibfnamefont {M.}~\bibnamefont {Eisenbach}},\ }\href@noop {} {\bibfield  {journal} {\bibinfo  {journal} {Journal of Physics: Condensed Matter}\ }\textbf {\bibinfo {volume} {33}} (\bibinfo {year} {2020})}\BibitemShut {NoStop}%
\bibitem [{\citenamefont {Medasani}\ \emph {et~al.}(2016)\citenamefont {Medasani}, \citenamefont {Gamst}, \citenamefont {Ding}, \citenamefont {Chen}, \citenamefont {Persson}, \citenamefont {Asta}, \citenamefont {Canning},\ and\ \citenamefont {Haranczyk}}]{medasani2016predicting}%
  \BibitemOpen
  \bibfield  {author} {\bibinfo {author} {\bibfnamefont {B.}~\bibnamefont {Medasani}}, \bibinfo {author} {\bibfnamefont {A.}~\bibnamefont {Gamst}}, \bibinfo {author} {\bibfnamefont {H.}~\bibnamefont {Ding}}, \bibinfo {author} {\bibfnamefont {W.}~\bibnamefont {Chen}}, \bibinfo {author} {\bibfnamefont {K.~A.}\ \bibnamefont {Persson}}, \bibinfo {author} {\bibfnamefont {M.}~\bibnamefont {Asta}}, \bibinfo {author} {\bibfnamefont {A.}~\bibnamefont {Canning}},\ and\ \bibinfo {author} {\bibfnamefont {M.}~\bibnamefont {Haranczyk}},\ }\href@noop {} {\bibfield  {journal} {\bibinfo  {journal} {npj Computational Materials}\ }\textbf {\bibinfo {volume} {2}},\ \bibinfo {pages} {1} (\bibinfo {year} {2016})}\BibitemShut {NoStop}%
\bibitem [{\citenamefont {Deshmukh}\ \emph {et~al.}(2024)\citenamefont {Deshmukh}, \citenamefont {Wichrowski}, \citenamefont {Evangelou}, \citenamefont {Ghanekar}, \citenamefont {Deshpande}, \citenamefont {Kevrekidis},\ and\ \citenamefont {Greeley}}]{greeley2024active}%
  \BibitemOpen
  \bibfield  {author} {\bibinfo {author} {\bibfnamefont {G.}~\bibnamefont {Deshmukh}}, \bibinfo {author} {\bibfnamefont {N.~J.}\ \bibnamefont {Wichrowski}}, \bibinfo {author} {\bibfnamefont {N.}~\bibnamefont {Evangelou}}, \bibinfo {author} {\bibfnamefont {P.~G.}\ \bibnamefont {Ghanekar}}, \bibinfo {author} {\bibfnamefont {S.}~\bibnamefont {Deshpande}}, \bibinfo {author} {\bibfnamefont {I.~G.}\ \bibnamefont {Kevrekidis}},\ and\ \bibinfo {author} {\bibfnamefont {J.}~\bibnamefont {Greeley}},\ }\href@noop {} {\bibfield  {journal} {\bibinfo  {journal} {npj Computational Materials}\ }\textbf {\bibinfo {volume} {10}} (\bibinfo {year} {2024})}\BibitemShut {NoStop}%
\bibitem [{\citenamefont {Freitas}\ \emph {et~al.}(2024)\citenamefont {Freitas}, \citenamefont {Cao},\ and\ \citenamefont {Sheriff}}]{freitas2024equivariant}%
  \BibitemOpen
  \bibfield  {author} {\bibinfo {author} {\bibfnamefont {R.}~\bibnamefont {Freitas}}, \bibinfo {author} {\bibfnamefont {Y.}~\bibnamefont {Cao}},\ and\ \bibinfo {author} {\bibfnamefont {K.}~\bibnamefont {Sheriff}},\ }\href@noop {} {\bibfield  {journal} {\bibinfo  {journal} {npj Computational Materials}\ }\textbf {\bibinfo {volume} {10}} (\bibinfo {year} {2024})}\BibitemShut {NoStop}%
\bibitem [{\citenamefont {Chen}\ \emph {et~al.}(2024)\citenamefont {Chen}, \citenamefont {Shang}, \citenamefont {Liu},\ and\ \citenamefont {Yang}}]{yang2024eutecticml}%
  \BibitemOpen
  \bibfield  {author} {\bibinfo {author} {\bibfnamefont {Z.}~\bibnamefont {Chen}}, \bibinfo {author} {\bibfnamefont {Y.}~\bibnamefont {Shang}}, \bibinfo {author} {\bibfnamefont {X.}~\bibnamefont {Liu}},\ and\ \bibinfo {author} {\bibfnamefont {Y.}~\bibnamefont {Yang}},\ }\href@noop {} {\bibfield  {journal} {\bibinfo  {journal} {npj Computational Materials}\ }\textbf {\bibinfo {volume} {10}} (\bibinfo {year} {2024})}\BibitemShut {NoStop}%
\bibitem [{\citenamefont {Vazquez}\ \emph {et~al.}(2023)\citenamefont {Vazquez}, \citenamefont {Chakravarty}, \citenamefont {Gurrola},\ and\ \citenamefont {Arróyave}}]{vazquez2023regressor}%
  \BibitemOpen
  \bibfield  {author} {\bibinfo {author} {\bibfnamefont {G.}~\bibnamefont {Vazquez}}, \bibinfo {author} {\bibfnamefont {S.}~\bibnamefont {Chakravarty}}, \bibinfo {author} {\bibfnamefont {R.}~\bibnamefont {Gurrola}},\ and\ \bibinfo {author} {\bibfnamefont {R.}~\bibnamefont {Arróyave}},\ }\href@noop {} {\bibfield  {journal} {\bibinfo  {journal} {npj Computational Materials}\ }\textbf {\bibinfo {volume} {9}} (\bibinfo {year} {2023})}\BibitemShut {NoStop}%
\bibitem [{\citenamefont {Giles}\ \emph {et~al.}(2022)\citenamefont {Giles}, \citenamefont {Sengupta}, \citenamefont {Broderick},\ and\ \citenamefont {Rajan}}]{giles2022rhea}%
  \BibitemOpen
  \bibfield  {author} {\bibinfo {author} {\bibfnamefont {S.~A.}\ \bibnamefont {Giles}}, \bibinfo {author} {\bibfnamefont {D.}~\bibnamefont {Sengupta}}, \bibinfo {author} {\bibfnamefont {S.~R.}\ \bibnamefont {Broderick}},\ and\ \bibinfo {author} {\bibfnamefont {K.}~\bibnamefont {Rajan}},\ }\href@noop {} {\bibfield  {journal} {\bibinfo  {journal} {npj Computational Materials}\ }\textbf {\bibinfo {volume} {8}} (\bibinfo {year} {2022})}\BibitemShut {NoStop}%
\bibitem [{\citenamefont {Zhang}\ \emph {et~al.}(2022{\natexlab{a}})\citenamefont {Zhang}, \citenamefont {Cai}, \citenamefont {Kim}, \citenamefont {Wang},\ and\ \citenamefont {Chen}}]{chen2022compdesign}%
  \BibitemOpen
  \bibfield  {author} {\bibinfo {author} {\bibfnamefont {J.}~\bibnamefont {Zhang}}, \bibinfo {author} {\bibfnamefont {C.}~\bibnamefont {Cai}}, \bibinfo {author} {\bibfnamefont {G.}~\bibnamefont {Kim}}, \bibinfo {author} {\bibfnamefont {Y.}~\bibnamefont {Wang}},\ and\ \bibinfo {author} {\bibfnamefont {W.}~\bibnamefont {Chen}},\ }\href@noop {} {\bibfield  {journal} {\bibinfo  {journal} {npj Computational Materials}\ }\textbf {\bibinfo {volume} {8}} (\bibinfo {year} {2022}{\natexlab{a}})}\BibitemShut {NoStop}%
\bibitem [{\citenamefont {Gao}\ \emph {et~al.}(2018)\citenamefont {Gao}, \citenamefont {Miracle}, \citenamefont {Maurice}, \citenamefont {Yan}, \citenamefont {Zhang},\ and\ \citenamefont {Hawk}}]{gao2018high}%
  \BibitemOpen
  \bibfield  {author} {\bibinfo {author} {\bibfnamefont {M.~C.}\ \bibnamefont {Gao}}, \bibinfo {author} {\bibfnamefont {D.~B.}\ \bibnamefont {Miracle}}, \bibinfo {author} {\bibfnamefont {D.}~\bibnamefont {Maurice}}, \bibinfo {author} {\bibfnamefont {X.}~\bibnamefont {Yan}}, \bibinfo {author} {\bibfnamefont {Y.}~\bibnamefont {Zhang}},\ and\ \bibinfo {author} {\bibfnamefont {J.~A.}\ \bibnamefont {Hawk}},\ }\href@noop {} {\bibfield  {journal} {\bibinfo  {journal} {Journal of Materials Research}\ }\textbf {\bibinfo {volume} {33}},\ \bibinfo {pages} {3138} (\bibinfo {year} {2018})}\BibitemShut {NoStop}%
\bibitem [{\citenamefont {Kumari}\ \emph {et~al.}(2022)\citenamefont {Kumari}, \citenamefont {Gupta}, \citenamefont {Mishra}, \citenamefont {Ahmad},\ and\ \citenamefont {Shahi}}]{kumari2022comprehensive}%
  \BibitemOpen
  \bibfield  {author} {\bibinfo {author} {\bibfnamefont {P.}~\bibnamefont {Kumari}}, \bibinfo {author} {\bibfnamefont {A.~K.}\ \bibnamefont {Gupta}}, \bibinfo {author} {\bibfnamefont {R.~K.}\ \bibnamefont {Mishra}}, \bibinfo {author} {\bibfnamefont {M.}~\bibnamefont {Ahmad}},\ and\ \bibinfo {author} {\bibfnamefont {R.~R.}\ \bibnamefont {Shahi}},\ }\href@noop {} {\bibfield  {journal} {\bibinfo  {journal} {Journal of Magnetism and Magnetic Materials}\ }\textbf {\bibinfo {volume} {554}},\ \bibinfo {pages} {169142} (\bibinfo {year} {2022})}\BibitemShut {NoStop}%
\bibitem [{\citenamefont {Dai}\ \emph {et~al.}(2020)\citenamefont {Dai}, \citenamefont {Wen}, \citenamefont {Sun}, \citenamefont {Xiang},\ and\ \citenamefont {Zhou}}]{dai2020theoretical}%
  \BibitemOpen
  \bibfield  {author} {\bibinfo {author} {\bibfnamefont {F.-Z.}\ \bibnamefont {Dai}}, \bibinfo {author} {\bibfnamefont {B.}~\bibnamefont {Wen}}, \bibinfo {author} {\bibfnamefont {Y.}~\bibnamefont {Sun}}, \bibinfo {author} {\bibfnamefont {H.}~\bibnamefont {Xiang}},\ and\ \bibinfo {author} {\bibfnamefont {Y.}~\bibnamefont {Zhou}},\ }\href@noop {} {\bibfield  {journal} {\bibinfo  {journal} {Journal of Materials Science \& Technology}\ }\textbf {\bibinfo {volume} {43}},\ \bibinfo {pages} {168} (\bibinfo {year} {2020})}\BibitemShut {NoStop}%
\bibitem [{\citenamefont {Deringer}\ \emph {et~al.}(2019)\citenamefont {Deringer}, \citenamefont {Caro},\ and\ \citenamefont {Cs{\'a}nyi}}]{deringer2019machine}%
  \BibitemOpen
  \bibfield  {author} {\bibinfo {author} {\bibfnamefont {V.~L.}\ \bibnamefont {Deringer}}, \bibinfo {author} {\bibfnamefont {M.~A.}\ \bibnamefont {Caro}},\ and\ \bibinfo {author} {\bibfnamefont {G.}~\bibnamefont {Cs{\'a}nyi}},\ }\href@noop {} {\bibfield  {journal} {\bibinfo  {journal} {Advanced Materials}\ }\textbf {\bibinfo {volume} {31}},\ \bibinfo {pages} {1902765} (\bibinfo {year} {2019})}\BibitemShut {NoStop}%
\bibitem [{\citenamefont {K{\"o}rmann}\ \emph {et~al.}(2021)\citenamefont {K{\"o}rmann}, \citenamefont {Kostiuchenko}, \citenamefont {Shapeev},\ and\ \citenamefont {Neugebauer}}]{kormann2021b2}%
  \BibitemOpen
  \bibfield  {author} {\bibinfo {author} {\bibfnamefont {F.}~\bibnamefont {K{\"o}rmann}}, \bibinfo {author} {\bibfnamefont {T.}~\bibnamefont {Kostiuchenko}}, \bibinfo {author} {\bibfnamefont {A.}~\bibnamefont {Shapeev}},\ and\ \bibinfo {author} {\bibfnamefont {J.}~\bibnamefont {Neugebauer}},\ }\href@noop {} {\bibfield  {journal} {\bibinfo  {journal} {Physical Review Materials}\ }\textbf {\bibinfo {volume} {5}},\ \bibinfo {pages} {053803} (\bibinfo {year} {2021})}\BibitemShut {NoStop}%
\bibitem [{\citenamefont {Byggm{\"a}star}\ \emph {et~al.}(2021)\citenamefont {Byggm{\"a}star}, \citenamefont {Nordlund},\ and\ \citenamefont {Djurabekova}}]{byggmastar2021modeling}%
  \BibitemOpen
  \bibfield  {author} {\bibinfo {author} {\bibfnamefont {J.}~\bibnamefont {Byggm{\"a}star}}, \bibinfo {author} {\bibfnamefont {K.}~\bibnamefont {Nordlund}},\ and\ \bibinfo {author} {\bibfnamefont {F.}~\bibnamefont {Djurabekova}},\ }\href@noop {} {\bibfield  {journal} {\bibinfo  {journal} {Physical Review B}\ }\textbf {\bibinfo {volume} {104}},\ \bibinfo {pages} {104101} (\bibinfo {year} {2021})}\BibitemShut {NoStop}%
\bibitem [{\citenamefont {Pandey}\ \emph {et~al.}(2022)\citenamefont {Pandey}, \citenamefont {Gigax},\ and\ \citenamefont {Pokharel}}]{pandey2022machine}%
  \BibitemOpen
  \bibfield  {author} {\bibinfo {author} {\bibfnamefont {A.}~\bibnamefont {Pandey}}, \bibinfo {author} {\bibfnamefont {J.}~\bibnamefont {Gigax}},\ and\ \bibinfo {author} {\bibfnamefont {R.}~\bibnamefont {Pokharel}},\ }\href@noop {} {\bibfield  {journal} {\bibinfo  {journal} {JOM}\ }\textbf {\bibinfo {volume} {74}},\ \bibinfo {pages} {2908} (\bibinfo {year} {2022})}\BibitemShut {NoStop}%
\bibitem [{\citenamefont {You}\ \emph {et~al.}(2024)\citenamefont {You}, \citenamefont {Zhang}, \citenamefont {Wu}, \citenamefont {Cao}, \citenamefont {Sun}, \citenamefont {Zhu},\ and\ \citenamefont {Wu}}]{you2024principal}%
  \BibitemOpen
  \bibfield  {author} {\bibinfo {author} {\bibfnamefont {Y.}~\bibnamefont {You}}, \bibinfo {author} {\bibfnamefont {D.}~\bibnamefont {Zhang}}, \bibinfo {author} {\bibfnamefont {F.}~\bibnamefont {Wu}}, \bibinfo {author} {\bibfnamefont {X.}~\bibnamefont {Cao}}, \bibinfo {author} {\bibfnamefont {Y.}~\bibnamefont {Sun}}, \bibinfo {author} {\bibfnamefont {Z.-Z.}\ \bibnamefont {Zhu}},\ and\ \bibinfo {author} {\bibfnamefont {S.}~\bibnamefont {Wu}},\ }\href@noop {} {\bibfield  {journal} {\bibinfo  {journal} {npj Computational Materials}\ }\textbf {\bibinfo {volume} {10}},\ \bibinfo {pages} {57} (\bibinfo {year} {2024})}\BibitemShut {NoStop}%
\bibitem [{\citenamefont {Li}\ \emph {et~al.}(2022)\citenamefont {Li}, \citenamefont {Chen},\ and\ \citenamefont {Jin}}]{li2022sigesn}%
  \BibitemOpen
  \bibfield  {author} {\bibinfo {author} {\bibfnamefont {T.}~\bibnamefont {Li}}, \bibinfo {author} {\bibfnamefont {S.}~\bibnamefont {Chen}},\ and\ \bibinfo {author} {\bibfnamefont {X.}~\bibnamefont {Jin}},\ }\href@noop {} {\bibfield  {journal} {\bibinfo  {journal} {Communications Materials}\ }\textbf {\bibinfo {volume} {3}} (\bibinfo {year} {2022})}\BibitemShut {NoStop}%
\bibitem [{\citenamefont {Cao}\ \emph {et~al.}(2020)\citenamefont {Cao}, \citenamefont {Chen}, \citenamefont {Jin}, \citenamefont {Liu},\ and\ \citenamefont {Li}}]{cao2020gesn}%
  \BibitemOpen
  \bibfield  {author} {\bibinfo {author} {\bibfnamefont {B.}~\bibnamefont {Cao}}, \bibinfo {author} {\bibfnamefont {S.}~\bibnamefont {Chen}}, \bibinfo {author} {\bibfnamefont {X.}~\bibnamefont {Jin}}, \bibinfo {author} {\bibfnamefont {J.}~\bibnamefont {Liu}},\ and\ \bibinfo {author} {\bibfnamefont {T.}~\bibnamefont {Li}},\ }\href@noop {} {\bibfield  {journal} {\bibinfo  {journal} {ACS Applied Materials \& Interfaces}\ }\textbf {\bibinfo {volume} {12}} (\bibinfo {year} {2020})}\BibitemShut {NoStop}%
\bibitem [{\citenamefont {Wirths}\ \emph {et~al.}(2016)\citenamefont {Wirths}, \citenamefont {Buca},\ and\ \citenamefont {Mantl}}]{wirths2016sigesn}%
  \BibitemOpen
  \bibfield  {author} {\bibinfo {author} {\bibfnamefont {S.}~\bibnamefont {Wirths}}, \bibinfo {author} {\bibfnamefont {D.}~\bibnamefont {Buca}},\ and\ \bibinfo {author} {\bibfnamefont {S.}~\bibnamefont {Mantl}},\ }\href@noop {} {\bibfield  {journal} {\bibinfo  {journal} {Progress in Crystal Growth and Characterization of Materials}\ }\textbf {\bibinfo {volume} {62}},\ \bibinfo {pages} {1} (\bibinfo {year} {2016})}\BibitemShut {NoStop}%
\bibitem [{\citenamefont {Zhuang}\ \emph {et~al.}(2019)\citenamefont {Zhuang}, \citenamefont {Wang}, \citenamefont {Liu},\ and\ \citenamefont {Huang}}]{zhuang2019sigesn}%
  \BibitemOpen
  \bibfield  {author} {\bibinfo {author} {\bibfnamefont {H.~L.}\ \bibnamefont {Zhuang}}, \bibinfo {author} {\bibfnamefont {D.}~\bibnamefont {Wang}}, \bibinfo {author} {\bibfnamefont {L.}~\bibnamefont {Liu}},\ and\ \bibinfo {author} {\bibfnamefont {W.}~\bibnamefont {Huang}},\ }\href@noop {} {\bibfield  {journal} {\bibinfo  {journal} {Journal of Applied Physics}\ }\textbf {\bibinfo {volume} {126}},\ \bibinfo {pages} {225703} (\bibinfo {year} {2019})}\BibitemShut {NoStop}%
\bibitem [{\citenamefont {Olesinski}\ and\ \citenamefont {Abbaschian}(1984)}]{olesinski1984sige}%
  \BibitemOpen
  \bibfield  {author} {\bibinfo {author} {\bibfnamefont {R.~W.}\ \bibnamefont {Olesinski}}\ and\ \bibinfo {author} {\bibfnamefont {G.}~\bibnamefont {Abbaschian}},\ }\href@noop {} {\bibfield  {journal} {\bibinfo  {journal} {Bulletin of Alloy Phase Diagrams}\ }\textbf {\bibinfo {volume} {5}},\ \bibinfo {pages} {180} (\bibinfo {year} {1984})}\BibitemShut {NoStop}%
\bibitem [{\citenamefont {Grützmacher}\ \emph {et~al.}(2023)\citenamefont {Grützmacher}, \citenamefont {Concepción}, \citenamefont {Zhao},\ and\ \citenamefont {Buca}}]{buca2023sigesn}%
  \BibitemOpen
  \bibfield  {author} {\bibinfo {author} {\bibfnamefont {D.}~\bibnamefont {Grützmacher}}, \bibinfo {author} {\bibfnamefont {O.}~\bibnamefont {Concepción}}, \bibinfo {author} {\bibfnamefont {Q.-T.}\ \bibnamefont {Zhao}},\ and\ \bibinfo {author} {\bibfnamefont {D.}~\bibnamefont {Buca}},\ }\href@noop {} {\bibfield  {journal} {\bibinfo  {journal} {Applied Physics A}\ }\textbf {\bibinfo {volume} {129}} (\bibinfo {year} {2023})}\BibitemShut {NoStop}%
\bibitem [{\citenamefont {Wang}\ \emph {et~al.}(2019)\citenamefont {Wang}, \citenamefont {Liu}, \citenamefont {Huang},\ and\ \citenamefont {Zhuang}}]{wang2019semiconducting}%
  \BibitemOpen
  \bibfield  {author} {\bibinfo {author} {\bibfnamefont {D.}~\bibnamefont {Wang}}, \bibinfo {author} {\bibfnamefont {L.}~\bibnamefont {Liu}}, \bibinfo {author} {\bibfnamefont {W.}~\bibnamefont {Huang}},\ and\ \bibinfo {author} {\bibfnamefont {H.~L.}\ \bibnamefont {Zhuang}},\ }\href@noop {} {\bibfield  {journal} {\bibinfo  {journal} {Journal of Applied Physics}\ }\textbf {\bibinfo {volume} {126}} (\bibinfo {year} {2019})}\BibitemShut {NoStop}%
\bibitem [{\citenamefont {Wang}\ \emph {et~al.}(2020)\citenamefont {Wang}, \citenamefont {Liu}, \citenamefont {Chen},\ and\ \citenamefont {Zhuang}}]{wang2020electrical}%
  \BibitemOpen
  \bibfield  {author} {\bibinfo {author} {\bibfnamefont {D.}~\bibnamefont {Wang}}, \bibinfo {author} {\bibfnamefont {L.}~\bibnamefont {Liu}}, \bibinfo {author} {\bibfnamefont {M.}~\bibnamefont {Chen}},\ and\ \bibinfo {author} {\bibfnamefont {H.}~\bibnamefont {Zhuang}},\ }\href@noop {} {\bibfield  {journal} {\bibinfo  {journal} {Acta Materialia}\ }\textbf {\bibinfo {volume} {199}},\ \bibinfo {pages} {443} (\bibinfo {year} {2020})}\BibitemShut {NoStop}%
\bibitem [{\citenamefont {Cantor}(2021{\natexlab{a}})}]{cantor2021multicomponent}%
  \BibitemOpen
  \bibfield  {author} {\bibinfo {author} {\bibfnamefont {B.}~\bibnamefont {Cantor}},\ }\href@noop {} {\bibfield  {journal} {\bibinfo  {journal} {Progress in Materials Science}\ }\textbf {\bibinfo {volume} {120}},\ \bibinfo {pages} {100754} (\bibinfo {year} {2021}{\natexlab{a}})}\BibitemShut {NoStop}%
\bibitem [{\citenamefont {He}\ \emph {et~al.}(2016)\citenamefont {He}, \citenamefont {Wang}, \citenamefont {Jin}, \citenamefont {Bei}, \citenamefont {Yasuda}, \citenamefont {Matsumura}, \citenamefont {Higashida},\ and\ \citenamefont {Robertson}}]{he2016irrad}%
  \BibitemOpen
  \bibfield  {author} {\bibinfo {author} {\bibfnamefont {M.-R.}\ \bibnamefont {He}}, \bibinfo {author} {\bibfnamefont {S.}~\bibnamefont {Wang}}, \bibinfo {author} {\bibfnamefont {K.}~\bibnamefont {Jin}}, \bibinfo {author} {\bibfnamefont {H.}~\bibnamefont {Bei}}, \bibinfo {author} {\bibfnamefont {K.}~\bibnamefont {Yasuda}}, \bibinfo {author} {\bibfnamefont {S.}~\bibnamefont {Matsumura}}, \bibinfo {author} {\bibfnamefont {K.}~\bibnamefont {Higashida}},\ and\ \bibinfo {author} {\bibfnamefont {I.}~\bibnamefont {Robertson}},\ }\href@noop {} {\bibfield  {journal} {\bibinfo  {journal} {Scripta Materialia}\ }\textbf {\bibinfo {volume} {125}} (\bibinfo {year} {2016})}\BibitemShut {NoStop}%
\bibitem [{\citenamefont {Lei}\ and\ \citenamefont {Medford}(2022)}]{lei2022universal}%
  \BibitemOpen
  \bibfield  {author} {\bibinfo {author} {\bibfnamefont {X.}~\bibnamefont {Lei}}\ and\ \bibinfo {author} {\bibfnamefont {A.~J.}\ \bibnamefont {Medford}},\ }\href@noop {} {\bibfield  {journal} {\bibinfo  {journal} {The Journal of Physical Chemistry Letters}\ }\textbf {\bibinfo {volume} {13}},\ \bibinfo {pages} {7911} (\bibinfo {year} {2022})}\BibitemShut {NoStop}%
\bibitem [{\citenamefont {Timmerman}\ \emph {et~al.}(2024)\citenamefont {Timmerman}, \citenamefont {Kumar}, \citenamefont {Suryanarayana},\ and\ \citenamefont {Medford}}]{Timmerman2024JCTC}%
  \BibitemOpen
  \bibfield  {author} {\bibinfo {author} {\bibfnamefont {L.~R.}\ \bibnamefont {Timmerman}}, \bibinfo {author} {\bibfnamefont {S.}~\bibnamefont {Kumar}}, \bibinfo {author} {\bibfnamefont {P.}~\bibnamefont {Suryanarayana}},\ and\ \bibinfo {author} {\bibfnamefont {A.~J.}\ \bibnamefont {Medford}},\ }\href {https://doi.org/10.1021/acs.jctc.4c00474} {\bibfield  {journal} {\bibinfo  {journal} {Journal of Chemical Theory and Computation}\ }\textbf {\bibinfo {volume} {20}},\ \bibinfo {pages} {5788} (\bibinfo {year} {2024})},\ \bibinfo {note} {pMID: 38975655},\ \Eprint {https://arxiv.org/abs/https://doi.org/10.1021/acs.jctc.4c00474} {https://doi.org/10.1021/acs.jctc.4c00474} \BibitemShut {NoStop}%
\bibitem [{\citenamefont {Nemani}\ \emph {et~al.}(2021)\citenamefont {Nemani}, \citenamefont {Zhang}, \citenamefont {Wyatt}, \citenamefont {Hood}, \citenamefont {Manna}, \citenamefont {Khaledialidusti}, \citenamefont {Hong}, \citenamefont {Sternberg}, \citenamefont {Sankaranarayanan},\ and\ \citenamefont {Anasori}}]{nemani2021high}%
  \BibitemOpen
  \bibfield  {author} {\bibinfo {author} {\bibfnamefont {S.~K.}\ \bibnamefont {Nemani}}, \bibinfo {author} {\bibfnamefont {B.}~\bibnamefont {Zhang}}, \bibinfo {author} {\bibfnamefont {B.~C.}\ \bibnamefont {Wyatt}}, \bibinfo {author} {\bibfnamefont {Z.~D.}\ \bibnamefont {Hood}}, \bibinfo {author} {\bibfnamefont {S.}~\bibnamefont {Manna}}, \bibinfo {author} {\bibfnamefont {R.}~\bibnamefont {Khaledialidusti}}, \bibinfo {author} {\bibfnamefont {W.}~\bibnamefont {Hong}}, \bibinfo {author} {\bibfnamefont {M.~G.}\ \bibnamefont {Sternberg}}, \bibinfo {author} {\bibfnamefont {S.~K.}\ \bibnamefont {Sankaranarayanan}},\ and\ \bibinfo {author} {\bibfnamefont {B.}~\bibnamefont {Anasori}},\ }\href@noop {} {\bibfield  {journal} {\bibinfo  {journal} {ACS nano}\ }\textbf {\bibinfo {volume} {15}},\ \bibinfo {pages} {12815} (\bibinfo {year} {2021})}\BibitemShut {NoStop}%
\bibitem [{\citenamefont {Nemani}\ \emph {et~al.}(2023)\citenamefont {Nemani}, \citenamefont {Torkamanzadeh}, \citenamefont {Wyatt}, \citenamefont {Presser},\ and\ \citenamefont {Anasori}}]{nemani2023functional}%
  \BibitemOpen
  \bibfield  {author} {\bibinfo {author} {\bibfnamefont {S.~K.}\ \bibnamefont {Nemani}}, \bibinfo {author} {\bibfnamefont {M.}~\bibnamefont {Torkamanzadeh}}, \bibinfo {author} {\bibfnamefont {B.~C.}\ \bibnamefont {Wyatt}}, \bibinfo {author} {\bibfnamefont {V.}~\bibnamefont {Presser}},\ and\ \bibinfo {author} {\bibfnamefont {B.}~\bibnamefont {Anasori}},\ }\href@noop {} {\bibfield  {journal} {\bibinfo  {journal} {Communications Materials}\ }\textbf {\bibinfo {volume} {4}},\ \bibinfo {pages} {16} (\bibinfo {year} {2023})}\BibitemShut {NoStop}%
\bibitem [{\citenamefont {Deshpande}\ \emph {et~al.}(2022)\citenamefont {Deshpande}, \citenamefont {Ratsch}, \citenamefont {Ciobanu},\ and\ \citenamefont {Kodambaka}}]{deshpande2022entropy}%
  \BibitemOpen
  \bibfield  {author} {\bibinfo {author} {\bibfnamefont {A.}~\bibnamefont {Deshpande}}, \bibinfo {author} {\bibfnamefont {C.}~\bibnamefont {Ratsch}}, \bibinfo {author} {\bibfnamefont {C.~V.}\ \bibnamefont {Ciobanu}},\ and\ \bibinfo {author} {\bibfnamefont {S.}~\bibnamefont {Kodambaka}},\ }\href@noop {} {\bibfield  {journal} {\bibinfo  {journal} {Journal of Applied Physics}\ }\textbf {\bibinfo {volume} {131}} (\bibinfo {year} {2022})}\BibitemShut {NoStop}%
\bibitem [{\citenamefont {Cantor}(2021{\natexlab{b}})}]{cantor2021crfeconi}%
  \BibitemOpen
  \bibfield  {author} {\bibinfo {author} {\bibfnamefont {B.}~\bibnamefont {Cantor}},\ }\href@noop {} {\bibfield  {journal} {\bibinfo  {journal} {Progress in Materials Science}\ }\textbf {\bibinfo {volume} {120}},\ \bibinfo {pages} {100754} (\bibinfo {year} {2021}{\natexlab{b}})}\BibitemShut {NoStop}%
\bibitem [{\citenamefont {He}\ \emph {et~al.}(2017)\citenamefont {He}, \citenamefont {Wang}, \citenamefont {Wu}, \citenamefont {Niu}, \citenamefont {Li}, \citenamefont {Wang},\ and\ \citenamefont {Liu}}]{wang2017crfeconi}%
  \BibitemOpen
  \bibfield  {author} {\bibinfo {author} {\bibfnamefont {F.}~\bibnamefont {He}}, \bibinfo {author} {\bibfnamefont {Z.}~\bibnamefont {Wang}}, \bibinfo {author} {\bibfnamefont {Q.}~\bibnamefont {Wu}}, \bibinfo {author} {\bibfnamefont {S.}~\bibnamefont {Niu}}, \bibinfo {author} {\bibfnamefont {J.}~\bibnamefont {Li}}, \bibinfo {author} {\bibfnamefont {J.}~\bibnamefont {Wang}},\ and\ \bibinfo {author} {\bibfnamefont {C.}~\bibnamefont {Liu}},\ }\href@noop {} {\bibfield  {journal} {\bibinfo  {journal} {Scripta Materialia}\ }\textbf {\bibinfo {volume} {131}},\ \bibinfo {pages} {42} (\bibinfo {year} {2017})}\BibitemShut {NoStop}%
\bibitem [{\citenamefont {Sen}\ \emph {et~al.}(2024)\citenamefont {Sen}, \citenamefont {Glienke}, \citenamefont {Yadav}, \citenamefont {Vaidya}, \citenamefont {Gururaj}, \citenamefont {Pradeep}, \citenamefont {Daum}, \citenamefont {Tas}, \citenamefont {Rogal}, \citenamefont {Wilde},\ and\ \citenamefont {Divinski}}]{sen2024crfeconi}%
  \BibitemOpen
  \bibfield  {author} {\bibinfo {author} {\bibfnamefont {S.}~\bibnamefont {Sen}}, \bibinfo {author} {\bibfnamefont {M.}~\bibnamefont {Glienke}}, \bibinfo {author} {\bibfnamefont {B.}~\bibnamefont {Yadav}}, \bibinfo {author} {\bibfnamefont {M.}~\bibnamefont {Vaidya}}, \bibinfo {author} {\bibfnamefont {K.}~\bibnamefont {Gururaj}}, \bibinfo {author} {\bibfnamefont {K.}~\bibnamefont {Pradeep}}, \bibinfo {author} {\bibfnamefont {L.}~\bibnamefont {Daum}}, \bibinfo {author} {\bibfnamefont {B.}~\bibnamefont {Tas}}, \bibinfo {author} {\bibfnamefont {L.}~\bibnamefont {Rogal}}, \bibinfo {author} {\bibfnamefont {G.}~\bibnamefont {Wilde}},\ and\ \bibinfo {author} {\bibfnamefont {S.~V.}\ \bibnamefont {Divinski}},\ }\href@noop {} {\bibfield  {journal} {\bibinfo  {journal} {Acta Materialia}\ }\textbf {\bibinfo {volume} {264}},\ \bibinfo {pages} {119588} (\bibinfo {year} {2024})}\BibitemShut {NoStop}%
\bibitem [{\citenamefont {Zhao}\ \emph {et~al.}(2022)\citenamefont {Zhao}, \citenamefont {Jiang}, \citenamefont {Yang}, \citenamefont {Wang}, \citenamefont {Zhang}, \citenamefont {Ji}, \citenamefont {Zhou}, \citenamefont {Curtin}, \citenamefont {Chen}, \citenamefont {Liaw}, \citenamefont {Chen},\ and\ \citenamefont {Wang}}]{chen2022crfeconi}%
  \BibitemOpen
  \bibfield  {author} {\bibinfo {author} {\bibfnamefont {L.}~\bibnamefont {Zhao}}, \bibinfo {author} {\bibfnamefont {L.}~\bibnamefont {Jiang}}, \bibinfo {author} {\bibfnamefont {L.}~\bibnamefont {Yang}}, \bibinfo {author} {\bibfnamefont {H.}~\bibnamefont {Wang}}, \bibinfo {author} {\bibfnamefont {W.}~\bibnamefont {Zhang}}, \bibinfo {author} {\bibfnamefont {G.}~\bibnamefont {Ji}}, \bibinfo {author} {\bibfnamefont {X.}~\bibnamefont {Zhou}}, \bibinfo {author} {\bibfnamefont {W.}~\bibnamefont {Curtin}}, \bibinfo {author} {\bibfnamefont {X.}~\bibnamefont {Chen}}, \bibinfo {author} {\bibfnamefont {P.}~\bibnamefont {Liaw}}, \bibinfo {author} {\bibfnamefont {S.}~\bibnamefont {Chen}},\ and\ \bibinfo {author} {\bibfnamefont {H.}~\bibnamefont {Wang}},\ }\href@noop {} {\bibfield  {journal} {\bibinfo  {journal} {Journal of Materials Science and Technology}\ }\textbf {\bibinfo {volume} {110}},\ \bibinfo {pages} {269} (\bibinfo {year} {2022})}\BibitemShut {NoStop}%
\bibitem [{\citenamefont {Zhang}\ \emph {et~al.}(2022{\natexlab{b}})\citenamefont {Zhang}, \citenamefont {Yao}, \citenamefont {Liu}, \citenamefont {Mou}, \citenamefont {Xin}, \citenamefont {Li},\ and\ \citenamefont {Cai}}]{cai2022crfeconi}%
  \BibitemOpen
  \bibfield  {author} {\bibinfo {author} {\bibfnamefont {Z.}~\bibnamefont {Zhang}}, \bibinfo {author} {\bibfnamefont {Y.}~\bibnamefont {Yao}}, \bibinfo {author} {\bibfnamefont {L.}~\bibnamefont {Liu}}, \bibinfo {author} {\bibfnamefont {T.}~\bibnamefont {Mou}}, \bibinfo {author} {\bibfnamefont {H.}~\bibnamefont {Xin}}, \bibinfo {author} {\bibfnamefont {L.}~\bibnamefont {Li}},\ and\ \bibinfo {author} {\bibfnamefont {W.}~\bibnamefont {Cai}},\ }\href@noop {} {\bibfield  {journal} {\bibinfo  {journal} {Journal of Materials Research}\ }\textbf {\bibinfo {volume} {37}},\ \bibinfo {pages} {2738–2748} (\bibinfo {year} {2022}{\natexlab{b}})}\BibitemShut {NoStop}%
\bibitem [{\citenamefont {Zhong}\ and\ \citenamefont {Yang}(2021)}]{zhong2021crfeconi}%
  \BibitemOpen
  \bibfield  {author} {\bibinfo {author} {\bibfnamefont {Y.}~\bibnamefont {Zhong}}\ and\ \bibinfo {author} {\bibfnamefont {S.}~\bibnamefont {Yang}},\ }\href@noop {} {\bibfield  {journal} {\bibinfo  {journal} {Journal of Phase Equilibria and Diffusion}\ }\textbf {\bibinfo {volume} {42}},\ \bibinfo {pages} {656–672} (\bibinfo {year} {2021})}\BibitemShut {NoStop}%
\bibitem [{\citenamefont {Zhao}\ \emph {et~al.}(2021)\citenamefont {Zhao}, \citenamefont {Park}, \citenamefont {Jang},\ and\ \citenamefont {Ramamurty}}]{ramamurty2021crfeconi}%
  \BibitemOpen
  \bibfield  {author} {\bibinfo {author} {\bibfnamefont {Y.}~\bibnamefont {Zhao}}, \bibinfo {author} {\bibfnamefont {J.-M.}\ \bibnamefont {Park}}, \bibinfo {author} {\bibfnamefont {J.-i.}\ \bibnamefont {Jang}},\ and\ \bibinfo {author} {\bibfnamefont {U.}~\bibnamefont {Ramamurty}},\ }\href@noop {} {\bibfield  {journal} {\bibinfo  {journal} {Acta Materialia}\ }\textbf {\bibinfo {volume} {202}},\ \bibinfo {pages} {124} (\bibinfo {year} {2021})}\BibitemShut {NoStop}%
\bibitem [{\citenamefont {Tuomisto}\ \emph {et~al.}(2020)\citenamefont {Tuomisto}, \citenamefont {Makkonen}, \citenamefont {Heikinheimo}, \citenamefont {Granberg}, \citenamefont {Djurabekova}, \citenamefont {Nordlund}, \citenamefont {Velisa}, \citenamefont {Bei}, \citenamefont {Xue}, \citenamefont {Weber},\ and\ \citenamefont {Zhang}}]{tuomisto2020crfeconi}%
  \BibitemOpen
  \bibfield  {author} {\bibinfo {author} {\bibfnamefont {F.}~\bibnamefont {Tuomisto}}, \bibinfo {author} {\bibfnamefont {I.}~\bibnamefont {Makkonen}}, \bibinfo {author} {\bibfnamefont {J.}~\bibnamefont {Heikinheimo}}, \bibinfo {author} {\bibfnamefont {F.}~\bibnamefont {Granberg}}, \bibinfo {author} {\bibfnamefont {F.}~\bibnamefont {Djurabekova}}, \bibinfo {author} {\bibfnamefont {K.}~\bibnamefont {Nordlund}}, \bibinfo {author} {\bibfnamefont {G.}~\bibnamefont {Velisa}}, \bibinfo {author} {\bibfnamefont {H.}~\bibnamefont {Bei}}, \bibinfo {author} {\bibfnamefont {H.}~\bibnamefont {Xue}}, \bibinfo {author} {\bibfnamefont {W.}~\bibnamefont {Weber}},\ and\ \bibinfo {author} {\bibfnamefont {Y.}~\bibnamefont {Zhang}},\ }\href@noop {} {\bibfield  {journal} {\bibinfo  {journal} {Acta Materialia}\ }\textbf {\bibinfo {volume} {196}},\ \bibinfo {pages} {44} (\bibinfo {year} {2020})}\BibitemShut {NoStop}%
\bibitem [{\citenamefont {Robarts}\ \emph {et~al.}(2020)\citenamefont {Robarts}, \citenamefont {Millichamp}, \citenamefont {Lagos}, \citenamefont {Laverock}, \citenamefont {Billington}, \citenamefont {Duffy}, \citenamefont {O’Neill}, \citenamefont {Giblin}, \citenamefont {Taylor}, \citenamefont {Kontrym-Sznajd}, \citenamefont {Samsel-Czekała}, \citenamefont {Bei}, \citenamefont {Mu}, \citenamefont {Samolyuk}, \citenamefont {Stocks},\ and\ \citenamefont {Dugdale}}]{dugdale2020crfeconi}%
  \BibitemOpen
  \bibfield  {author} {\bibinfo {author} {\bibfnamefont {H.~C.}\ \bibnamefont {Robarts}}, \bibinfo {author} {\bibfnamefont {T.~E.}\ \bibnamefont {Millichamp}}, \bibinfo {author} {\bibfnamefont {D.~A.}\ \bibnamefont {Lagos}}, \bibinfo {author} {\bibfnamefont {J.}~\bibnamefont {Laverock}}, \bibinfo {author} {\bibfnamefont {D.}~\bibnamefont {Billington}}, \bibinfo {author} {\bibfnamefont {J.~A.}\ \bibnamefont {Duffy}}, \bibinfo {author} {\bibfnamefont {D.}~\bibnamefont {O’Neill}}, \bibinfo {author} {\bibfnamefont {S.~R.}\ \bibnamefont {Giblin}}, \bibinfo {author} {\bibfnamefont {J.~W.}\ \bibnamefont {Taylor}}, \bibinfo {author} {\bibfnamefont {G.}~\bibnamefont {Kontrym-Sznajd}}, \bibinfo {author} {\bibfnamefont {M.}~\bibnamefont {Samsel-Czekała}}, \bibinfo {author} {\bibfnamefont {H.}~\bibnamefont {Bei}}, \bibinfo {author} {\bibfnamefont {S.}~\bibnamefont {Mu}}, \bibinfo {author} {\bibfnamefont {G.~D.}\ \bibnamefont {Samolyuk}}, \bibinfo {author} {\bibfnamefont {G.~M.}\ \bibnamefont {Stocks}},\ and\ \bibinfo
  {author} {\bibfnamefont {S.~B.}\ \bibnamefont {Dugdale}},\ }\href@noop {} {\bibfield  {journal} {\bibinfo  {journal} {Physical Review Letters}\ }\textbf {\bibinfo {volume} {124}},\ \bibinfo {pages} {046402} (\bibinfo {year} {2020})}\BibitemShut {NoStop}%
\bibitem [{\citenamefont {Wei}\ \emph {et~al.}(2019)\citenamefont {Wei}, \citenamefont {Li}, \citenamefont {Heng}, \citenamefont {Koizumi}, \citenamefont {He}, \citenamefont {Choi}, \citenamefont {Lee}, \citenamefont {Kim}, \citenamefont {Kato},\ and\ \citenamefont {Chiba}}]{kim2019crfeconi}%
  \BibitemOpen
  \bibfield  {author} {\bibinfo {author} {\bibfnamefont {D.}~\bibnamefont {Wei}}, \bibinfo {author} {\bibfnamefont {X.}~\bibnamefont {Li}}, \bibinfo {author} {\bibfnamefont {W.}~\bibnamefont {Heng}}, \bibinfo {author} {\bibfnamefont {Y.}~\bibnamefont {Koizumi}}, \bibinfo {author} {\bibfnamefont {F.}~\bibnamefont {He}}, \bibinfo {author} {\bibfnamefont {W.-M.}\ \bibnamefont {Choi}}, \bibinfo {author} {\bibfnamefont {B.-J.}\ \bibnamefont {Lee}}, \bibinfo {author} {\bibfnamefont {H.~S.}\ \bibnamefont {Kim}}, \bibinfo {author} {\bibfnamefont {H.}~\bibnamefont {Kato}},\ and\ \bibinfo {author} {\bibfnamefont {A.}~\bibnamefont {Chiba}},\ }\href@noop {} {\bibfield  {journal} {\bibinfo  {journal} {Materials Research Letters}\ }\textbf {\bibinfo {volume} {7}} (\bibinfo {year} {2019})}\BibitemShut {NoStop}%
\bibitem [{\citenamefont {Bae}\ \emph {et~al.}(2018)\citenamefont {Bae}, \citenamefont {Seol}, \citenamefont {Moon}, \citenamefont {Sohn}, \citenamefont {Jang}, \citenamefont {Um}, \citenamefont {Lee},\ and\ \citenamefont {Kim}}]{kim2018crfeconi}%
  \BibitemOpen
  \bibfield  {author} {\bibinfo {author} {\bibfnamefont {J.~W.}\ \bibnamefont {Bae}}, \bibinfo {author} {\bibfnamefont {J.~B.}\ \bibnamefont {Seol}}, \bibinfo {author} {\bibfnamefont {J.}~\bibnamefont {Moon}}, \bibinfo {author} {\bibfnamefont {S.~S.}\ \bibnamefont {Sohn}}, \bibinfo {author} {\bibfnamefont {M.~J.}\ \bibnamefont {Jang}}, \bibinfo {author} {\bibfnamefont {H.~Y.}\ \bibnamefont {Um}}, \bibinfo {author} {\bibfnamefont {B.-J.}\ \bibnamefont {Lee}},\ and\ \bibinfo {author} {\bibfnamefont {H.~S.}\ \bibnamefont {Kim}},\ }\href@noop {} {\bibfield  {journal} {\bibinfo  {journal} {Acta Materialia}\ }\textbf {\bibinfo {volume} {161}},\ \bibinfo {pages} {388} (\bibinfo {year} {2018})}\BibitemShut {NoStop}%
\bibitem [{\citenamefont {Niu}\ \emph {et~al.}(2016)\citenamefont {Niu}, \citenamefont {Zaddach}, \citenamefont {Koch},\ and\ \citenamefont {Irving}}]{irving2016crfeconi}%
  \BibitemOpen
  \bibfield  {author} {\bibinfo {author} {\bibfnamefont {C.}~\bibnamefont {Niu}}, \bibinfo {author} {\bibfnamefont {A.}~\bibnamefont {Zaddach}}, \bibinfo {author} {\bibfnamefont {C.}~\bibnamefont {Koch}},\ and\ \bibinfo {author} {\bibfnamefont {D.}~\bibnamefont {Irving}},\ }\href@noop {} {\bibfield  {journal} {\bibinfo  {journal} {Journal of Alloys and Compounds}\ }\textbf {\bibinfo {volume} {672}},\ \bibinfo {pages} {510} (\bibinfo {year} {2016})}\BibitemShut {NoStop}%
\bibitem [{\citenamefont {Middleburgh}\ \emph {et~al.}(2014)\citenamefont {Middleburgh}, \citenamefont {King}, \citenamefont {Lumpkin}, \citenamefont {Cortie},\ and\ \citenamefont {Edwards}}]{middleburgh2014segregatio}%
  \BibitemOpen
  \bibfield  {author} {\bibinfo {author} {\bibfnamefont {S.}~\bibnamefont {Middleburgh}}, \bibinfo {author} {\bibfnamefont {D.}~\bibnamefont {King}}, \bibinfo {author} {\bibfnamefont {G.}~\bibnamefont {Lumpkin}}, \bibinfo {author} {\bibfnamefont {M.}~\bibnamefont {Cortie}},\ and\ \bibinfo {author} {\bibfnamefont {L.}~\bibnamefont {Edwards}},\ }\href@noop {} {\bibfield  {journal} {\bibinfo  {journal} {Journal of alloys and compounds}\ }\textbf {\bibinfo {volume} {599}},\ \bibinfo {pages} {179} (\bibinfo {year} {2014})}\BibitemShut {NoStop}%
\bibitem [{\citenamefont {Ghosh}\ and\ \citenamefont {Suryanarayana}(2017{\natexlab{a}})}]{ghosh2017sparc_2}%
  \BibitemOpen
  \bibfield  {author} {\bibinfo {author} {\bibfnamefont {S.}~\bibnamefont {Ghosh}}\ and\ \bibinfo {author} {\bibfnamefont {P.}~\bibnamefont {Suryanarayana}},\ }\href@noop {} {\bibfield  {journal} {\bibinfo  {journal} {Computer Physics Communications}\ }\textbf {\bibinfo {volume} {216}},\ \bibinfo {pages} {109} (\bibinfo {year} {2017}{\natexlab{a}})}\BibitemShut {NoStop}%
\bibitem [{\citenamefont {Suryanarayana}(2017)}]{suryanarayana2017nearsightedness}%
  \BibitemOpen
  \bibfield  {author} {\bibinfo {author} {\bibfnamefont {P.}~\bibnamefont {Suryanarayana}},\ }\href@noop {} {\bibfield  {journal} {\bibinfo  {journal} {Chemical Physics Letters}\ }\textbf {\bibinfo {volume} {679}},\ \bibinfo {pages} {146} (\bibinfo {year} {2017})}\BibitemShut {NoStop}%
\bibitem [{\citenamefont {Sauer}(2019)}]{sauer2019ab}%
  \BibitemOpen
  \bibfield  {author} {\bibinfo {author} {\bibfnamefont {J.}~\bibnamefont {Sauer}},\ }\href@noop {} {\bibfield  {journal} {\bibinfo  {journal} {Accounts of chemical research}\ }\textbf {\bibinfo {volume} {52}},\ \bibinfo {pages} {3502} (\bibinfo {year} {2019})}\BibitemShut {NoStop}%
\bibitem [{\citenamefont {Xu}\ \emph {et~al.}(2018)\citenamefont {Xu}, \citenamefont {Suryanarayana},\ and\ \citenamefont {Pask}}]{xu2018discrete}%
  \BibitemOpen
  \bibfield  {author} {\bibinfo {author} {\bibfnamefont {Q.}~\bibnamefont {Xu}}, \bibinfo {author} {\bibfnamefont {P.}~\bibnamefont {Suryanarayana}},\ and\ \bibinfo {author} {\bibfnamefont {J.~E.}\ \bibnamefont {Pask}},\ }\href@noop {} {\bibfield  {journal} {\bibinfo  {journal} {The Journal of chemical physics}\ }\textbf {\bibinfo {volume} {149}} (\bibinfo {year} {2018})}\BibitemShut {NoStop}%
\bibitem [{\citenamefont {Willand}\ \emph {et~al.}(2013)\citenamefont {Willand}, \citenamefont {Kvashnin}, \citenamefont {Genovese}, \citenamefont {V{\'a}zquez-Mayagoitia}, \citenamefont {Deb}, \citenamefont {Sadeghi}, \citenamefont {Deutsch},\ and\ \citenamefont {Goedecker}}]{willand2013norm}%
  \BibitemOpen
  \bibfield  {author} {\bibinfo {author} {\bibfnamefont {A.}~\bibnamefont {Willand}}, \bibinfo {author} {\bibfnamefont {Y.~O.}\ \bibnamefont {Kvashnin}}, \bibinfo {author} {\bibfnamefont {L.}~\bibnamefont {Genovese}}, \bibinfo {author} {\bibfnamefont {{\'A}.}~\bibnamefont {V{\'a}zquez-Mayagoitia}}, \bibinfo {author} {\bibfnamefont {A.~K.}\ \bibnamefont {Deb}}, \bibinfo {author} {\bibfnamefont {A.}~\bibnamefont {Sadeghi}}, \bibinfo {author} {\bibfnamefont {T.}~\bibnamefont {Deutsch}},\ and\ \bibinfo {author} {\bibfnamefont {S.}~\bibnamefont {Goedecker}},\ }\href@noop {} {\bibfield  {journal} {\bibinfo  {journal} {The Journal of chemical physics}\ }\textbf {\bibinfo {volume} {138}} (\bibinfo {year} {2013})}\BibitemShut {NoStop}%
\bibitem [{\citenamefont {et~al.}(2016)}]{science2016reproducibility}%
  \BibitemOpen
  \bibfield  {author} {\bibinfo {author} {\bibfnamefont {L.}~\bibnamefont {et~al.}},\ }\href@noop {} {\bibfield  {journal} {\bibinfo  {journal} {Science}\ }\textbf {\bibinfo {volume} {351}} (\bibinfo {year} {2016})}\BibitemShut {NoStop}%
\bibitem [{\citenamefont {Liu}\ \emph {et~al.}(2023)\citenamefont {Liu}, \citenamefont {He},\ and\ \citenamefont {Mo}}]{liu2023errors}%
  \BibitemOpen
  \bibfield  {author} {\bibinfo {author} {\bibfnamefont {Y.}~\bibnamefont {Liu}}, \bibinfo {author} {\bibfnamefont {X.}~\bibnamefont {He}},\ and\ \bibinfo {author} {\bibfnamefont {Y.}~\bibnamefont {Mo}},\ }\href@noop {} {\bibfield  {journal} {\bibinfo  {journal} {npj Computational Materials}\ }\textbf {\bibinfo {volume} {9}} (\bibinfo {year} {2023})}\BibitemShut {NoStop}%
\bibitem [{\citenamefont {Deng}\ \emph {et~al.}(2025)\citenamefont {Deng}, \citenamefont {Choi}, \citenamefont {Zhong}, \citenamefont {Riebesell}, \citenamefont {Anand}, \citenamefont {Li}, \citenamefont {Jun}, \citenamefont {Persson},\ and\ \citenamefont {Ceder}}]{deng2025systematic}%
  \BibitemOpen
  \bibfield  {author} {\bibinfo {author} {\bibfnamefont {B.}~\bibnamefont {Deng}}, \bibinfo {author} {\bibfnamefont {Y.}~\bibnamefont {Choi}}, \bibinfo {author} {\bibfnamefont {P.}~\bibnamefont {Zhong}}, \bibinfo {author} {\bibfnamefont {J.}~\bibnamefont {Riebesell}}, \bibinfo {author} {\bibfnamefont {S.}~\bibnamefont {Anand}}, \bibinfo {author} {\bibfnamefont {Z.}~\bibnamefont {Li}}, \bibinfo {author} {\bibfnamefont {K.}~\bibnamefont {Jun}}, \bibinfo {author} {\bibfnamefont {K.~A.}\ \bibnamefont {Persson}},\ and\ \bibinfo {author} {\bibfnamefont {G.}~\bibnamefont {Ceder}},\ }\href@noop {} {\bibfield  {journal} {\bibinfo  {journal} {npj Computational Materials}\ }\textbf {\bibinfo {volume} {11}},\ \bibinfo {pages} {1} (\bibinfo {year} {2025})}\BibitemShut {NoStop}%
\bibitem [{\citenamefont {Xu}\ \emph {et~al.}(2021)\citenamefont {Xu}, \citenamefont {Sharma}, \citenamefont {Comer}, \citenamefont {Huang}, \citenamefont {Chow}, \citenamefont {Medford}, \citenamefont {Pask},\ and\ \citenamefont {Suryanarayana}}]{xu2021sparc}%
  \BibitemOpen
  \bibfield  {author} {\bibinfo {author} {\bibfnamefont {Q.}~\bibnamefont {Xu}}, \bibinfo {author} {\bibfnamefont {A.}~\bibnamefont {Sharma}}, \bibinfo {author} {\bibfnamefont {B.}~\bibnamefont {Comer}}, \bibinfo {author} {\bibfnamefont {H.}~\bibnamefont {Huang}}, \bibinfo {author} {\bibfnamefont {E.}~\bibnamefont {Chow}}, \bibinfo {author} {\bibfnamefont {A.~J.}\ \bibnamefont {Medford}}, \bibinfo {author} {\bibfnamefont {J.~E.}\ \bibnamefont {Pask}},\ and\ \bibinfo {author} {\bibfnamefont {P.}~\bibnamefont {Suryanarayana}},\ }\href@noop {} {\bibfield  {journal} {\bibinfo  {journal} {SoftwareX}\ }\textbf {\bibinfo {volume} {15}},\ \bibinfo {pages} {100709} (\bibinfo {year} {2021})}\BibitemShut {NoStop}%
\bibitem [{\citenamefont {Zhang}\ \emph {et~al.}(2024)\citenamefont {Zhang}, \citenamefont {Jing}, \citenamefont {Xu}, \citenamefont {Kumar}, \citenamefont {Sharma}, \citenamefont {Erlandson}, \citenamefont {Sahoo}, \citenamefont {Chow}, \citenamefont {Medford}, \citenamefont {Pask} \emph {et~al.}}]{zhang2024sparc}%
  \BibitemOpen
  \bibfield  {author} {\bibinfo {author} {\bibfnamefont {B.}~\bibnamefont {Zhang}}, \bibinfo {author} {\bibfnamefont {X.}~\bibnamefont {Jing}}, \bibinfo {author} {\bibfnamefont {Q.}~\bibnamefont {Xu}}, \bibinfo {author} {\bibfnamefont {S.}~\bibnamefont {Kumar}}, \bibinfo {author} {\bibfnamefont {A.}~\bibnamefont {Sharma}}, \bibinfo {author} {\bibfnamefont {L.}~\bibnamefont {Erlandson}}, \bibinfo {author} {\bibfnamefont {S.~J.}\ \bibnamefont {Sahoo}}, \bibinfo {author} {\bibfnamefont {E.}~\bibnamefont {Chow}}, \bibinfo {author} {\bibfnamefont {A.~J.}\ \bibnamefont {Medford}}, \bibinfo {author} {\bibfnamefont {J.~E.}\ \bibnamefont {Pask}}, \emph {et~al.},\ }\href@noop {} {\bibfield  {journal} {\bibinfo  {journal} {Software Impacts}\ }\textbf {\bibinfo {volume} {20}},\ \bibinfo {pages} {100649} (\bibinfo {year} {2024})}\BibitemShut {NoStop}%
\bibitem [{\citenamefont {Ghosh}\ and\ \citenamefont {Suryanarayana}(2017{\natexlab{b}})}]{ghosh2017sparc}%
  \BibitemOpen
  \bibfield  {author} {\bibinfo {author} {\bibfnamefont {S.}~\bibnamefont {Ghosh}}\ and\ \bibinfo {author} {\bibfnamefont {P.}~\bibnamefont {Suryanarayana}},\ }\href@noop {} {\bibfield  {journal} {\bibinfo  {journal} {Computer Physics Communications}\ }\textbf {\bibinfo {volume} {212}},\ \bibinfo {pages} {189} (\bibinfo {year} {2017}{\natexlab{b}})}\BibitemShut {NoStop}%
\bibitem [{\citenamefont {Hamann}(2013)}]{hamann2013optimized}%
  \BibitemOpen
  \bibfield  {author} {\bibinfo {author} {\bibfnamefont {D.}~\bibnamefont {Hamann}},\ }\href@noop {} {\bibfield  {journal} {\bibinfo  {journal} {Physical Review B}\ }\textbf {\bibinfo {volume} {88}},\ \bibinfo {pages} {085117} (\bibinfo {year} {2013})}\BibitemShut {NoStop}%
\bibitem [{\citenamefont {Perdew}\ \emph {et~al.}(1996)\citenamefont {Perdew}, \citenamefont {Burke},\ and\ \citenamefont {Ernzerhof}}]{perdew1996generalized}%
  \BibitemOpen
  \bibfield  {author} {\bibinfo {author} {\bibfnamefont {J.~P.}\ \bibnamefont {Perdew}}, \bibinfo {author} {\bibfnamefont {K.}~\bibnamefont {Burke}},\ and\ \bibinfo {author} {\bibfnamefont {M.}~\bibnamefont {Ernzerhof}},\ }\href@noop {} {\bibfield  {journal} {\bibinfo  {journal} {Physical review letters}\ }\textbf {\bibinfo {volume} {77}},\ \bibinfo {pages} {3865} (\bibinfo {year} {1996})}\BibitemShut {NoStop}%
\bibitem [{\citenamefont {Banerjee}\ \emph {et~al.}(2016{\natexlab{b}})\citenamefont {Banerjee}, \citenamefont {Suryanarayana},\ and\ \citenamefont {Pask}}]{banerjee2016periodic}%
  \BibitemOpen
  \bibfield  {author} {\bibinfo {author} {\bibfnamefont {A.~S.}\ \bibnamefont {Banerjee}}, \bibinfo {author} {\bibfnamefont {P.}~\bibnamefont {Suryanarayana}},\ and\ \bibinfo {author} {\bibfnamefont {J.~E.}\ \bibnamefont {Pask}},\ }\href@noop {} {\bibfield  {journal} {\bibinfo  {journal} {Chemical Physics Letters}\ }\textbf {\bibinfo {volume} {647}},\ \bibinfo {pages} {31} (\bibinfo {year} {2016}{\natexlab{b}})}\BibitemShut {NoStop}%
\bibitem [{\citenamefont {Chen}\ \emph {et~al.}(2021)\citenamefont {Chen}, \citenamefont {Zuo}, \citenamefont {Ye}, \citenamefont {Li},\ and\ \citenamefont {Ong}}]{chen2021learning}%
  \BibitemOpen
  \bibfield  {author} {\bibinfo {author} {\bibfnamefont {C.}~\bibnamefont {Chen}}, \bibinfo {author} {\bibfnamefont {Y.}~\bibnamefont {Zuo}}, \bibinfo {author} {\bibfnamefont {W.}~\bibnamefont {Ye}}, \bibinfo {author} {\bibfnamefont {X.}~\bibnamefont {Li}},\ and\ \bibinfo {author} {\bibfnamefont {S.~P.}\ \bibnamefont {Ong}},\ }\href@noop {} {\bibfield  {journal} {\bibinfo  {journal} {Nature Computational Science}\ }\textbf {\bibinfo {volume} {1}},\ \bibinfo {pages} {46} (\bibinfo {year} {2021})}\BibitemShut {NoStop}%
\bibitem [{\citenamefont {Chen}\ and\ \citenamefont {Ong}(2022)}]{chen2022universal}%
  \BibitemOpen
  \bibfield  {author} {\bibinfo {author} {\bibfnamefont {C.}~\bibnamefont {Chen}}\ and\ \bibinfo {author} {\bibfnamefont {S.~P.}\ \bibnamefont {Ong}},\ }\href@noop {} {\bibfield  {journal} {\bibinfo  {journal} {Nature Computational Science}\ }\textbf {\bibinfo {volume} {2}},\ \bibinfo {pages} {718} (\bibinfo {year} {2022})}\BibitemShut {NoStop}%
\bibitem [{\citenamefont {Thompson}\ \emph {et~al.}(2015)\citenamefont {Thompson}, \citenamefont {Swiler}, \citenamefont {Trott}, \citenamefont {Foiles},\ and\ \citenamefont {Tucker}}]{thompson2015spectral}%
  \BibitemOpen
  \bibfield  {author} {\bibinfo {author} {\bibfnamefont {A.~P.}\ \bibnamefont {Thompson}}, \bibinfo {author} {\bibfnamefont {L.~P.}\ \bibnamefont {Swiler}}, \bibinfo {author} {\bibfnamefont {C.~R.}\ \bibnamefont {Trott}}, \bibinfo {author} {\bibfnamefont {S.~M.}\ \bibnamefont {Foiles}},\ and\ \bibinfo {author} {\bibfnamefont {G.~J.}\ \bibnamefont {Tucker}},\ }\href@noop {} {\bibfield  {journal} {\bibinfo  {journal} {Journal of Computational Physics}\ }\textbf {\bibinfo {volume} {285}},\ \bibinfo {pages} {316} (\bibinfo {year} {2015})}\BibitemShut {NoStop}%
\bibitem [{\citenamefont {Kohn}(1996)}]{kohn1996density}%
  \BibitemOpen
  \bibfield  {author} {\bibinfo {author} {\bibfnamefont {W.}~\bibnamefont {Kohn}},\ }\href@noop {} {\bibfield  {journal} {\bibinfo  {journal} {Physical Review Letters}\ }\textbf {\bibinfo {volume} {76}},\ \bibinfo {pages} {3168} (\bibinfo {year} {1996})}\BibitemShut {NoStop}%
\bibitem [{\citenamefont {Prodan}\ and\ \citenamefont {Kohn}(2005)}]{prodan2005nearsightedness}%
  \BibitemOpen
  \bibfield  {author} {\bibinfo {author} {\bibfnamefont {E.}~\bibnamefont {Prodan}}\ and\ \bibinfo {author} {\bibfnamefont {W.}~\bibnamefont {Kohn}},\ }\href@noop {} {\bibfield  {journal} {\bibinfo  {journal} {Proceedings of the National Academy of Sciences}\ }\textbf {\bibinfo {volume} {102}},\ \bibinfo {pages} {11635} (\bibinfo {year} {2005})}\BibitemShut {NoStop}%
\bibitem [{\citenamefont {Zeng}\ \emph {et~al.}(2023)\citenamefont {Zeng}, \citenamefont {Zhang}, \citenamefont {Lu}, \citenamefont {Mo}, \citenamefont {Li}, \citenamefont {Chen}, \citenamefont {Rynik}, \citenamefont {Huang}, \citenamefont {Li}, \citenamefont {Shi} \emph {et~al.}}]{zeng2023deepmd}%
  \BibitemOpen
  \bibfield  {author} {\bibinfo {author} {\bibfnamefont {J.}~\bibnamefont {Zeng}}, \bibinfo {author} {\bibfnamefont {D.}~\bibnamefont {Zhang}}, \bibinfo {author} {\bibfnamefont {D.}~\bibnamefont {Lu}}, \bibinfo {author} {\bibfnamefont {P.}~\bibnamefont {Mo}}, \bibinfo {author} {\bibfnamefont {Z.}~\bibnamefont {Li}}, \bibinfo {author} {\bibfnamefont {Y.}~\bibnamefont {Chen}}, \bibinfo {author} {\bibfnamefont {M.}~\bibnamefont {Rynik}}, \bibinfo {author} {\bibfnamefont {L.}~\bibnamefont {Huang}}, \bibinfo {author} {\bibfnamefont {Z.}~\bibnamefont {Li}}, \bibinfo {author} {\bibfnamefont {S.}~\bibnamefont {Shi}}, \emph {et~al.},\ }\href@noop {} {\bibfield  {journal} {\bibinfo  {journal} {The Journal of Chemical Physics}\ }\textbf {\bibinfo {volume} {159}} (\bibinfo {year} {2023})}\BibitemShut {NoStop}%
\bibitem [{\citenamefont {Ellis}\ \emph {et~al.}(2021{\natexlab{a}})\citenamefont {Ellis}, \citenamefont {Fiedler}, \citenamefont {Popoola}, \citenamefont {Modine}, \citenamefont {Stephens}, \citenamefont {Thompson}, \citenamefont {Cangi},\ and\ \citenamefont {Rajamanickam}}]{MalaGit}%
  \BibitemOpen
  \bibfield  {author} {\bibinfo {author} {\bibfnamefont {J.~A.}\ \bibnamefont {Ellis}}, \bibinfo {author} {\bibfnamefont {L.}~\bibnamefont {Fiedler}}, \bibinfo {author} {\bibfnamefont {G.~A.}\ \bibnamefont {Popoola}}, \bibinfo {author} {\bibfnamefont {N.~A.}\ \bibnamefont {Modine}}, \bibinfo {author} {\bibfnamefont {J.~A.}\ \bibnamefont {Stephens}}, \bibinfo {author} {\bibfnamefont {A.~P.}\ \bibnamefont {Thompson}}, \bibinfo {author} {\bibfnamefont {A.}~\bibnamefont {Cangi}},\ and\ \bibinfo {author} {\bibfnamefont {S.}~\bibnamefont {Rajamanickam}},\ }\href@noop {} {\bibinfo {title} {mala-project}},\ \bibinfo {howpublished} {\url{https://github.com/mala-project/mala}} (\bibinfo {year} {2021}{\natexlab{a}})\BibitemShut {NoStop}%
\bibitem [{\citenamefont {Ellis}\ \emph {et~al.}(2021{\natexlab{b}})\citenamefont {Ellis}, \citenamefont {Fiedler}, \citenamefont {Popoola}, \citenamefont {Modine}, \citenamefont {Stephens}, \citenamefont {Thompson}, \citenamefont {Cangi},\ and\ \citenamefont {Rajamanickam}}]{ellis2021accelerating}%
  \BibitemOpen
  \bibfield  {author} {\bibinfo {author} {\bibfnamefont {J.~A.}\ \bibnamefont {Ellis}}, \bibinfo {author} {\bibfnamefont {L.}~\bibnamefont {Fiedler}}, \bibinfo {author} {\bibfnamefont {G.~A.}\ \bibnamefont {Popoola}}, \bibinfo {author} {\bibfnamefont {N.~A.}\ \bibnamefont {Modine}}, \bibinfo {author} {\bibfnamefont {J.~A.}\ \bibnamefont {Stephens}}, \bibinfo {author} {\bibfnamefont {A.~P.}\ \bibnamefont {Thompson}}, \bibinfo {author} {\bibfnamefont {A.}~\bibnamefont {Cangi}},\ and\ \bibinfo {author} {\bibfnamefont {S.}~\bibnamefont {Rajamanickam}},\ }\href@noop {} {\bibfield  {journal} {\bibinfo  {journal} {Physical Review B}\ }\textbf {\bibinfo {volume} {104}},\ \bibinfo {pages} {035120} (\bibinfo {year} {2021}{\natexlab{b}})}\BibitemShut {NoStop}%
\bibitem [{\citenamefont {Ashcroft}\ and\ \citenamefont {Mermin}(2022)}]{ashcroft2022solid}%
  \BibitemOpen
  \bibfield  {author} {\bibinfo {author} {\bibfnamefont {N.~W.}\ \bibnamefont {Ashcroft}}\ and\ \bibinfo {author} {\bibfnamefont {N.~D.}\ \bibnamefont {Mermin}},\ }\href@noop {} {\emph {\bibinfo {title} {Solid state physics}}}\ (\bibinfo  {publisher} {Cengage Learning},\ \bibinfo {year} {2022})\BibitemShut {NoStop}%
\bibitem [{\citenamefont {Barnard}\ \emph {et~al.}(2023)\citenamefont {Barnard}, \citenamefont {Tseng}, \citenamefont {Darby}, \citenamefont {Bartók}, \citenamefont {Broo},\ and\ \citenamefont {Sosso}}]{Barnard_RSC2023}%
  \BibitemOpen
  \bibfield  {author} {\bibinfo {author} {\bibfnamefont {T.}~\bibnamefont {Barnard}}, \bibinfo {author} {\bibfnamefont {S.}~\bibnamefont {Tseng}}, \bibinfo {author} {\bibfnamefont {J.~P.}\ \bibnamefont {Darby}}, \bibinfo {author} {\bibfnamefont {A.~P.}\ \bibnamefont {Bartók}}, \bibinfo {author} {\bibfnamefont {A.}~\bibnamefont {Broo}},\ and\ \bibinfo {author} {\bibfnamefont {G.~C.}\ \bibnamefont {Sosso}},\ }\href {https://doi.org/10.1039/D2ME00149G} {\bibfield  {journal} {\bibinfo  {journal} {Mol. Syst. Des. Eng.}\ }\textbf {\bibinfo {volume} {8}},\ \bibinfo {pages} {300} (\bibinfo {year} {2023})}\BibitemShut {NoStop}%
\bibitem [{\citenamefont {Hamer}\ and\ \citenamefont {Dupont}(2021)}]{Hamer_Dupont2021JMRL}%
  \BibitemOpen
  \bibfield  {author} {\bibinfo {author} {\bibfnamefont {V.}~\bibnamefont {Hamer}}\ and\ \bibinfo {author} {\bibfnamefont {P.}~\bibnamefont {Dupont}},\ }\href@noop {} {\bibfield  {journal} {\bibinfo  {journal} {J. Mach. Learn. Res.}\ }\textbf {\bibinfo {volume} {22}} (\bibinfo {year} {2021})}\BibitemShut {NoStop}%
\bibitem [{\citenamefont {Guyon}\ and\ \citenamefont {Elisseeff}(2003)}]{Guyon_Elisseeff2003JMRL}%
  \BibitemOpen
  \bibfield  {author} {\bibinfo {author} {\bibfnamefont {I.}~\bibnamefont {Guyon}}\ and\ \bibinfo {author} {\bibfnamefont {A.}~\bibnamefont {Elisseeff}},\ }\href@noop {} {\bibfield  {journal} {\bibinfo  {journal} {J. Mach. Learn. Res.}\ }\textbf {\bibinfo {volume} {3}},\ \bibinfo {pages} {1157–1182} (\bibinfo {year} {2003})}\BibitemShut {NoStop}%
\bibitem [{\citenamefont {Bishop}(2006)}]{bishop2006pattern}%
  \BibitemOpen
  \bibfield  {author} {\bibinfo {author} {\bibfnamefont {C.~M.}\ \bibnamefont {Bishop}},\ }\href@noop {} {\bibfield  {journal} {\bibinfo  {journal} {Machine learning}\ }\textbf {\bibinfo {volume} {128}} (\bibinfo {year} {2006})}\BibitemShut {NoStop}%
\bibitem [{\citenamefont {Yadav}\ \emph {et~al.}(2021)\citenamefont {Yadav}, \citenamefont {Pathrudkar},\ and\ \citenamefont {Ghosh}}]{yadav2021interpretable}%
  \BibitemOpen
  \bibfield  {author} {\bibinfo {author} {\bibfnamefont {U.}~\bibnamefont {Yadav}}, \bibinfo {author} {\bibfnamefont {S.}~\bibnamefont {Pathrudkar}},\ and\ \bibinfo {author} {\bibfnamefont {S.}~\bibnamefont {Ghosh}},\ }\href@noop {} {\bibfield  {journal} {\bibinfo  {journal} {Physical Review B}\ }\textbf {\bibinfo {volume} {103}},\ \bibinfo {pages} {035407} (\bibinfo {year} {2021})}\BibitemShut {NoStop}%
\bibitem [{\citenamefont {Gastegger}\ \emph {et~al.}(2018)\citenamefont {Gastegger}, \citenamefont {Schwiedrzik}, \citenamefont {Bittermann}, \citenamefont {Berzsenyi},\ and\ \citenamefont {Marquetand}}]{gastegger2018wacsf}%
  \BibitemOpen
  \bibfield  {author} {\bibinfo {author} {\bibfnamefont {M.}~\bibnamefont {Gastegger}}, \bibinfo {author} {\bibfnamefont {L.}~\bibnamefont {Schwiedrzik}}, \bibinfo {author} {\bibfnamefont {M.}~\bibnamefont {Bittermann}}, \bibinfo {author} {\bibfnamefont {F.}~\bibnamefont {Berzsenyi}},\ and\ \bibinfo {author} {\bibfnamefont {P.}~\bibnamefont {Marquetand}},\ }\href@noop {} {\bibfield  {journal} {\bibinfo  {journal} {The Journal of chemical physics}\ }\textbf {\bibinfo {volume} {148}} (\bibinfo {year} {2018})}\BibitemShut {NoStop}%
\bibitem [{\citenamefont {Imbalzano}\ \emph {et~al.}(2018)\citenamefont {Imbalzano}, \citenamefont {Anelli}, \citenamefont {Giofr{\'e}}, \citenamefont {Klees}, \citenamefont {Behler},\ and\ \citenamefont {Ceriotti}}]{imbalzano2018automatic}%
  \BibitemOpen
  \bibfield  {author} {\bibinfo {author} {\bibfnamefont {G.}~\bibnamefont {Imbalzano}}, \bibinfo {author} {\bibfnamefont {A.}~\bibnamefont {Anelli}}, \bibinfo {author} {\bibfnamefont {D.}~\bibnamefont {Giofr{\'e}}}, \bibinfo {author} {\bibfnamefont {S.}~\bibnamefont {Klees}}, \bibinfo {author} {\bibfnamefont {J.}~\bibnamefont {Behler}},\ and\ \bibinfo {author} {\bibfnamefont {M.}~\bibnamefont {Ceriotti}},\ }\href@noop {} {\bibfield  {journal} {\bibinfo  {journal} {The Journal of chemical physics}\ }\textbf {\bibinfo {volume} {148}} (\bibinfo {year} {2018})}\BibitemShut {NoStop}%
\bibitem [{\citenamefont {Bart{\'o}k}\ \emph {et~al.}(2013)\citenamefont {Bart{\'o}k}, \citenamefont {Kondor},\ and\ \citenamefont {Cs{\'a}nyi}}]{bartok2013representing}%
  \BibitemOpen
  \bibfield  {author} {\bibinfo {author} {\bibfnamefont {A.~P.}\ \bibnamefont {Bart{\'o}k}}, \bibinfo {author} {\bibfnamefont {R.}~\bibnamefont {Kondor}},\ and\ \bibinfo {author} {\bibfnamefont {G.}~\bibnamefont {Cs{\'a}nyi}},\ }\href@noop {} {\bibfield  {journal} {\bibinfo  {journal} {Physical Review B}\ }\textbf {\bibinfo {volume} {87}},\ \bibinfo {pages} {184115} (\bibinfo {year} {2013})}\BibitemShut {NoStop}%
\bibitem [{\citenamefont {Thomas}\ \emph {et~al.}(2018)\citenamefont {Thomas}, \citenamefont {Smidt}, \citenamefont {Kearnes}, \citenamefont {Yang}, \citenamefont {Li}, \citenamefont {Kohlhoff},\ and\ \citenamefont {Riley}}]{thomas2018tensor}%
  \BibitemOpen
  \bibfield  {author} {\bibinfo {author} {\bibfnamefont {N.}~\bibnamefont {Thomas}}, \bibinfo {author} {\bibfnamefont {T.}~\bibnamefont {Smidt}}, \bibinfo {author} {\bibfnamefont {S.}~\bibnamefont {Kearnes}}, \bibinfo {author} {\bibfnamefont {L.}~\bibnamefont {Yang}}, \bibinfo {author} {\bibfnamefont {L.}~\bibnamefont {Li}}, \bibinfo {author} {\bibfnamefont {K.}~\bibnamefont {Kohlhoff}},\ and\ \bibinfo {author} {\bibfnamefont {P.}~\bibnamefont {Riley}},\ }\href@noop {} {\bibfield  {journal} {\bibinfo  {journal} {arXiv preprint arXiv:1802.08219}\ } (\bibinfo {year} {2018})}\BibitemShut {NoStop}%
\bibitem [{\citenamefont {Hinton}\ and\ \citenamefont {Van~Camp}(1993)}]{hinton1993keeping}%
  \BibitemOpen
  \bibfield  {author} {\bibinfo {author} {\bibfnamefont {G.~E.}\ \bibnamefont {Hinton}}\ and\ \bibinfo {author} {\bibfnamefont {D.}~\bibnamefont {Van~Camp}},\ }in\ \href@noop {} {\emph {\bibinfo {booktitle} {Proceedings of the sixth annual conference on Computational learning theory}}}\ (\bibinfo {year} {1993})\ pp.\ \bibinfo {pages} {5--13}\BibitemShut {NoStop}%
\bibitem [{\citenamefont {Graves}(2011)}]{graves2011practical}%
  \BibitemOpen
  \bibfield  {author} {\bibinfo {author} {\bibfnamefont {A.}~\bibnamefont {Graves}},\ }\href@noop {} {\bibfield  {journal} {\bibinfo  {journal} {Advances in neural information processing systems}\ }\textbf {\bibinfo {volume} {24}} (\bibinfo {year} {2011})}\BibitemShut {NoStop}%
\bibitem [{\citenamefont {Blundell}\ \emph {et~al.}(2015)\citenamefont {Blundell}, \citenamefont {Cornebise}, \citenamefont {Kavukcuoglu},\ and\ \citenamefont {Wierstra}}]{blundell2015weight}%
  \BibitemOpen
  \bibfield  {author} {\bibinfo {author} {\bibfnamefont {C.}~\bibnamefont {Blundell}}, \bibinfo {author} {\bibfnamefont {J.}~\bibnamefont {Cornebise}}, \bibinfo {author} {\bibfnamefont {K.}~\bibnamefont {Kavukcuoglu}},\ and\ \bibinfo {author} {\bibfnamefont {D.}~\bibnamefont {Wierstra}},\ }in\ \href@noop {} {\emph {\bibinfo {booktitle} {International conference on machine learning}}}\ (\bibinfo {organization} {Proceedings of Machine Learning Research},\ \bibinfo {year} {2015})\ pp.\ \bibinfo {pages} {1613--1622}\BibitemShut {NoStop}%
\bibitem [{\citenamefont {Thiagarajan}\ \emph {et~al.}(2021)\citenamefont {Thiagarajan}, \citenamefont {Khairnar},\ and\ \citenamefont {Ghosh}}]{thiagarajan2021explanation}%
  \BibitemOpen
  \bibfield  {author} {\bibinfo {author} {\bibfnamefont {P.}~\bibnamefont {Thiagarajan}}, \bibinfo {author} {\bibfnamefont {P.}~\bibnamefont {Khairnar}},\ and\ \bibinfo {author} {\bibfnamefont {S.}~\bibnamefont {Ghosh}},\ }\href@noop {} {\bibfield  {journal} {\bibinfo  {journal} {IEEE Transactions on Medical Imaging}\ }\textbf {\bibinfo {volume} {41}},\ \bibinfo {pages} {815} (\bibinfo {year} {2021})}\BibitemShut {NoStop}%
\bibitem [{\citenamefont {Thiagarajan}\ and\ \citenamefont {Ghosh}(2025)}]{thiagarajan2025jensen}%
  \BibitemOpen
  \bibfield  {author} {\bibinfo {author} {\bibfnamefont {P.}~\bibnamefont {Thiagarajan}}\ and\ \bibinfo {author} {\bibfnamefont {S.}~\bibnamefont {Ghosh}},\ }\href@noop {} {\bibfield  {journal} {\bibinfo  {journal} {Neurocomputing}\ }\textbf {\bibinfo {volume} {618}},\ \bibinfo {pages} {129115} (\bibinfo {year} {2025})}\BibitemShut {NoStop}%
\bibitem [{\citenamefont {Kendall}\ and\ \citenamefont {Gal}(2017)}]{kendall2017uncertainties}%
  \BibitemOpen
  \bibfield  {author} {\bibinfo {author} {\bibfnamefont {A.}~\bibnamefont {Kendall}}\ and\ \bibinfo {author} {\bibfnamefont {Y.}~\bibnamefont {Gal}},\ }\href@noop {} {\bibfield  {journal} {\bibinfo  {journal} {Advances in neural information processing systems}\ }\textbf {\bibinfo {volume} {30}} (\bibinfo {year} {2017})}\BibitemShut {NoStop}%
\bibitem [{\citenamefont {Busk}\ \emph {et~al.}(2021)\citenamefont {Busk}, \citenamefont {J{\o}rgensen}, \citenamefont {Bhowmik}, \citenamefont {Schmidt}, \citenamefont {Winther},\ and\ \citenamefont {Vegge}}]{busk2021calibrated}%
  \BibitemOpen
  \bibfield  {author} {\bibinfo {author} {\bibfnamefont {J.}~\bibnamefont {Busk}}, \bibinfo {author} {\bibfnamefont {P.~B.}\ \bibnamefont {J{\o}rgensen}}, \bibinfo {author} {\bibfnamefont {A.}~\bibnamefont {Bhowmik}}, \bibinfo {author} {\bibfnamefont {M.~N.}\ \bibnamefont {Schmidt}}, \bibinfo {author} {\bibfnamefont {O.}~\bibnamefont {Winther}},\ and\ \bibinfo {author} {\bibfnamefont {T.}~\bibnamefont {Vegge}},\ }\href@noop {} {\bibfield  {journal} {\bibinfo  {journal} {Machine Learning: Science and Technology}\ }\textbf {\bibinfo {volume} {3}},\ \bibinfo {pages} {015012} (\bibinfo {year} {2021})}\BibitemShut {NoStop}%
\bibitem [{\citenamefont {Gruich}\ \emph {et~al.}(2023)\citenamefont {Gruich}, \citenamefont {Madhavan}, \citenamefont {Wang},\ and\ \citenamefont {Goldsmith}}]{gruich2023clarifying}%
  \BibitemOpen
  \bibfield  {author} {\bibinfo {author} {\bibfnamefont {C.~J.}\ \bibnamefont {Gruich}}, \bibinfo {author} {\bibfnamefont {V.}~\bibnamefont {Madhavan}}, \bibinfo {author} {\bibfnamefont {Y.}~\bibnamefont {Wang}},\ and\ \bibinfo {author} {\bibfnamefont {B.~R.}\ \bibnamefont {Goldsmith}},\ }\href@noop {} {\bibfield  {journal} {\bibinfo  {journal} {Machine Learning: Science and Technology}\ }\textbf {\bibinfo {volume} {4}},\ \bibinfo {pages} {025019} (\bibinfo {year} {2023})}\BibitemShut {NoStop}%
\bibitem [{\citenamefont {Gabbrielli}\ \emph {et~al.}(2012)\citenamefont {Gabbrielli}, \citenamefont {Jiao},\ and\ \citenamefont {Torquato}}]{gabbrielli2012families}%
  \BibitemOpen
  \bibfield  {author} {\bibinfo {author} {\bibfnamefont {R.}~\bibnamefont {Gabbrielli}}, \bibinfo {author} {\bibfnamefont {Y.}~\bibnamefont {Jiao}},\ and\ \bibinfo {author} {\bibfnamefont {S.}~\bibnamefont {Torquato}},\ }\href@noop {} {\bibfield  {journal} {\bibinfo  {journal} {Physical Review E—Statistical, Nonlinear, and Soft Matter Physics}\ }\textbf {\bibinfo {volume} {86}},\ \bibinfo {pages} {041141} (\bibinfo {year} {2012})}\BibitemShut {NoStop}%
\bibitem [{\citenamefont {Alred}\ \emph {et~al.}(2018)\citenamefont {Alred}, \citenamefont {Bets}, \citenamefont {Xie},\ and\ \citenamefont {Yakobson}}]{alred2018machine}%
  \BibitemOpen
  \bibfield  {author} {\bibinfo {author} {\bibfnamefont {J.~M.}\ \bibnamefont {Alred}}, \bibinfo {author} {\bibfnamefont {K.~V.}\ \bibnamefont {Bets}}, \bibinfo {author} {\bibfnamefont {Y.}~\bibnamefont {Xie}},\ and\ \bibinfo {author} {\bibfnamefont {B.~I.}\ \bibnamefont {Yakobson}},\ }\href@noop {} {\bibfield  {journal} {\bibinfo  {journal} {Composites Science and Technology}\ }\textbf {\bibinfo {volume} {166}},\ \bibinfo {pages} {3} (\bibinfo {year} {2018})}\BibitemShut {NoStop}%
\bibitem [{\citenamefont {Xu}\ \emph {et~al.}(2020)\citenamefont {Xu}, \citenamefont {Sharma},\ and\ \citenamefont {Suryanarayana}}]{xu2020m}%
  \BibitemOpen
  \bibfield  {author} {\bibinfo {author} {\bibfnamefont {Q.}~\bibnamefont {Xu}}, \bibinfo {author} {\bibfnamefont {A.}~\bibnamefont {Sharma}},\ and\ \bibinfo {author} {\bibfnamefont {P.}~\bibnamefont {Suryanarayana}},\ }\href@noop {} {\bibfield  {journal} {\bibinfo  {journal} {SoftwareX}\ }\textbf {\bibinfo {volume} {11}},\ \bibinfo {pages} {100423} (\bibinfo {year} {2020})}\BibitemShut {NoStop}%
\bibitem [{\citenamefont {Zhang}\ \emph {et~al.}(2023{\natexlab{c}})\citenamefont {Zhang}, \citenamefont {Jing}, \citenamefont {Kumar},\ and\ \citenamefont {Suryanarayana}}]{zhang2023version}%
  \BibitemOpen
  \bibfield  {author} {\bibinfo {author} {\bibfnamefont {B.}~\bibnamefont {Zhang}}, \bibinfo {author} {\bibfnamefont {X.}~\bibnamefont {Jing}}, \bibinfo {author} {\bibfnamefont {S.}~\bibnamefont {Kumar}},\ and\ \bibinfo {author} {\bibfnamefont {P.}~\bibnamefont {Suryanarayana}},\ }\href@noop {} {\bibfield  {journal} {\bibinfo  {journal} {SoftwareX}\ }\textbf {\bibinfo {volume} {21}},\ \bibinfo {pages} {101295} (\bibinfo {year} {2023}{\natexlab{c}})}\BibitemShut {NoStop}%
\bibitem [{\citenamefont {Harris}(1985)}]{harris1985simplified}%
  \BibitemOpen
  \bibfield  {author} {\bibinfo {author} {\bibfnamefont {J.}~\bibnamefont {Harris}},\ }\href@noop {} {\bibfield  {journal} {\bibinfo  {journal} {Physical Review B}\ }\textbf {\bibinfo {volume} {31}},\ \bibinfo {pages} {1770} (\bibinfo {year} {1985})}\BibitemShut {NoStop}%
\bibitem [{\citenamefont {Foulkes}\ and\ \citenamefont {Haydock}(1989)}]{foulkes1989tight}%
  \BibitemOpen
  \bibfield  {author} {\bibinfo {author} {\bibfnamefont {W.~M.~C.}\ \bibnamefont {Foulkes}}\ and\ \bibinfo {author} {\bibfnamefont {R.}~\bibnamefont {Haydock}},\ }\href@noop {} {\bibfield  {journal} {\bibinfo  {journal} {Physical review B}\ }\textbf {\bibinfo {volume} {39}},\ \bibinfo {pages} {12520} (\bibinfo {year} {1989})}\BibitemShut {NoStop}%
\bibitem [{\citenamefont {Foulkes}(1993)}]{foulkes1993accuracy}%
  \BibitemOpen
  \bibfield  {author} {\bibinfo {author} {\bibfnamefont {W.}~\bibnamefont {Foulkes}},\ }\href@noop {} {\bibfield  {journal} {\bibinfo  {journal} {Physical Review B}\ }\textbf {\bibinfo {volume} {48}},\ \bibinfo {pages} {14216} (\bibinfo {year} {1993})}\BibitemShut {NoStop}%
\bibitem [{\citenamefont {Momma}\ and\ \citenamefont {Izumi}(2008)}]{momma2008vesta}%
  \BibitemOpen
  \bibfield  {author} {\bibinfo {author} {\bibfnamefont {K.}~\bibnamefont {Momma}}\ and\ \bibinfo {author} {\bibfnamefont {F.}~\bibnamefont {Izumi}},\ }\href@noop {} {\bibfield  {journal} {\bibinfo  {journal} {Journal of Applied crystallography}\ }\textbf {\bibinfo {volume} {41}},\ \bibinfo {pages} {653} (\bibinfo {year} {2008})}\BibitemShut {NoStop}%
\bibitem [{\citenamefont {Jacob}\ \emph {et~al.}(2007)\citenamefont {Jacob}, \citenamefont {Raj},\ and\ \citenamefont {Rannesh}}]{jacob2007vegard}%
  \BibitemOpen
  \bibfield  {author} {\bibinfo {author} {\bibfnamefont {K.}~\bibnamefont {Jacob}}, \bibinfo {author} {\bibfnamefont {S.}~\bibnamefont {Raj}},\ and\ \bibinfo {author} {\bibfnamefont {L.}~\bibnamefont {Rannesh}},\ }\href@noop {} {\bibfield  {journal} {\bibinfo  {journal} {International Journal of Materials Research}\ }\textbf {\bibinfo {volume} {98}},\ \bibinfo {pages} {776} (\bibinfo {year} {2007})}\BibitemShut {NoStop}%
\bibitem [{\citenamefont {Barrett}\ and\ \citenamefont {Massalski}(1966)}]{Barrett1966}%
  \BibitemOpen
  \bibfield  {author} {\bibinfo {author} {\bibfnamefont {C.}~\bibnamefont {Barrett}}\ and\ \bibinfo {author} {\bibfnamefont {T.}~\bibnamefont {Massalski}},\ }\href@noop {} {\emph {\bibinfo {title} {Structure of Metals}}}\ (\bibinfo  {publisher} {McGraw-Hill},\ \bibinfo {year} {1966})\ pp.\ \bibinfo {pages} {Appendix VII, pg. 552}\BibitemShut {NoStop}%
\bibitem [{\citenamefont {Hirel}(2015)}]{hirel2015atomsk}%
  \BibitemOpen
  \bibfield  {author} {\bibinfo {author} {\bibfnamefont {P.}~\bibnamefont {Hirel}},\ }\href@noop {} {\bibfield  {journal} {\bibinfo  {journal} {Computer Physics Communications}\ }\textbf {\bibinfo {volume} {197}},\ \bibinfo {pages} {212} (\bibinfo {year} {2015})}\BibitemShut {NoStop}%
\bibitem [{\citenamefont {Read}\ and\ \citenamefont {Needs}(1989)}]{read1989tests}%
  \BibitemOpen
  \bibfield  {author} {\bibinfo {author} {\bibfnamefont {A.}~\bibnamefont {Read}}\ and\ \bibinfo {author} {\bibfnamefont {R.}~\bibnamefont {Needs}},\ }\href@noop {} {\bibfield  {journal} {\bibinfo  {journal} {Journal of Physics: Condensed Matter}\ }\textbf {\bibinfo {volume} {1}},\ \bibinfo {pages} {7565} (\bibinfo {year} {1989})}\BibitemShut {NoStop}%
\bibitem [{\citenamefont {Bellchambers}\ and\ \citenamefont {Manby}(2011)}]{bellchambers2011approximate}%
  \BibitemOpen
  \bibfield  {author} {\bibinfo {author} {\bibfnamefont {G.~D.}\ \bibnamefont {Bellchambers}}\ and\ \bibinfo {author} {\bibfnamefont {F.}~\bibnamefont {Manby}},\ }\href@noop {} {\bibfield  {journal} {\bibinfo  {journal} {The Journal of chemical physics}\ }\textbf {\bibinfo {volume} {135}} (\bibinfo {year} {2011})}\BibitemShut {NoStop}%
\end{thebibliography}%

\newpage
\clearpage

\begin{figure}[htbp]
    \centering
    \includegraphics[width=0.95\linewidth]{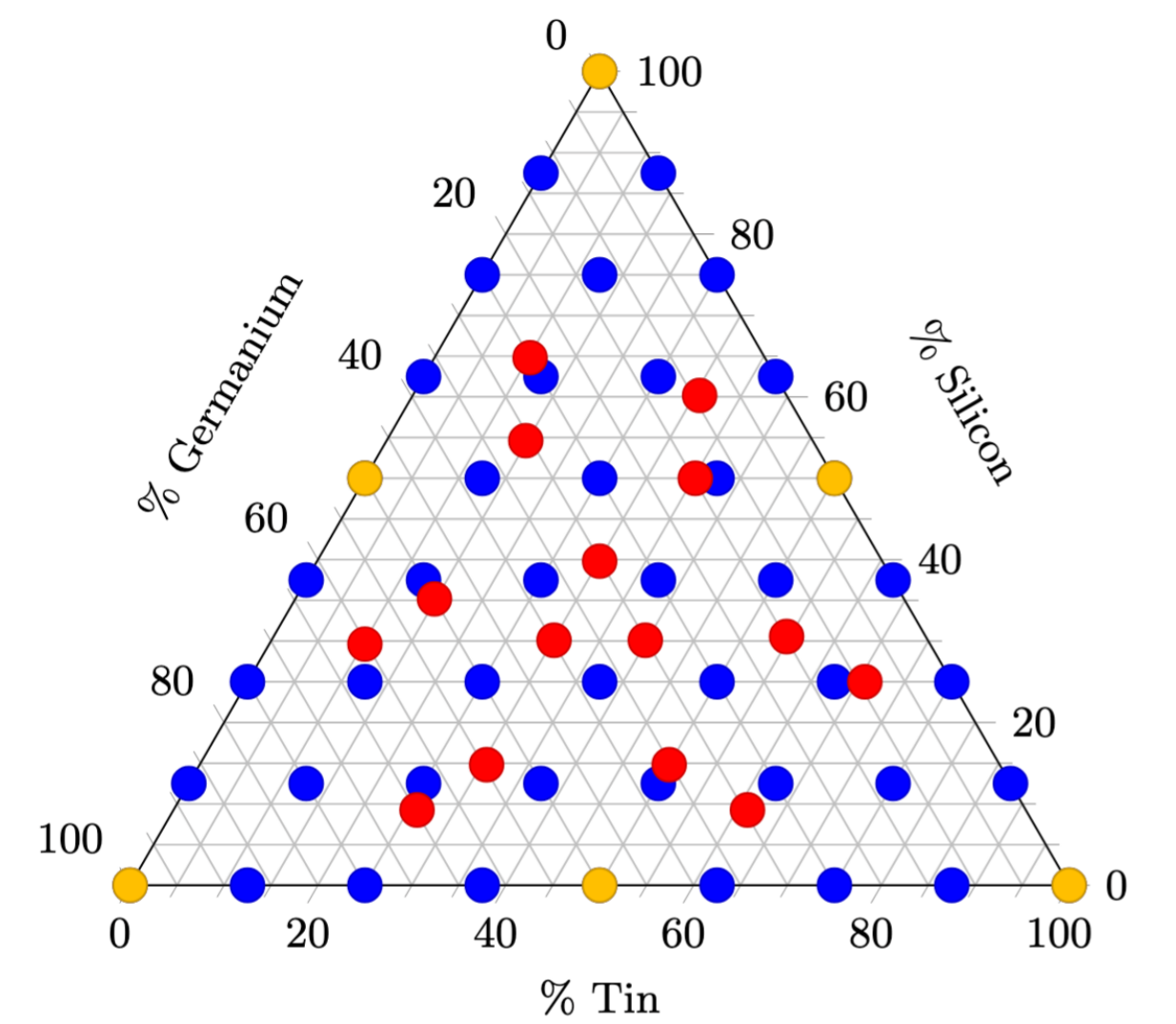}
    \caption{\textbf{Training and Testing compositions in the \ce{SiGeSn} data set.} Points in blue indicate the 64-atom data set. Note that, only a subset of compositions are used for training. Points in red indicate the 216-atom data set (not used in training). Points in yellow indicate compositions present in both the 64-atom and 216-atom data sets. See \textbf{Tables \ref{tab:SiGeSn_64atom_complist}} and \textbf{\ref{tab:SiGeSn_216atom_complist}} for the full list of values.
    }

    \label{fig:compSiGeSn}
\end{figure}

\begin{figure}[htbp]
    \centering
    \includegraphics[width=0.9\linewidth]{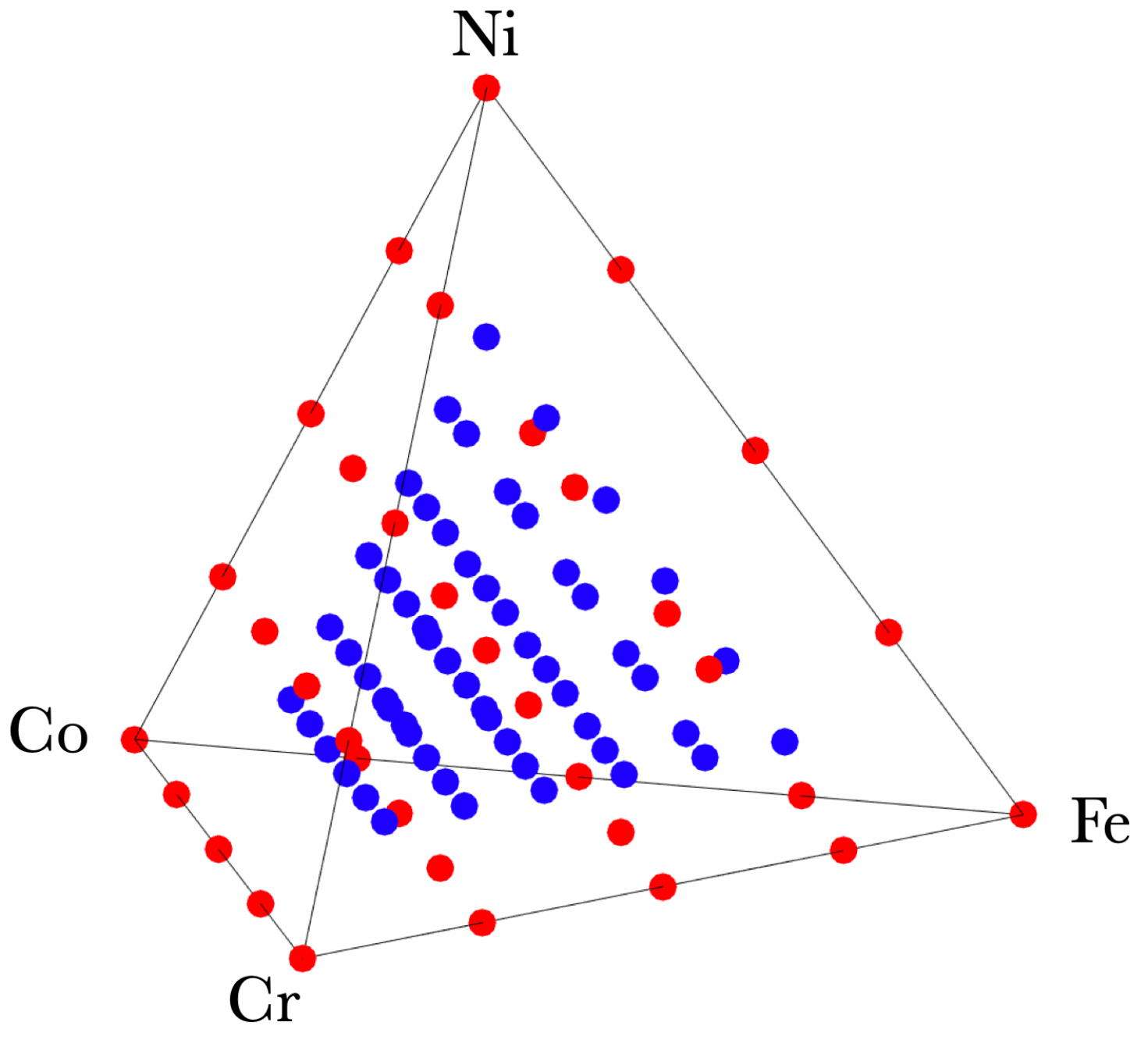}
    \caption{\textbf{Training and testing compositions in the 32-atom \ce{CrFeCoNi} data set}. Points in blue indicate the true quaternary compositions, while points in red indicate ternary, binary, and unary derivatives. Note that, only a subset of compositions are used for training. See \textbf{Table \ref{tab:CrFeCoNi_32atom_complist}} for the full list of values.}
    \label{fig:quatcomp}
\end{figure}

\begin{figure}[htbp]
    \centering
    \includegraphics[width=1\linewidth]{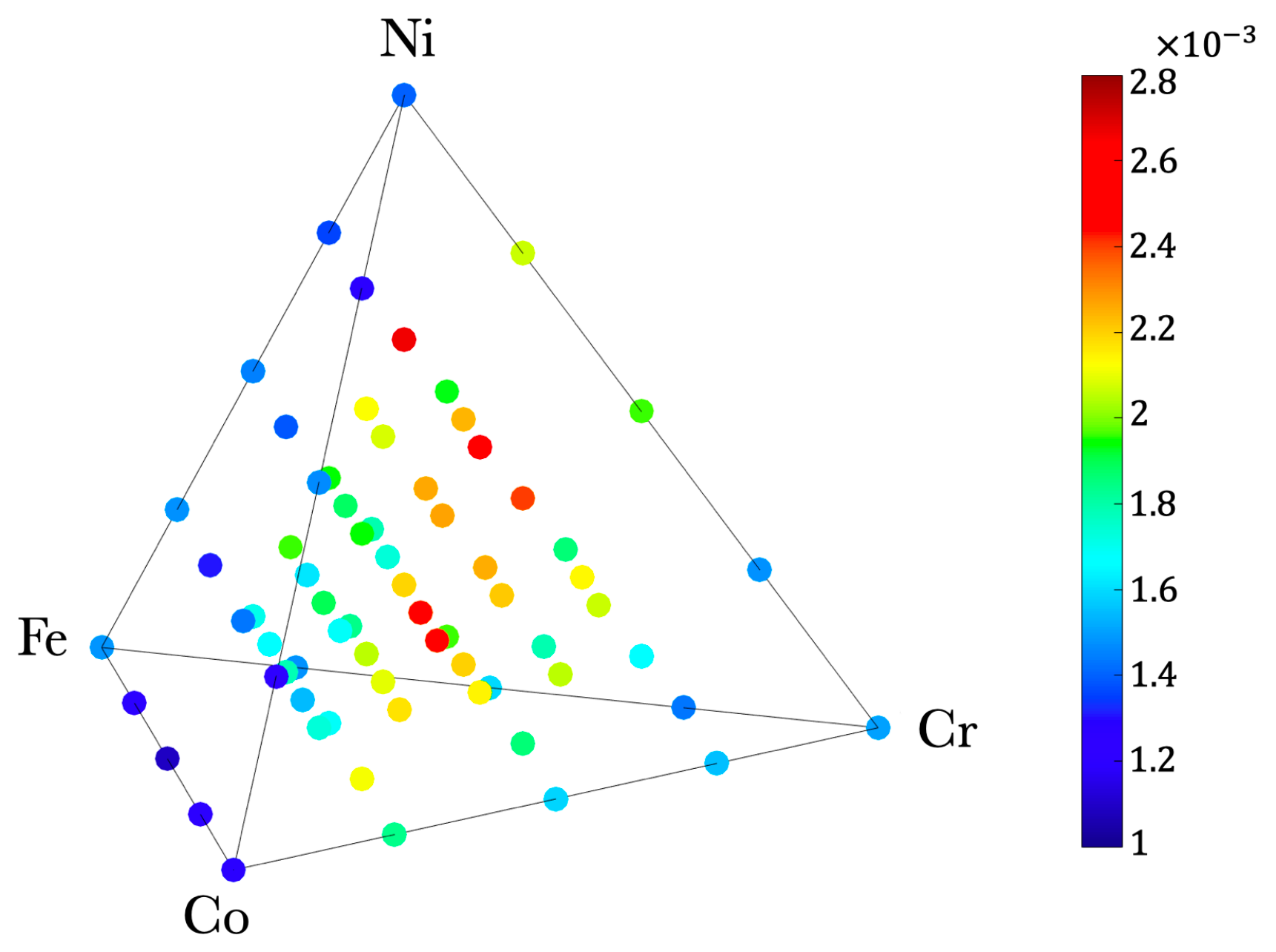}
    \caption{\textbf{Error in energy for the quaternary \ce{CrFeCoNi} system for the model trained with charge density $\rho$ instead of $\delta\rho$.} The plot shows the average error in energy at test compositions for the pristine 32-atom \ce{CrFeCoNi} data set, expressed in Hartree per atom (Ha/atom). The order of magnitude of the colorbar is $10^{-3}$.}
    \label{fig:Quat_EnergyErrors}
\end{figure}

\begin{figure*}[htbp]
    \centering
    \begin{minipage}{0.45\linewidth}
    \centering
    \begin{tikzpicture}
    \begin{ternaryaxis}[colorbar, colormap/jet,
     xmin=0,
     xmax=100,
     ymin=0,
     ymax=100,
     zmin=0,
     zmax=100, 
     xlabel=\% Silicon,
     ylabel=\% Germanium,
     zlabel=\% Tin,
     label style={sloped},
     minor tick num=3,
     grid=both,
     point meta min=0.0001,
     point meta max=0.0015,
     colorbar style={
        ytick distance=0.0002,
        width = 8pt,
        ticklabel style={
            /pgf/number format/precision=3,
            max space between ticks=20pt
        },
    },
    width=0.9\linewidth
    ]
        \addplot3+[only marks, 
        point meta=\thisrow{myvalue}, 
        mark size = 3pt,
         nodes near coords*={\tiny{}}, 
         visualization depends on={\thisrow{myvalue} \as \myvalue} 
         ] table {
    x       y       z       myvalue
    9.259   29.63   61.111  6.02E-04
    9.259   64.815  25.926  7.10E-04
    29.63   60.185  10.185  7.03E-04        
    64.815  25      10.185  5.81E-04
    60.185  9.259   30.556  6.32E-04
    30.093  30.093  39.815  1.34E-03            
    30.093  39.815  30.093  7.00E-04            
    25      9.259   65.741  8.76E-04
    39.815  30.093  30.093  1.31E-03            
    30.556  14.815  54.63   6.24E-04            
    50      14.815  35.185  5.14E-04
    54.63   30.556  14.815  5.35E-04
    0       50      50      9.68E-04                        
    50      0       50      7.95E-04
    50      50      0       3.77E-04
    100     0       0       1.19E-03                        
    0       100     0       6.85E-04                            
    0       0       100     5.30E-04                        
    35.185  50      14.815  7.33E-04            
    14.815  54.63   30.556  6.94E-04            
    14.815  35.185  50      7.73E-04               
    };
    \end{ternaryaxis}
    \end{tikzpicture}
    \end{minipage}%
    \hspace{0.25cm} 
    \begin{minipage}{0.45\linewidth}
    \centering
    \begin{tikzpicture}
    \begin{ternaryaxis}[colorbar, colormap/jet,
     xmin=0,
     xmax=100,
     ymin=0,
     ymax=100,
     zmin=0,
     zmax=100, 
     xlabel=\% Silicon,
     ylabel=\% Germanium,
     zlabel=\% Tin,
     label style={sloped},
     minor tick num=3,
     grid=both,
     point meta min=1.74E-02,
     point meta max=1.86E-02,
     colorbar style={
     width = 8pt,
        ticklabel style={
            /pgf/number format/precision=4,
            max space between ticks=30pt
        },
    },
    width=0.9\linewidth
    ]
        \addplot3+[only marks, 
        point meta=\thisrow{myvalue}, 
        mark size = 3pt,
         nodes near coords*={\tiny{}}, 
         visualization depends on={\thisrow{myvalue} \as \myvalue} 
         ] table {
    x       y       z       myvalue
    9.259   29.63   61.111  1.82E-02
    9.259   64.815  25.926  1.78E-02
    29.63   60.185  10.185  1.80E-02   
    64.815  25      10.185  1.81E-02
    60.185  9.259   30.556  1.83E-02
    30.093  30.093  39.815  1.80E-02    
    30.093  39.815  30.093  1.80E-02     
    25      9.259   65.741  1.81E-02
    39.815  30.093  30.093  1.80E-02       
    30.556  14.815  54.63   1.80E-02   
    50      14.815  35.185  1.79E-02
    54.63   30.556  14.815  1.79E-02
    0       50      50      1.79E-02               
    50      0       50      1.82E-02
    50      50      0       1.79E-02
    100     0       0       1.81E-02         
    0       100     0       1.77E-02            
    0       0       100     1.84E-02            
    35.185  50      14.815  1.83E-02
    14.815  54.63   30.556  1.80E-02      
    14.815  35.185  50      1.78E-02
    };
    \end{ternaryaxis}
    \end{tikzpicture}
    \end{minipage}
    
    \caption{\textbf{Error for ternary system beyond training compositions and training system size.} With a 64 atom ternary system, a limited number of test compositions is possible. We used a bigger 216-atom system to obtain intermediate test compositions, not possible with the 64 atom system. \textit{Left:} Average errors in energy at test compositions for the pristine 216-atom \ce{SiGeSn} data set, using the AL2 model. Units: Hartree/atom. Note that the order of magnitude of the colorbar is $10^{-3}$. \textit{Right:} Average error in density at test compositions for the pristine 216-atom \ce{SiGeSn} data set, in terms of relative L1, using the AL2 model. Note that the order of magnitude of the colorbar is $10^{-2}$.}
    \label{fig:errorSiGeSn_216atom}
\end{figure*}
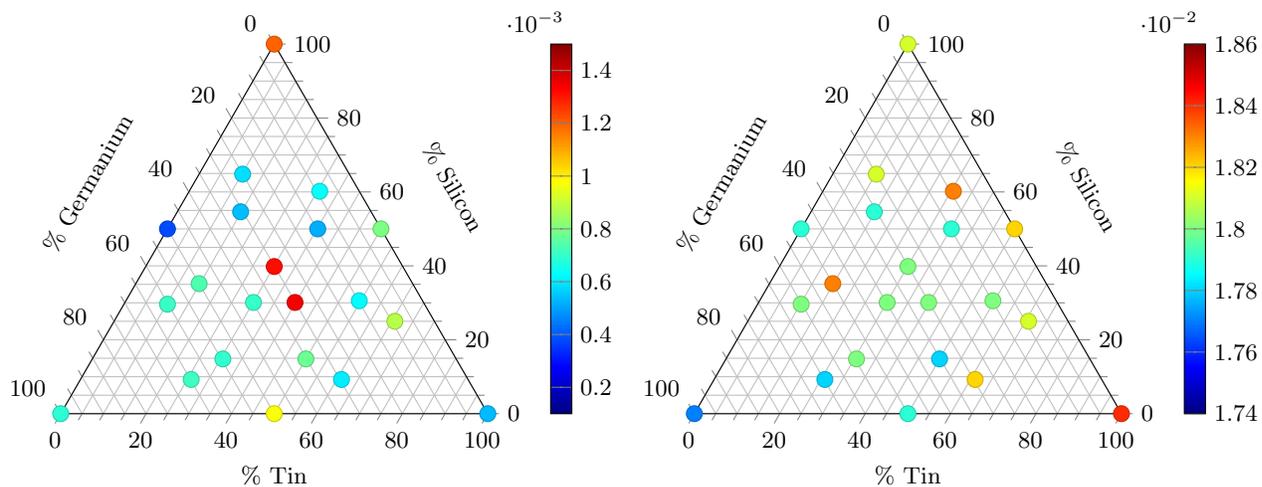

\begin{figure*}[htbp]
    \centering
    \includegraphics[width=0.95\linewidth]{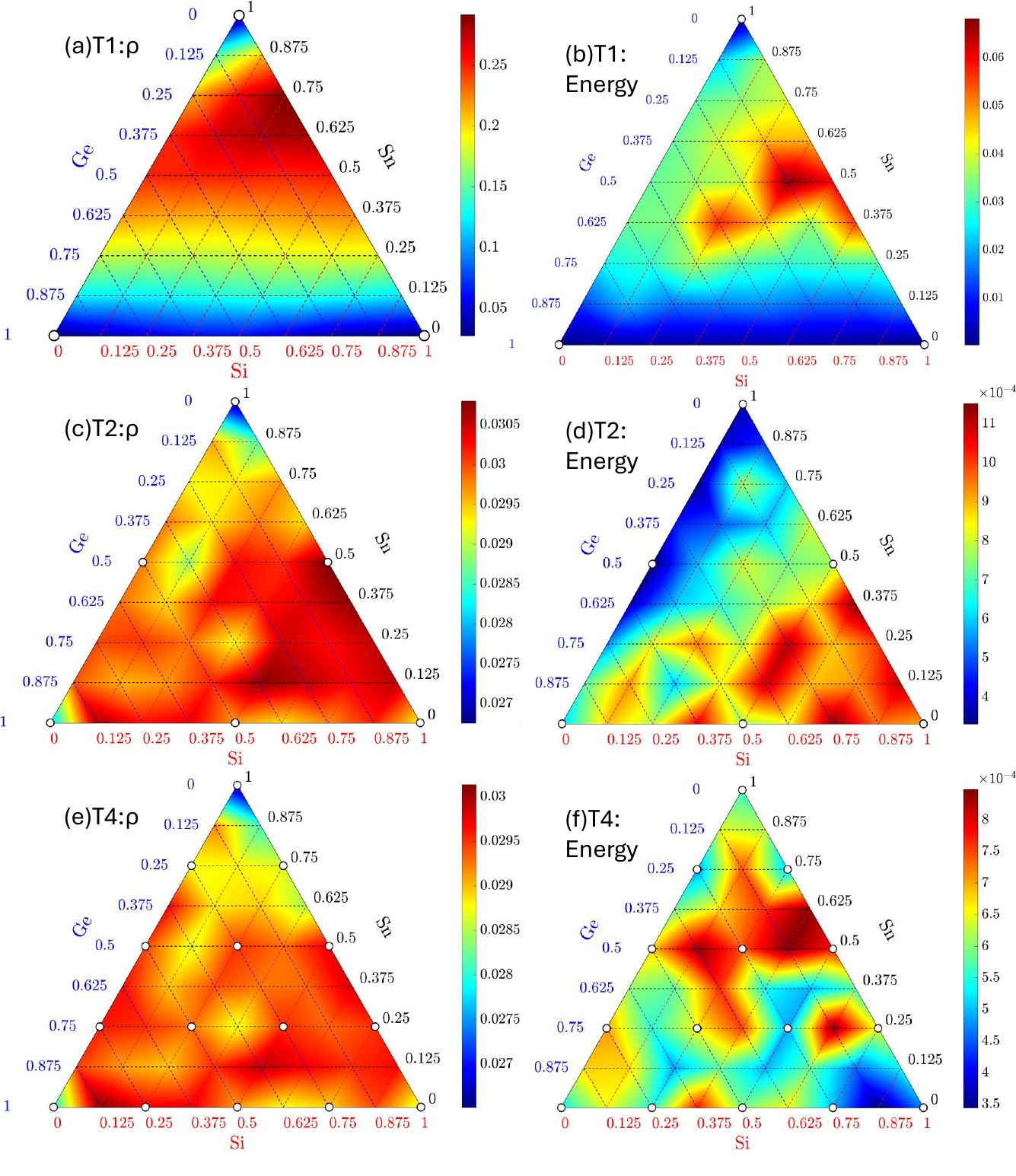}
    \caption{\textbf{Tessellation based approach to iteratively select training compositions to accurately predict across composition space of Ternary alloy.} (a) NRMSE across the composition space after 1st iteration of  Tessellation based iterative learning. The model is trained using only 3 pure compositions shown using white circles in the figure. This model is termed as T1. (b) Energy prediction error across the composition space after 1st iteration of Tessellation. This model, termed as T1, is trained only for 3 pure compositions. (c) NRMSE across the composition space after 2nd iteration of Tessellation. Three additional training points. This model is termed as T2. (d) Error in energy prediction across composition space. The unit of energy error is Ha/atom. The predicted energy is obtained from $\rho$ predictions from T2. The energy error is within chemical accuracy across the composition space. (e) NRMSE across the composition space after 4th iteration of Tessellation, resulting in nine additional training points from T2. This model is termed as T2. (f) Energy prediction error across the composition space after 4th iteration of Tessellation. This model, termed as T4, is trained only for 15 compositions.}
    \label{fig:Fig26_HEAML_Ternary_Collage_colorbar}
\end{figure*}

\begin{figure*}[htbp]
    \centering
    \includegraphics[width=0.95\linewidth]{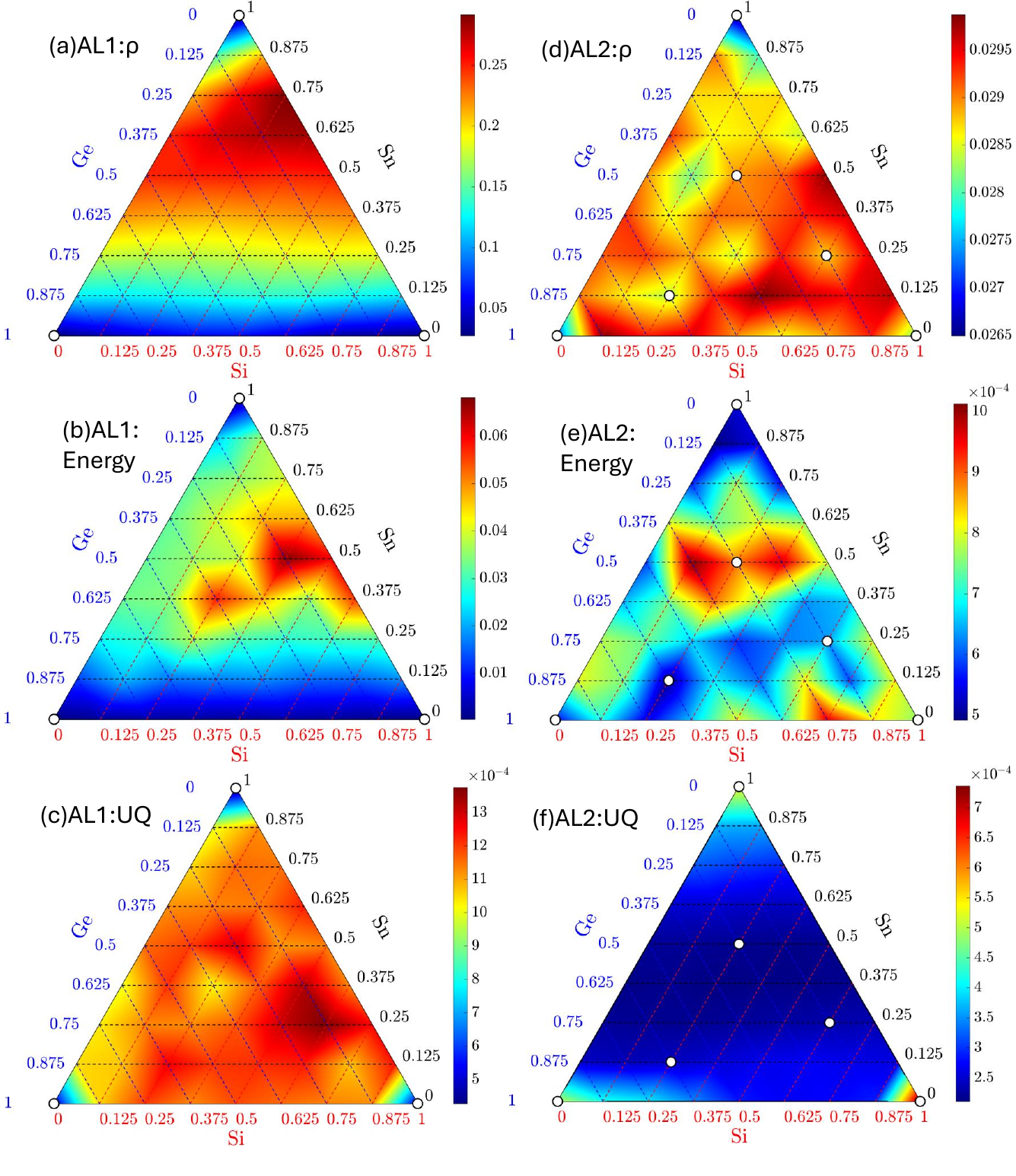}
    \caption{\textbf{Bayesian Active Learning to iteratively select training compositions to accurately predict across composition space of Ternary alloy.} (a) NRMSE across the composition space after 1st iteration of Active Learning, termed as AL1, trained using only 3 pure compositions shown using white circles. (b) Energy prediction error for model AL1 with 3 pure composition. (c) Epistemic Uncertainty in $\rho$ prediction across composition space after prediction with model AL1. Query points (additional training points) for the next iteration of Bayesian Active Learning are selected based on highest uncertainty regions shown in `f'. (d) NRMSE across the composition space after 2nd iteration of Active Learning. 3 additional training points are added as per the uncertainty contour in subfigure, `c'. This model is termed as AL2. We observe that the NRMSE is low and consistent across the composition space showing the effectiveness of query points selection through uncertainty. (e) Error in energy prediction across composition space. The unit of energy error is Ha/atom. The predicted energy is obtained from $\rho$ predictions from AL2. The energy error is within chemical accuracy across the composition space. (f) Epistemic Uncertainty in $\rho$ prediction across composition space after prediction with model AL2.}

    \label{fig:AL4figs_true_scale}
\end{figure*}

\begin{figure*}[h]
     \centering
     {\includegraphics[width=0.3275\linewidth]{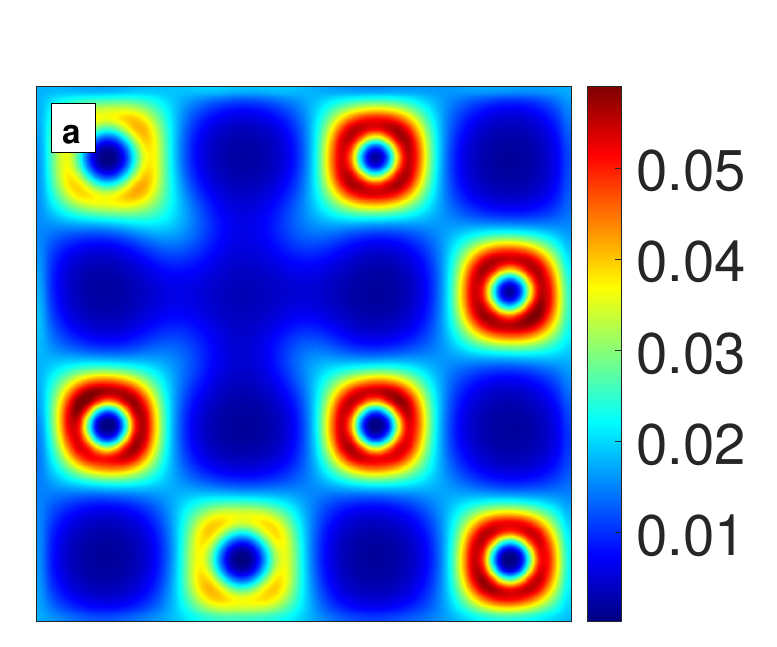}}
     {\includegraphics[width=0.3275\linewidth]{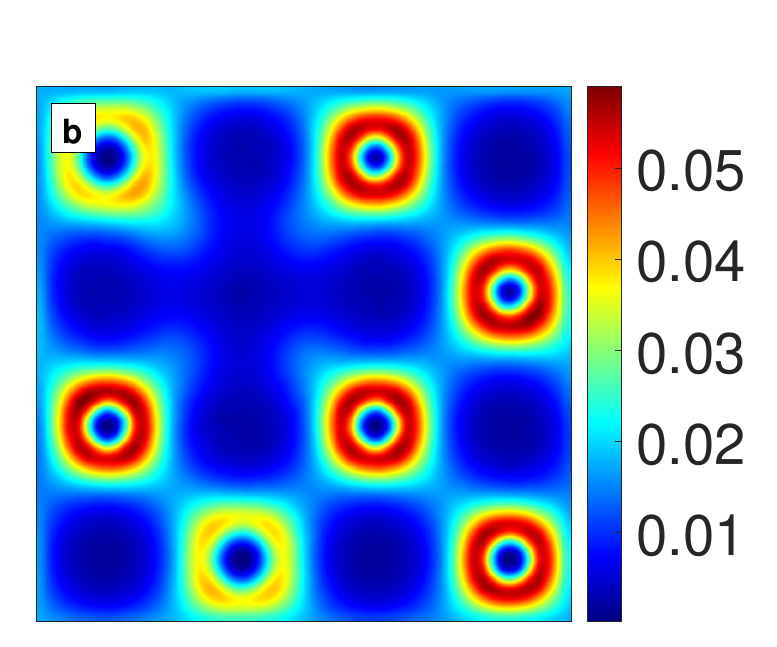}}
     {\includegraphics[width=0.3275\linewidth]{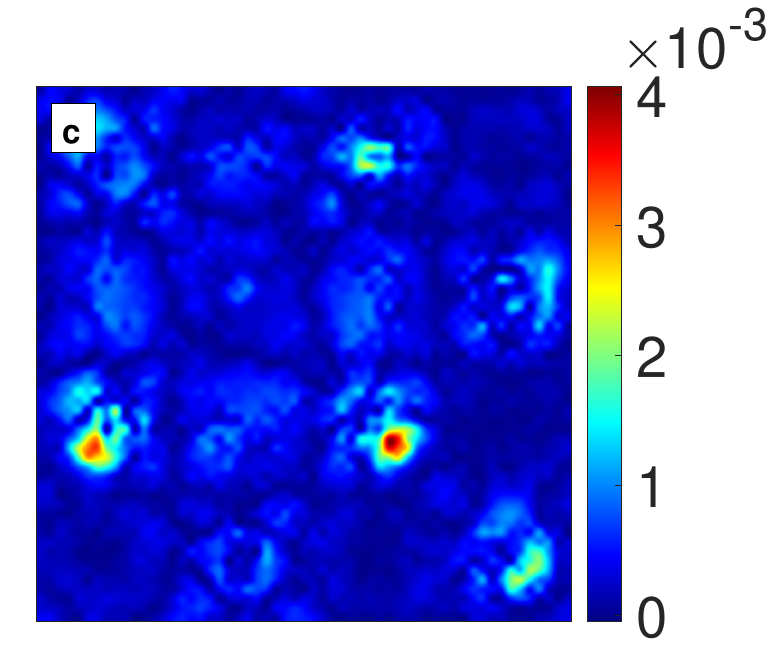}}
     
     \caption{Electron densities (a) calculated from KS-DFT and (b) predicted by ML and the absolute difference between them (c) for a vacancy defect for \ce{SiGeSn}. The snapshot corresponds to 64 atom Si\textsubscript{29.7}Ge\textsubscript{29.7}Sn\textsubscript{40.6} simulation cell at 2400K with an Sn vacancy. The ternary AL2 model was used here.}
     \label{fig:defect_slices}
\end{figure*}

\begin{figure*}[h]
     \centering
     
     {\includegraphics[width=0.85\linewidth]{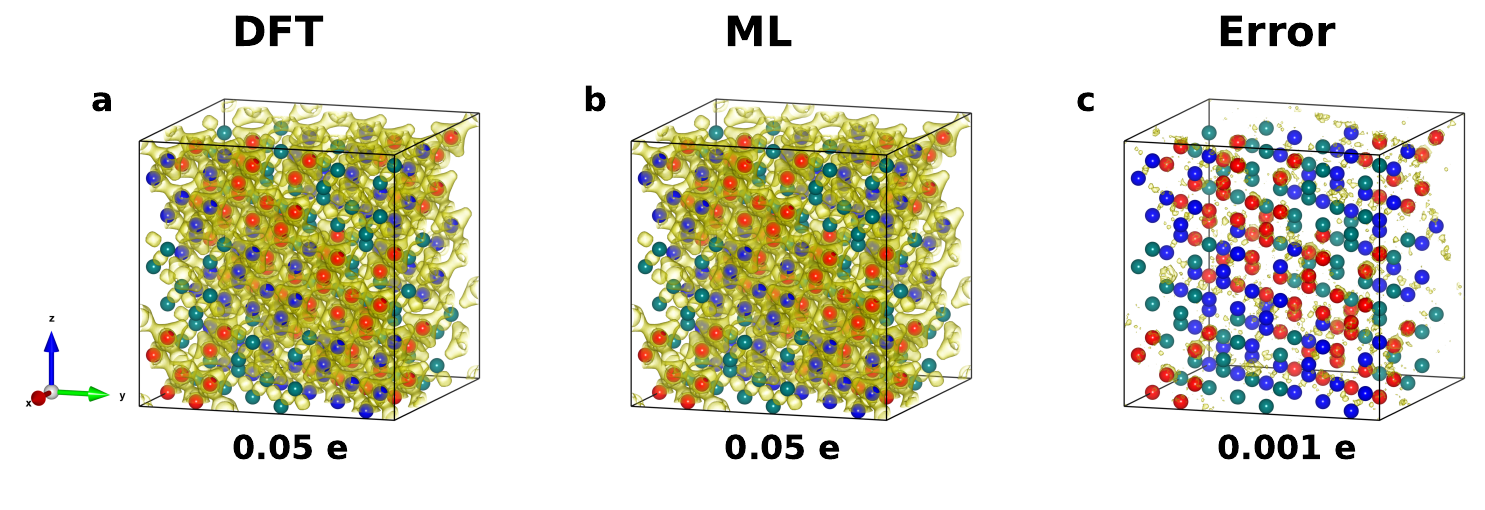}}\vspace{-1.65em}
     {\includegraphics[width=0.85\linewidth]{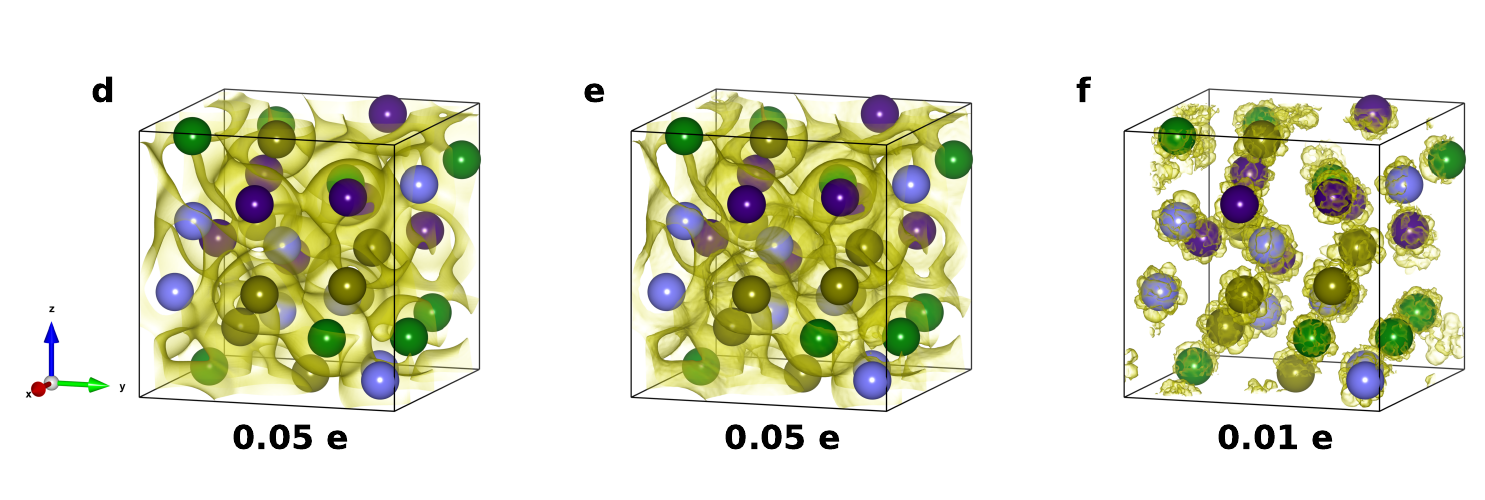}}\vspace{-1.65em}

     \caption{Electron densities (a, d) calculated by  KS-DFT and (b, e) predicted by  ML and the Error (absolute difference) between them (c, f) for
     \ce{SiGeSn} (a, b, c) and CrFeCoNi (d, e, f), using the AL2 model. Subplots (a, b, c) correspond to a 216-atom Si\textsubscript{33.3}Ge\textsubscript{33.3}Sn\textsubscript{33.3} simulation cell at 2400K for the handcrafted systems featuring species segregation.
     Subplots (d, e, f) are 32-atom simulation cells at 5000K corresponding to  Cr\textsubscript{25}Fe\textsubscript{25}Ni\textsubscript{25}Co\textsubscript{25} for the $\rho$ model. The values refer to the iso-surface values.}
     \label{fig:cubeplots_2}
\end{figure*}

\begin{figure*}[h]
     \centering
     {\includegraphics[width=0.85\linewidth]{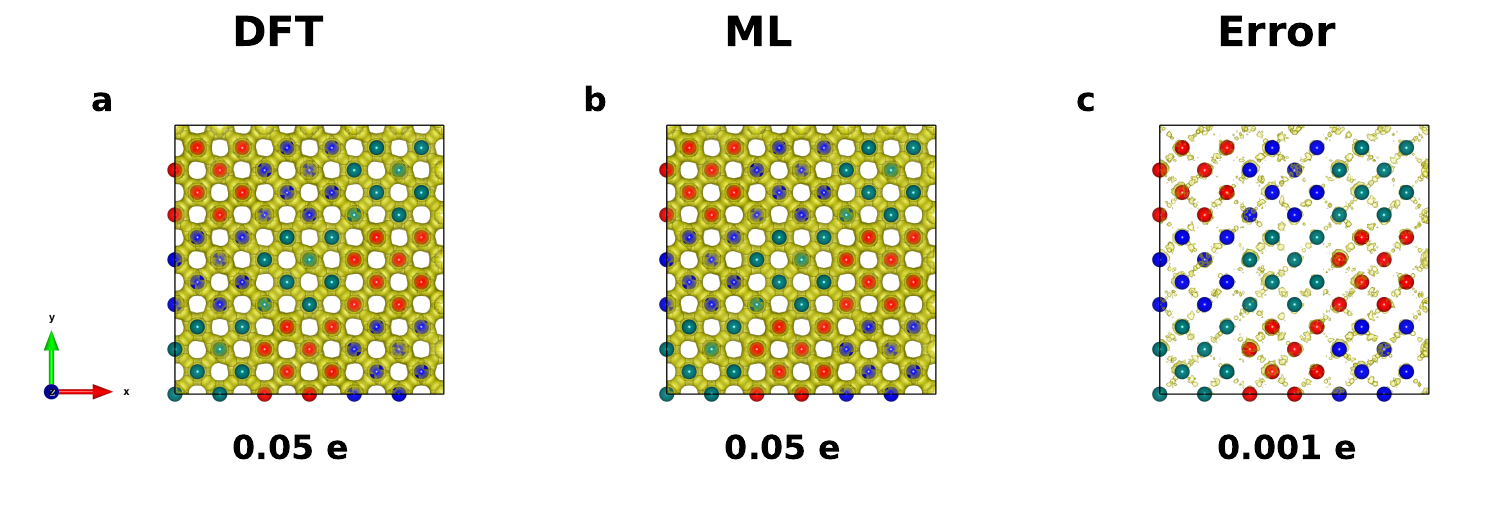}}\vspace{-1.65em}
     \caption{Electron densities (a) calculated by  KS-DFT and (b) predicted by  ML and the Error (absolute difference) between them (c) for
     \ce{SiGeSn} (a, b, c) using the AL2 model. Subplots (a, b, c) correspond to a 216-atom Si\textsubscript{33.3}Ge\textsubscript{33.3}Sn\textsubscript{33.3} simulation cell at 2400K same as Figure \ref{fig:cubeplots_2} for the handcrafted systems featuring species segregation visualized in the xy plane.
     The values refer to the iso-surface values.   
     Blue, red and turquoise spheres represents Si, Ge, and Sn atoms, respectively. }
     \label{fig:cubeplots_3}
\end{figure*}

\begin{figure*}[htbp]
    \centering
    \includegraphics[width=0.95\linewidth]{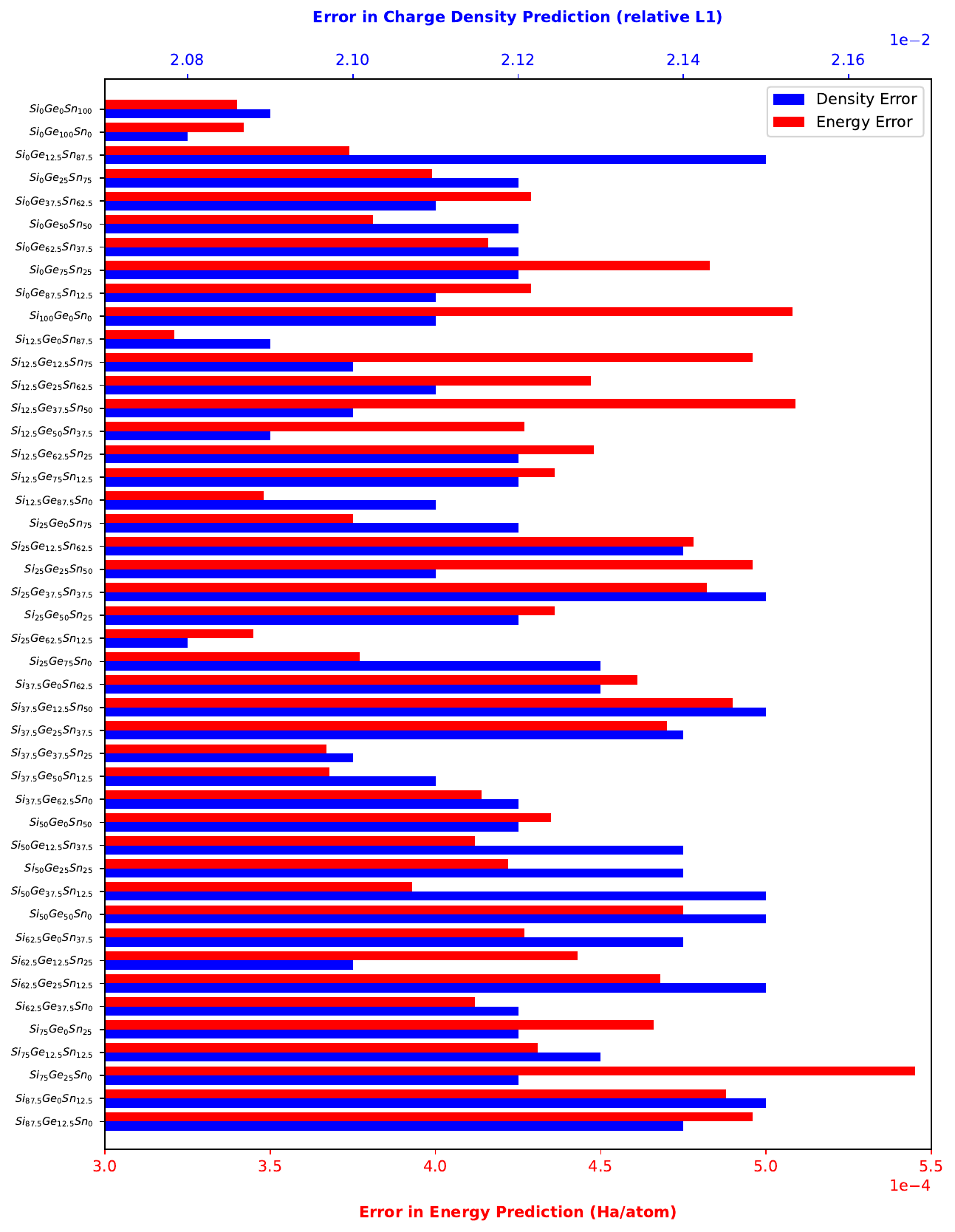}
    \caption{Plot of average energy and density errors, by composition, for the pristine $64$-atom \ce{SiGeSn} test data. The ternary AL2 model was used here. The results are averaged over all snapshots available for each given composition. \mysquare{dens_error}: Density Error, \mysquare{e_error}: Energy Error}
    \label{fig:SiGeSn_DensEnergyError}
\end{figure*}


\begin{figure*}[htbp]
    \centering
    \includegraphics[width=0.95\linewidth]{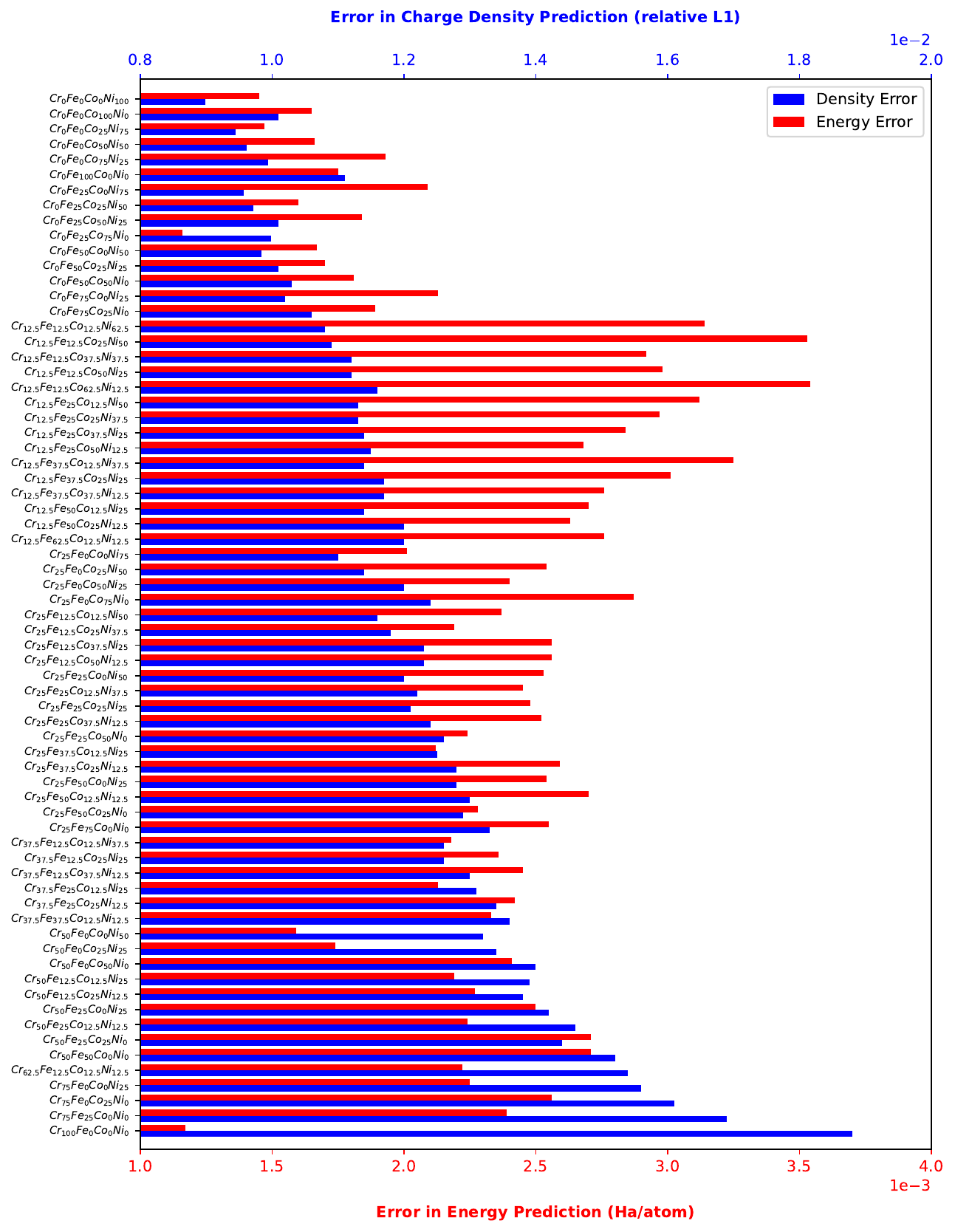}
    \caption{Plot of average energy and density errors, by composition, for the pristine 32-atom \ce{CrFeCoNi} test data (trained on the difference between the charge density field and the atomic densities). The quaternary AL3 model was used here. \mysquare{dens_error}: Density Error, \mysquare{e_error}: Energy Error}
    \label{fig:DensEnergyError_CrFeCoNi}
\end{figure*}

\begin{figure*}[htbp]
    \centering
    
    \includegraphics[width=0.99\linewidth]{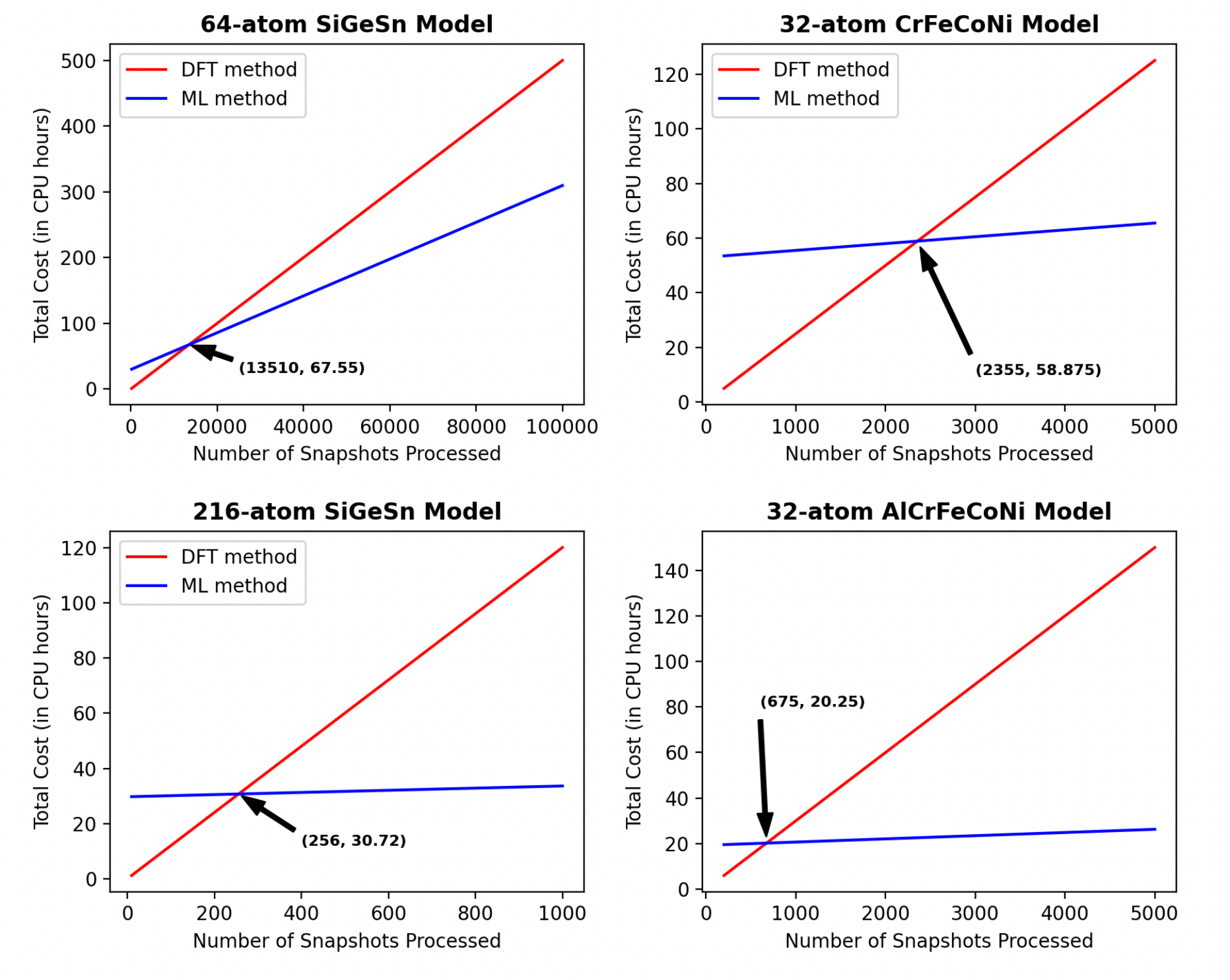}
 
    \caption{{\textbf{Snapshots vs. Total Cost.} These subplots show the crossover point where the ML method becomes less computationally costly compared to the KS-DFT method. The primary cost of the ML method is the upfront cost of training the model. The crossover point arises since the cost of generating a prediction from the model is much cheaper that the cost of performing a full KS-DFT calculation. Each system will have a different crossover point, depending on the expense of a KS-DFT calculation for that system and on the setup choices for the ML model. The crossover points shown here are for four example systems that were considered in this work; the assumptions made to produce these subplots are shown in Table \ref{tab:costDemonstration}. The crossover point for the 64-atom SiGeSn system is higher due to the low cost of the KS-DFT calculation; the number of valence electrons considered was minimal due to the p-block location of the elements. The number of snapshots shown on the x-axis of these subplots is obtained by multiplying the number of compositions by the number of configurational snapshots at each composition. An overview of the number of compositions that would be needed to explore the composition space in different increment sizes is shown in Table \ref{tab:countingUniqueCompositions}. Additionally, for a direct comparison of all four systems, Figure \ref{fig:consolidatedPlot} consolidates these four subplots into a single plot. Note that, for the sake of a fair comparison, all computations relevant to obtaining these costs were carried out  on CPUs (for both  KS-DFT and ML approaches).} \fullblue: ML method, \fullred: DFT method}
    \label{fig:efficiencyGain}
\end{figure*}

\begin{figure*}[htbp]
    \centering
    
    \includegraphics[width=0.99\linewidth]{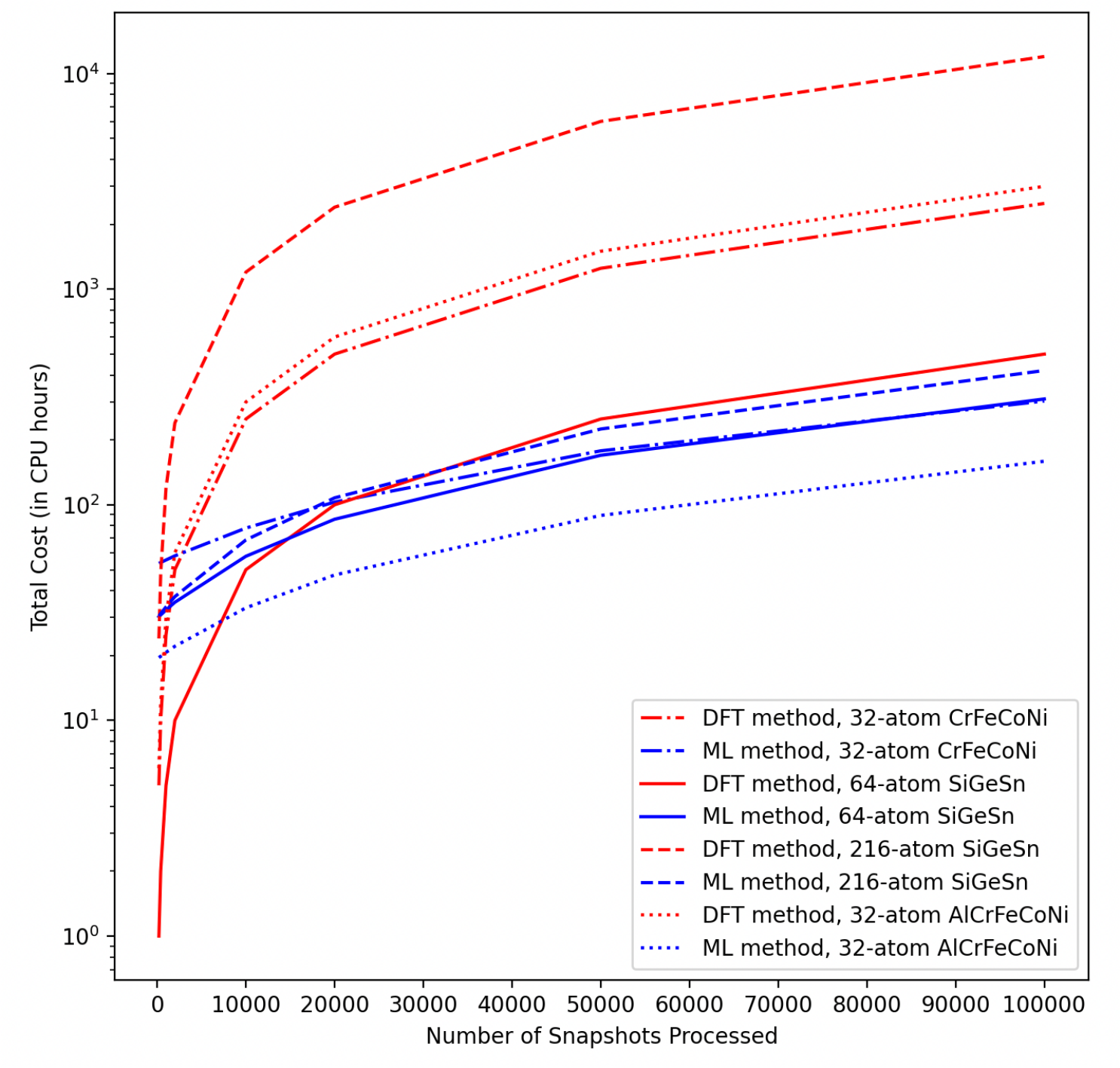}
 
    \caption{{\textbf{Consolidated Plot.} This graph presents the data from Figure \ref{fig:efficiencyGain} into a single logarithmic plot to allow for more direct comparison between systems. Notice that, in each case, the ML costs (shown in blue) have a higher starting value, but they soon overtake the KS-DFT costs (shown in red) in terms of efficiency as the number of snapshots to obtain electron densities for increases. Note that, for the sake of a fair comparison, all computations relevant to obtaining these costs were carried out  on CPUs (for both  KS-DFT and ML approaches).} \chainred: DFT method, 32-atom \ce{CrFeCoNi}, \chainblue: ML method, 32-atom \ce{CrFeCoNi}, \fullred: DFT method, 64-atom \ce{SiGeSn}, \fullblue: ML method, 64-atom \ce{SiGeSn}, \dashedred: DFT method, 216-atom \ce{SiGeSn}, \dashedblue: ML method, 216-atom \ce{SiGeSn}, \dottedred: DFT method, 32-atom \ce{AlCrFeCoNi}, \dottedblue: ML method, 32-atom \ce{AlCrFeCoNi}.}
    \label{fig:consolidatedPlot}
\end{figure*}
\end{document}